\documentclass{aa}  

\usepackage{graphicx}
\usepackage{txfonts}
\usepackage{multirow}
\usepackage{lscape}
\usepackage{pdflscape}
\usepackage{longtable}
\usepackage{wasysym}
\usepackage{float}
\usepackage{amsmath}
\usepackage{relsize}
\usepackage{color}
\usepackage[breaklinks=true]{hyperref} 
\usepackage{float}
\usepackage{breqn}
\definecolor{darkblue}{rgb}{0.0,0.0,0.4}
\definecolor{darkgreen}{rgb}{0.0,0.4,0.0}
\definecolor{violet}{rgb}{0.5,0.,0.5}
\hypersetup{colorlinks,breaklinks, linkcolor=darkblue,urlcolor=darkblue, anchorcolor=darkblue,citecolor=darkblue}

\def\kms{km.s$^{-1}$}         
\def\ms{\hbox{m.s$^{-1}$}}         
\def\cmss{\hbox{cm.s$^{-2}$}}       
\def\kms{\hbox{km.s$^{-1}$}}       
\def\vsini{\hbox{$\upsilon \sin i_{\star}$}}      
\def\Msun{\hbox{$\mathrm{M}_{\astrosun}$}}             
\def\Rsun{\hbox{$\mathrm{R}_{\astrosun}$}}
\def\Mjup{\hbox{$\mathrm{M}_{\jupiter}$}}
\def\Rjup{\hbox{$\mathrm{R}_{\jupiter}$}}
\def\Mearth{\hbox{$\mathrm{M}_{\oplus}$}}
\def\Rearth{\hbox{$\mathrm{R}_{\oplus}$}}
\def\degr{\hbox{$^\circ$}}
\def\teff{T$_{\rm eff}$}
\def\logg{log~{\it g}}
\def\met{[Fe/H]}

\def\figw{\columnwidth}

\begin{document} 

   \title{SOPHIE velocimetry of \textit{Kepler} transit candidates}

   \subtitle{XVII. The physical properties of giant exoplanets within 400 days of period\thanks{Based on observations made with SOPHIE on the 1.93-m telescope at Observatoire de Haute-Provence (CNRS), France}}

   \author{A.~Santerne\inst{\ref{IA}}\fnmsep\inst{\ref{lam}}
          \and C.~Moutou \inst{\ref{lam}}\fnmsep\inst{\ref{cfht}}
          \and M.~Tsantaki\inst{\ref{IA}}
          \and F.~Bouchy \inst{\ref{lam}}\fnmsep\inst{\ref{Geneva}}
          \and G.~H\'ebrard \inst{\ref{iap}}\fnmsep\inst{\ref{ohp}}
          \and V.~Adibekyan\inst{\ref{IA}}
          \and J.-M.~Almenara\inst{\ref{UGrenoble}}\fnmsep\inst{\ref{IPAG}}
          \and L.~Amard\inst{\ref{LUPM}}\fnmsep\inst{\ref{Geneva}}
          \and S.~C.~C.~Barros\inst{\ref{IA}}\fnmsep\inst{\ref{lam}}
          \and I.~Boisse \inst{\ref{lam}}
          \and A.~S.~Bonomo \inst{\ref{torino}}
          \and G.~Bruno \inst{\ref{lam}}
          \and B.~Courcol \inst{\ref{lam}}
          \and M.~Deleuil \inst{\ref{lam}}
          \and O.~Demangeon \inst{\ref{lam}}
          \and R.~F.~D\'iaz\inst{\ref{Geneva}}
          \and T.~Guillot \inst{\ref{oca}}
          \and M.~Havel \inst{\ref{oca}}\fnmsep\inst{\ref{Columbia}}
          \and G.~Montagnier \inst{\ref{iap}}\fnmsep\inst{\ref{ohp}}
          \and A.~S.~Rajpurohit \inst{\ref{lam}}
          \and J.~Rey\inst{\ref{Geneva}}
          \and N.~C.~Santos \inst{\ref{IA}}\fnmsep\inst{\ref{porto}}
          }

   \institute{
   Instituto de Astrof\'isica e Ci\^{e}ncias do Espa\c co, Universidade do Porto, CAUP, Rua das Estrelas, P-4150-762 Porto, Portugal\label{IA} 
	\and Aix Marseille Universit\'e, CNRS, LAM (Laboratoire d'Astrophysique de Marseille) UMR 7326, F-13388, Marseille, France\label{lam}
	\and CNRS, Canada-France-Hawaii Telescope Corporation, 65-1238 Mamalahoa Hwy., Kamuela, HI-96743, USA\label{cfht}
	\and Observatoire Astronomique de l'Universit\'e de Gen\`eve, 51 chemin des Maillettes, CH-1290 Versoix, Switzerland\label{Geneva}
	\and Institut d'Astrophysique de Paris, UMR7095 CNRS, Universit\'e Pierre \& Marie Curie, 98bis boulevard Arago, F-75014 Paris, France\label{iap}
	\and Observatoire de Haute-Provence, Universit\'e d'Aix-Marseille \& CNRS, F-04870 Saint Michel l'Observatoire, France\label{ohp}
  	\and Univ. Grenoble Alpes, IPAG, F-38000 Grenoble, France\label{UGrenoble}
        	\and CNRS, IPAG, F-38000 Grenoble, France\label{IPAG}
	\and LUPM, Universit\'e Montpellier II, CNRS, UMR 5299, Place E. Bataillon, F-34095 Montpellier, France\label{LUPM}
	\and INAF -- Osservatorio Astrofisico di Torino, via Osservatorio 20, I-10025 Pino Torinese, Italy\label{torino}
	\and Laboratoire Lagrange, UMR7239, Universit\'e de Nice Sophia-Antipolis, CNRS, Observatoire de la Cote d'Azur, F-06300 Nice, France \label{oca}
	\and Department of Astronomy, Columbia University, Pupin Physics Laboratory, 550 West, 120th Street, New York, NY 10027, USA \label{Columbia}
	\and Departamento de F\'isica e Astronomia, Faculdade de Ci\^encias, Universidade do Porto, Rua do Campo Alegre, P-4169-007 Porto, Portugal\label{porto}
}	

   \date{Received TBD; accepted TBD}

 
  \abstract
   {While giant extrasolar planets have been studied for more than two decades now, there are still some open questions such as their dominant formation and migration process, as well as their atmospheric evolution in different stellar environments. In this paper, we study a sample of giant transiting exoplanets detected by the \textit{Kepler} telescope with orbital periods up to 400 days. We first defined a sample of 129 giant-planet candidates that we followed up with the SOPHIE spectrograph (OHP, France) in a 6-year radial velocity campaign. This allow us to unveil the nature of these candidates and to measure a false-positive rate of 54.6$\pm$6.5 \% for giant-planet candidates orbiting within 400 days of period. Based on a sample of confirmed or likely planets, we then derive the occurrence rates of giant planets in different ranges of orbital periods. The overall occurrence rate of giant planets within 400 days is 4.6$\pm$0.6\%. We recovered, for the first time in the \textit{Kepler} data, the different populations of giant planets reported by radial velocity surveys. Comparing these rates with other yields, we find that the occurrence rate of giant planets is lower only for hot jupiters but not for the longer period planets. We also derive a first measurement on the occurrence rate of brown dwarfs in the brown-dwarf desert with a value of 0.29$\pm$0.17\%. Finally, we discuss the physical properties of the giant planets in our sample. We confirm that giant planets receiving a moderate irradiation are not inflated but we find that they are in average smaller than predicted by formation and evolution models. In this regime of low-irradiated giant planets, we find a possible correlation between their bulk density and the Iron abundance of the host star, which needs more detections to be confirmed.}


   \maketitle


\section{Introduction}
\label{intro}

Twenty years after the discovery of the first extrasolar giant planet around a main sequence star \citep{1995Natur.378..355M}, not all questions about extrasolar giant planets (EGPs) have been answered. Their formation, migration and evolution are far from being fully understood. As an example, both the well-adopted core -- accretion model \citep[e.g.][]{2009A&A...501.1161M} and the latest results from the disk -- instability model \citep[e.g.][]{2014arXiv1411.5264N, 2015arXiv150207585N} are able to reproduce the observed correlation of giant-planet formation rates with the metallicity of host star \citep{2001A&A...373.1019S}, hence reopening the question about their dominant formation process. Another example is the inflation of some giant, highly irradiated planets that could not be modelled with reasonable physical ingredients \citep[e.g.][]{2015A&A...575A..71A}. Different physical processes are currently proposed to explain their large inflation \citep[see e.g.][for a review]{2014prpl.conf..763B} but this question is still not completely solved. Even the definition of what is a giant planet is still an open question, in both borders: towards the lower mass planets \citep{2015arXiv150605097H} and the brown dwarf regime \citep{2011A&A...532A..79S, 2014prpl.conf..619C}. When the orbital obliquity is put in the picture, it raises even more questions and complexity in the planet formation and evolution \citep{2010ApJ...718L.145W, 2010A&A...516A..95H, 2011A&A...534L...6T, 2014ApJ...790L..31D}. 

At a time when small planets in the habitable zone are found \citep[e.g.][]{2015AJ....150...56J}, the characterisation of EGPs is still of high importance to answer the aforementioned questions. Moreover, their formation process being tightly connected, it is important to understand the formation processes of the large planets before exploring the one of the smallest ones. A lot of constraints about EGPs have already been brought by radial velocity (RV) surveys \citep[e.g.][]{2001A&A...373.1019S, 2003A&A...407..369U, 2010Sci...330..653H, 2011arXiv1109.2497M, 2013A&A...560A..51A, 2013ApJ...767L..24D}. However, these planets do not have radius measurement \citep[except for a few of them, e.g.][]{2009A&A...498L...5M}, which does not allow one to understand their density diversity, nor their atmospheric physical properties. 

The regime of transiting EGPs receiving a moderate or low irradiation is still poorly explored, with only five objects well characterised (mass and radius significantly measured) with orbital periods longer than a month \citep{2014A&A...571A..37S}. These planets rarely seen in transit\footnote{Their transit probability is at the level of 1\% or below.}, paving the way between the hot jupiters and the solar system giants can bring unprecedented constraints to understand the physics of the atmosphere, the formation, and the migration of such planets. In this context, the \textit{Kepler} space telescope \citep{2009Sci...325..709B} has detected giant-planet candidates with orbital periods as long as several hundreds of days (Coughlin et al., in prep.), hence probing this population of low-irradiation planets. 

Giant transiting exoplanets are easily mimicked by false positives \citep[e.g.][]{2003ApJ...593L.125B, 2005ApJ...619..558T, 2012Natur.492...48C}, making difficult the interpretation of the candidates without the establishment of their nature. Spectroscopic follow-up can easily reveal blended multiple stellar systems \citep{2012AA...545A..76S, 2015AJ....149...18K}, and high-resolution imaging \citep[e.g.][]{2014A&A...566A.103L} can unveil close-by companions. However, to firmly establish their planetary nature, one has to detect their Doppler signature or use statistical (also known as planet-validation) methods \citep[see][for an illustration of both methods]{2014A&A...571A..37S}. To correctly interpret the transit detections it is therefore needed to performed follow-up observations, especially for the population of giant exoplanets.

In this paper, we present the result and the interpretation of a 6-year RV campaign with the SOPHIE spectrograph (Observatoire de Haute-Provence, France) of a complete sample of giant transiting candidates detected by \textit{Kepler} within 400 days of orbital periods. This paper completes and extends the work presented in \citet{2012AA...545A..76S}. In Section \ref{sample}, we define the giant-planet candidates sample detected by \textit{Kepler} and selected for our RV follow-up programme. In Section \ref{nature}, we present the performed spectroscopic observations, their analysis and the nature of the candidates that are discussed case by case in the Appendix \ref{SpectroObs}. In Section \ref{FPR}, we computed the false-positive rate of \textit{Kepler} exoplanet giant-planet (EGP) candidates within 400 days and compare it with previous estimations. In Section \ref{rates}, we measure the occurrence rates of EGPs and brown dwarfs (BD) in different ranges of orbital periods that we compare with the values determined in other stellar populations (e.g. the solar neighborhood). In Section \ref{physics}, we discuss some physical properties of these EGPs and the ones of their host stars. Finally, we make a summary of the main results of this paper and draw our conclusions in Section \ref{conclusion}. The spectroscopic data are listed in the appendices \ref{sophiedata} and \ref{tableHD}.


\section{The giant-planet candidates sample}
\label{sample}

To select the EGP candidates, we used the list of \textit{Kepler} objects of interest (KOI) successively published in \citet{2011ApJ...728..117B}, \citet{2011ApJ...736...19B}, \citet{2013ApJS..204...24B}, \citet{2014ApJS..210...19B}, \citet{2015ApJS..217...16R}, \citet{2015arXiv150202038M}, and Coughlin et al. (in prep.). The latest release corresponds to the candidates detected based on the full dataset of the \textit{Kepler} prime mission (from quarter Q1 to Q17). These candidates are listed in the NASA exoplanet archive\footnote{\url{http://exoplanetarchive.ipac.caltech.edu}}, together with their orbital and transit parameters. We used the cumulative KOI table as of 2015-06-05. In this table, there are 8826 KOIs. We first removed all the KOIs that were already identified as false positives using the \textit{Kepler} data. These false positives are mostly background eclipsing binaries (EBs) and background transiting planets that produce an in-transit astrometric signal, called centroid effect \citep{2010ApJ...713L.103B, 2013PASP..125..889B}. Among all the KOIs, 4661 are not obvious false positives and are labelled as planet candidates in the catalog. These candidates have a host star magnitude in the Kepler bandpass (K$_{p}$) ranging from 8.2 to 19.5, with a median of 14.6.

From this list of 4661 candidates, we kept only the 2481 ones that transit a host star with a magnitude K$_{p} <$ 14.7. This was chosen to match the maximum magnitude for which the SOPHIE spectrograph (see Section \ref{nature}) could reach a RV photon noise better than 20 \ms\, for slow-rotating stars, in a maximum of 3600s of exposure time \citep{2013sf2a.conf..555S}. Such precision is the minimum needed to significantly detect the RV signal of a Jupiter-mass planet with orbital periods of up to a few tens of days \citep[e.g.][]{2011A&A...536A..70S,2014A&A...571A..37S}.

To select the candidates that are compatible with an EGP, we kept the KOIs that have a reported transit depth ($\delta$) between 0.4\% and 3\%. Very few EGPs have been found so far with a transit depth below 0.4\% and most of them are transiting evolved stars, e.g. KOI-428 \citep{2011A&A...528A..63S}, WASP-72 \citep{2013A&A...552A..82G}, WASP-73 \citep{2014A&A...563A.143D}. On the other limit, only one EGP has been found with a transit depth greater than 3\%, KOI-254 \citep{2012AJ....143..111J}, which M dwarf host represents a small fraction of the KOIs \citep{2013ApJ...767...95D}. We are therefore confident that those criteria select the majority of the EGPs transiting FGK dwarfs. We did not select the giant-planet candidates based on their estimated radius because this value strongly depends on the 40\% uncertain estimated radius of the host \citep[][]{2014ApJS..211....2H}. The complete selection of candidates based on their estimated radius is therefore uncertain. 
Moreover, the transit depth is a directly measured observable, and as such more reliable than the estimated planetary radius. By selecting candidates based on their transit depth, however, one might have some contamination from low-mass EBs or small planets transiting small stars.

\begin{figure}[h]
\begin{center}
\includegraphics[width=\figw]{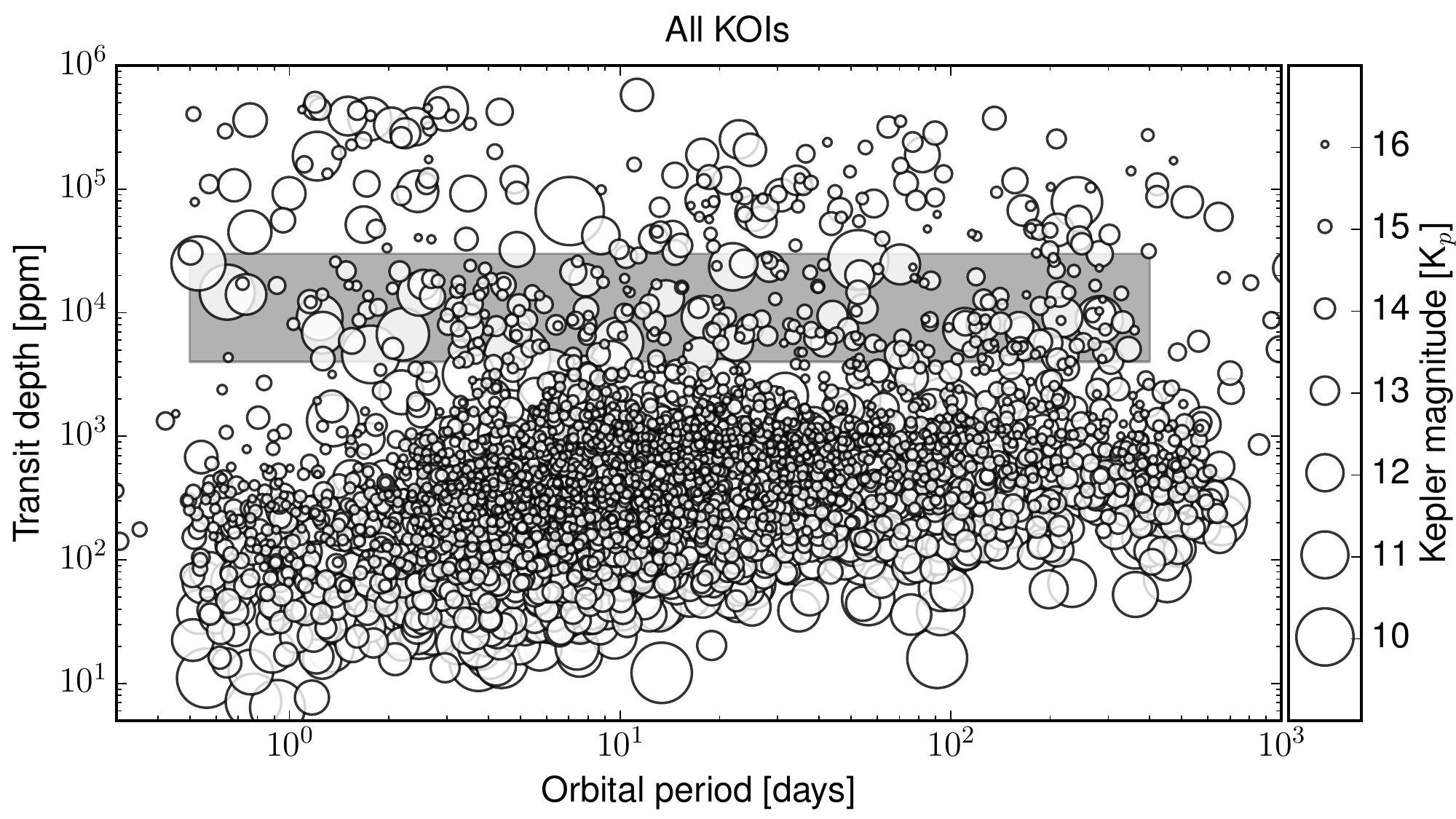}\\
\includegraphics[width=\figw]{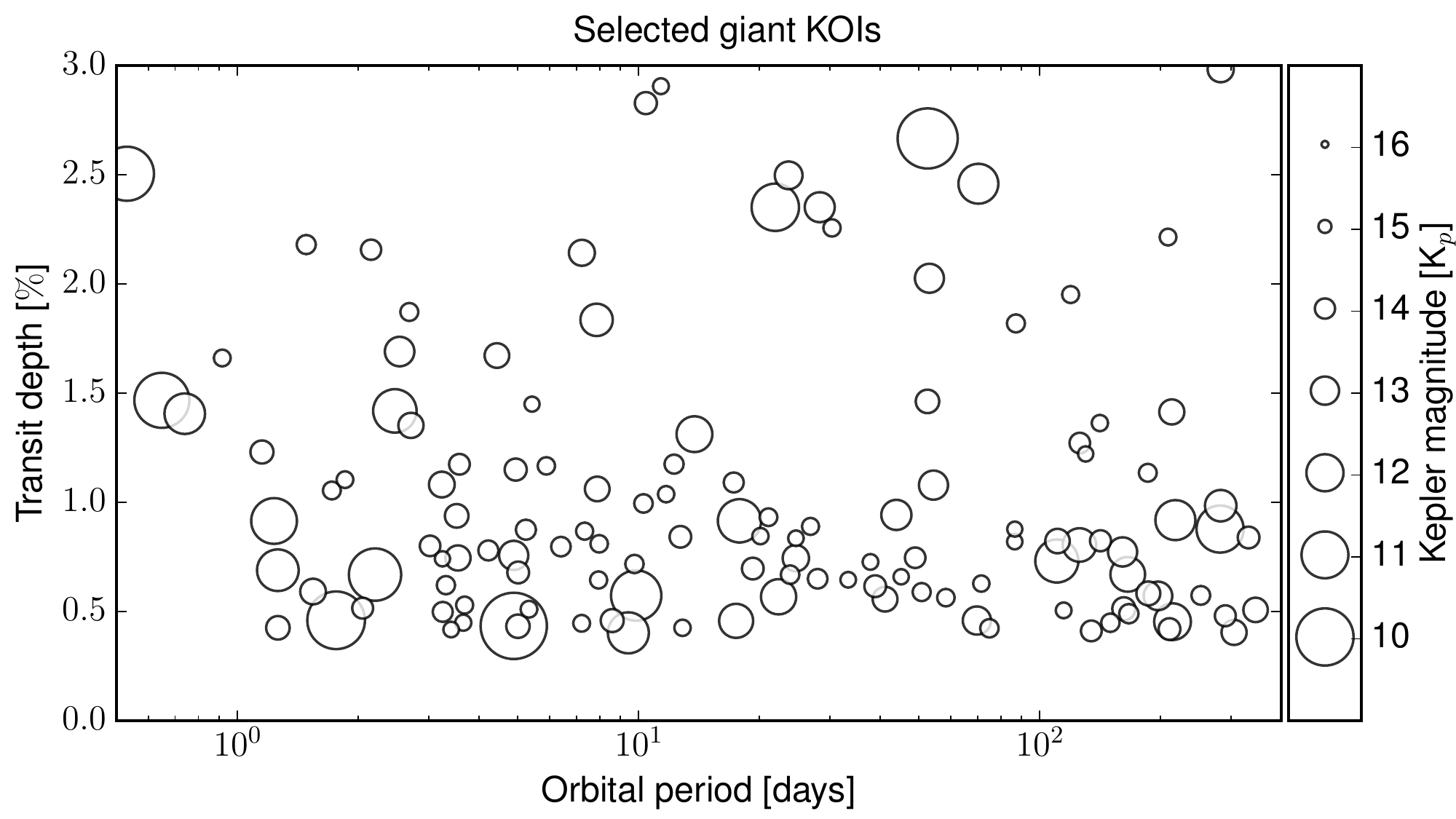}
\caption{Planet candidates detected by the \textit{Kepler} telescope in the Q1 -- Q17 dataset. Their transit depth is displayed here as a function of their orbital period. The size of the marker is relative to the magnitude of the host. The grey region in the upper panel represents the selection criteria used to define the giant-planet candidate sample (see text). The lower panel is a zoom to this selected population of candidates.}
\label{KOIsSelection}
\end{center}
\end{figure}

We finally select among the giant-planet candidates all the ones with an orbital period (P) of less than 400 days. This insures that at least three transits were observed during the entire duration of the \textit{Kepler} prime mission. By applying the three selection criteria (K$_{p} <$ 14.7, 0.4\% $< \delta <$ 3\%, and P $<$ 400 d), we find 129 KOIs on 125 target stars. They are displayed in Fig. \ref{KOIsSelection} and listed in Table \ref{BigTable}, together with their various ID, their main orbital and transit parameters (period, depth, and scaled distance to star), as well as their host properties as determined by \citet{2014ApJS..211....2H}.

\citet{2012AA...545A..76S} used the same criteria in terms of magnitude limits and transit depths, but the candidate periods were limited to 25 days. The new sample extends the sample for periods up to 400 days and contains three times more candidates than the previous study.

\section{Unveiling the nature of the candidates}
\label{nature}

\subsection{SOPHIE observations and reduction}

We observed the candidate sample with the SOPHIE spectrograph \citep{2009A&A...505..853B} mounted on the 1.93 m telescope at the Observatoire de Haute-Provence (France). SOPHIE is a fibre-fed high-resolution stable spectrograph dedicated to high-precision radial velocity (RV) measurements \citep{2008SPIE.7014E..17P,2009A&A...505..853B, 2013A&A...549A..49B}. The observations were done as part of a large programme dedicated to \textit{Kepler} targets and funded by the French Program of Planetology\footnote{programme IDs: 10A.PNP.CONS, 10B.PNP.MOUT, 11A.PNP.MOUT, 11B.PNP.MOUT, 12A.PNP.MOUT, 12B.PNP.MOUT, 13A.PNP.MOUT, 13B.PNP.HEBR, 14A.PNP.HEBR, 14B.PNP.HEBR, 15A.PNP.HEBR.} from 2010-07-14 to 2015-07-15. 

During these six observing campaigns, this programme collected more than 1000 spectra on 154 different targets, spread over more than 370 night, cumulating more than 640 hours of open-shutter time. Each target was observed between two and 51 different epochs, with a typical precision of about 20 \ms.

Most observations were performed using the high-efficiency (HE) mode of SOPHIE with an instrumental resolution of $\sim$ 39,000. For a few targets brighter than K$_{p} = 12$, we observed them using the high-resolution mode (HR), which has an instrumental resolution of $\sim$ 75,000 and a better light scrambling \citep{2011SPIE.8151E..37P}, providing a better precision. All spectra were reduced using the online pipeline. We computed the weighted cross-correlation function (CCF) using a G2 mask \citep{1996A&AS..119..373B,2002A&A...388..632P}. This mask has been optimised for solar-type stars which is the main population observed by \textit{Kepler}.

When necessary, we corrected the CCFs affected by the Moon background light following the procedure described in \citet{1996A&AS..119..373B}. We then measured the RV, bisector span and full width half maximum (FWHM). All the measurements are reported in Tables \ref{RVsingletable}, \ref{RVSB2table}, and \ref{RVSB3table} and analysed in the Appendix \ref{SpectroObs}. The errors on the RV are estimated using the method explained in \citet{2001A&A...374..733B} and in the appendix A of \citet{2010A&A...523A..88B}. For the bisector and FWHM, we used the photon noise factors listed in \citet{2015MNRAS.451.2337S}. These spectroscopic diagnostics are used to reveal the presence of contaminating stars, therefore likely false positives that might be the source of the transit event \citep{2002A&A...392..215S, 2005ApJ...619..558T}. Several stars presenting a $\sim$ 100\ms\ scatter in FWHM, including the RV constant star HD185144 \citep{2014A&A...571A..37S}, we concluded it is due to the insufficient thermal control of the instrument which introduces slight changes in focus \citep{2015arXiv150607144C}. For this reason, we used the FWHM as a vetting tool only if the variation is much larger than 100 \ms.

We corrected the RV from the CCD charge transfer inefficiency \citep{2009EAS....37..247B} using the calibration described in \citet{2012AA...545A..76S}. Following \citet{2014A&A...571A..37S}, we also correct instrumental drifts in the RV using the ones measured on the constant star HD185144 on the same nights. The RV we used for this correction are listed in Table \ref{table185144}. This allowed us to reach a $rms$ down to 13 \ms\, over more than two years, on stars as faint as K$_{p}$ = 14.5, which is equivalent to the photon noise.

\subsection{Stellar atmospheric analyses}

\subsubsection{Stellar atmospheric parameters}

To support the determination of the nature of the candidates showing no significant RV variation (within 3-$\sigma$, see section \ref{NoVar}), we performed a detailed spectral analysis of the targets\footnote{The spectral analysis of \textit{bona-fide} exoplanets are presented in dedicated papers.}. This allowed us to improve the upper limits on the candidate mass and to identify evolved stars, that are hosts of false positives. Some spectra have a signal-to-noise ratio (S/N) too low for a detailed spectral analysis. Among the 125 candidates hosts, we selected 12 stars with no significant RV variation and a S/N high enough to analyse their SOPHIE spectra. We derived the atmospheric parameters of those 12 stars after correcting for their RV shifts and the cosmic-ray impacts. We subtracted the sky contamination (using the spectra of fiber B) from the target spectra (in fiber A), after correcting for the relative efficiency of the two fibers. To derive the atmospheric parameters, namely the effective temperature (\teff), surface gravity (\logg), metallicity (\met), and microturbulence ($\xi_{t}$), we followed the methodology described in  \citet{2008A&A...487..373S} and \citet{2013A&A...555A.150T}. This method relies on the measurement of the equivalent widths (EWs) of \ion{Fe}{I} and \ion{Fe}{II} lines and by imposing excitation and ionization equilibrium.

The analysis was performed assuming local thermodynamic equilibrium using a grid of model atmospheres \citep{1993KurCD..13.....K} and the radiative transfer code \texttt{MOOG} \citep{1973PhDT.......180S}. The iron lines lists for this analysis were taken from \citet{2008A&A...487..373S} for the hotter stars ($>$5200\,K) and from \cite{2013A&A...555A.150T} for the cooler ones. The EWs were measured automatically with the \texttt{ARES~2.0} code \citep{2015A&A...577A..67S}. To ensure accurate measurements of the EWs, we excluded any lines with errors larger than 20\% of their absolute values. We corrected the observed \logg\, using the asteroseismic calibration of \citet{2014A&A...572A..95M}. The derived parameters are reported in Table~\ref{SpectroResults} and discussed case by case in the Appendix \ref{SpectroObs}. We finally updated the stellar fundamental parameters using the Dartmouth stellar evolution tracks of \citet{2008ApJS..178...89D}. 

In Table \ref{SpectroResults}, we also list the spectroscopic parameters of 25 planet hosts derived by our team and published in previous papers \citep[e.g.][]{2015A&A...575A..71A,2015A&A...575A..85B}. These stellar parameters were derived using either the MOOG (as described above) or the VWA software. Comparison between the two on some targets have shown no significant differences \citep[e.g.][]{2014A&A...571A..37S}. These stellar parameters are also available in SWEET-Cat\footnote{The SWEET-Cat is available at: \url{http://www.astro.up.pt/resources/sweet-cat/}} \citep{2013A&A...556A.150S}. For the other candidates or planet hosts, we used the spectroscopic parameters found in the literature \citep[e.g.][]{2014ApJS..211....2H}. 
For some targets we used an ESPaDOnS\footnote{CFHT programme 12BF24 (PI: Deleuil)} \citep{2015A&A...575A..85B} or HARPS-N\footnote{OPTICON programme IDs: OPT12B\_13, OPT13A\_8, OPT13B\_30 (PI: H\'ebrard) ; TNG programme IDs: A28DD2 (PI: Santerne)} co-added  spectrum \citep{2014A&A...572A..93H}. In total, we thus have 37 stars from our sample for which we could derive precise parameters from a spectroscopic analysis.

We determined the \vsini\, of the single-line spectra using the average width of the SOPHIE CCF and the relations in the Appendix B of \citet{2010A&A...523A..88B}. We estimated the (B-V) of the host stars based on their atmospheric parameters reported by \citet{2014ApJS..211....2H} and the calibration from \citet{2013ApJS..208....9P}. We did not use the observed (B-V) because it is affected by unknown interstellar extinction, which would introduce systematic noise. The method of \citet{2010A&A...523A..88B} finds an uncertainty of 1~\kms\, that we conservatively increased by 20\% to account for the errors in the \teff\, and in the (B-V) calibration. For fast rotating stars (\vsini\, $\gtrsim$ 10 \kms), we fitted the CCF with a rotation profile as described in \citet{2012A&A...544L..12S} to determine their \vsini. We list their measured values and uncertainties in the Table \ref{vsini}.

\subsubsection{Comparison with \citet{2014ApJS..211....2H}}
\label{CompHuber}

\begin{figure*}[h!]
\begin{center}
\includegraphics[width=0.9\textwidth]{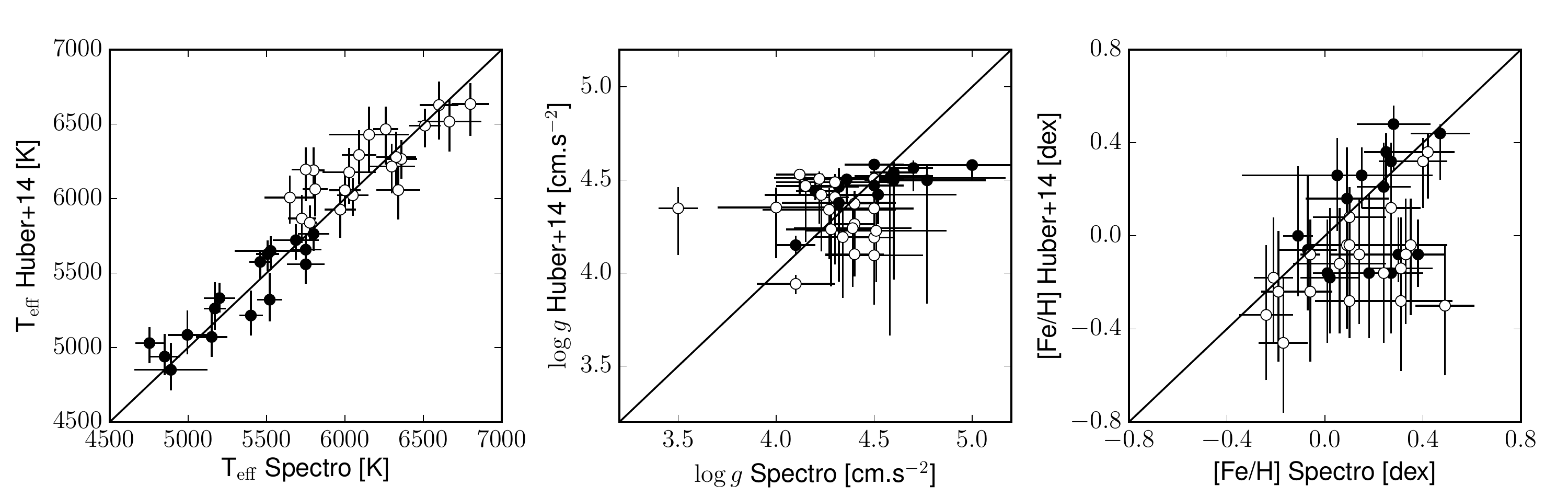}
\caption{Comparison of the \teff, \logg, and \met\, we derived by spectroscopy with the ones derived photometrically by \citet{2014ApJS..211....2H} for the 37 targets listed in Table \ref{SpectroResults}. Open and filled circles are for stars hotter and cooler than the Sun (respectively). The \logg\, of the giant host KOI-5976 is not displayed here for the clarity of the plot.}
\label{ComparSpectro}
\end{center}
\end{figure*}

We compared the results from the spectral analyses we performed in the context of this spectroscopic follow-up of \textit{Kepler} giant-planet candidates with the ones of \citet{2014ApJS..211....2H}, derived based on color photometry. In Fig. \ref{ComparSpectro}, we compare the \teff, \logg, and \met\, of the 37 stars, derived by spectroscopy with the ones independently reported by \citet{2014ApJS..211....2H}. 

We find an agreement between the spectroscopic and photometric \teff\, with a systematic offset of $\Delta$\teff\, = \teff$^{\rm Spectro}$ $-$ \teff$^{\rm Huber+14} = - 51$  $\pm$ 29\footnote{The values and errors reported in this paragraph correspond to the mean and its uncertainty computed as $\sigma/\sqrt{N-k_{f}}$, with $\sigma$ the standard deviation, $N$ the number of points, and $k_{f}$ the number of free parameters \citep{2001ApJ...549....1G}. Here, $k_{f} = 1$.} K. The \logg\, values are very noisy and no systematic offset is found, with $\Delta$\logg\, = \logg$^{\rm Spectro}$ $-$ \logg$^{\rm Huber+14}$ = -0.01 $\pm$ 0.04 \cmss. However, for the \met\, some stars seems to have a lower photometric metallicity compared with the spectroscopic one. The systematic offset is $\Delta$\met\, = \met$^{\rm Spectro}$ $-$ \met$^{\rm Huber+14}$ = 0.17 $\pm$ 0.04 dex. This systematic offset perfectly agrees with the value found by the massive low-resolution spectroscopic survey of the \textit{Kepler} fields performed with LAMOST \citep{2014ApJ...789L...3D}. We find that this offset in the stellar metallicity seems to depend on the stellar effective temperature. If we divided our list of targets in two subsamples (see Fig. \ref{ComparSpectro}), one for stars cooler or hotter than the Sun, this systematic offset is $\Delta$\met$^{cool}$ = 0.09 $\pm$ 0.06 dex and $\Delta$\met$^{hot}$ = 0.22 $\pm$ 0.05 dex. Stellar rotation, higher for the hot stars, might be one of the reasons of this discrepancy. 

This \teff\, -- \met\, trend might also be an artifact of the spectroscopic method, either used for our analyses or to calibrate the photometric values in \citet{2014ApJS..211....2H}. \citet{2012ApJ...757..161T} already pointed out some systematic effects in the determination of the spectroscopic parameters, especially for stars hotter than 6000 K. We expect these systematics to be particularly strong at relatively low S/N (typically $<$ 50), which is the regime of S/N for the spectra of most of the \textit{Kepler} targets. 

Note that \citet{2015AJ....149...14W} proposed a correction of the metallicities from the \textit{Kepler} input catalog \citep{2011AJ....142..112B} using the spectroscopic data from \citet{2014Natur.509..593B}. Since \citet{2014ApJS..211....2H} used the same spectroscopic data to calibrate their metallicities, this correction is no longer valid.

\subsection{Nature of the candidates}
\label{NatCand}

\begin{figure}
\begin{center}
\includegraphics[width=\figw]{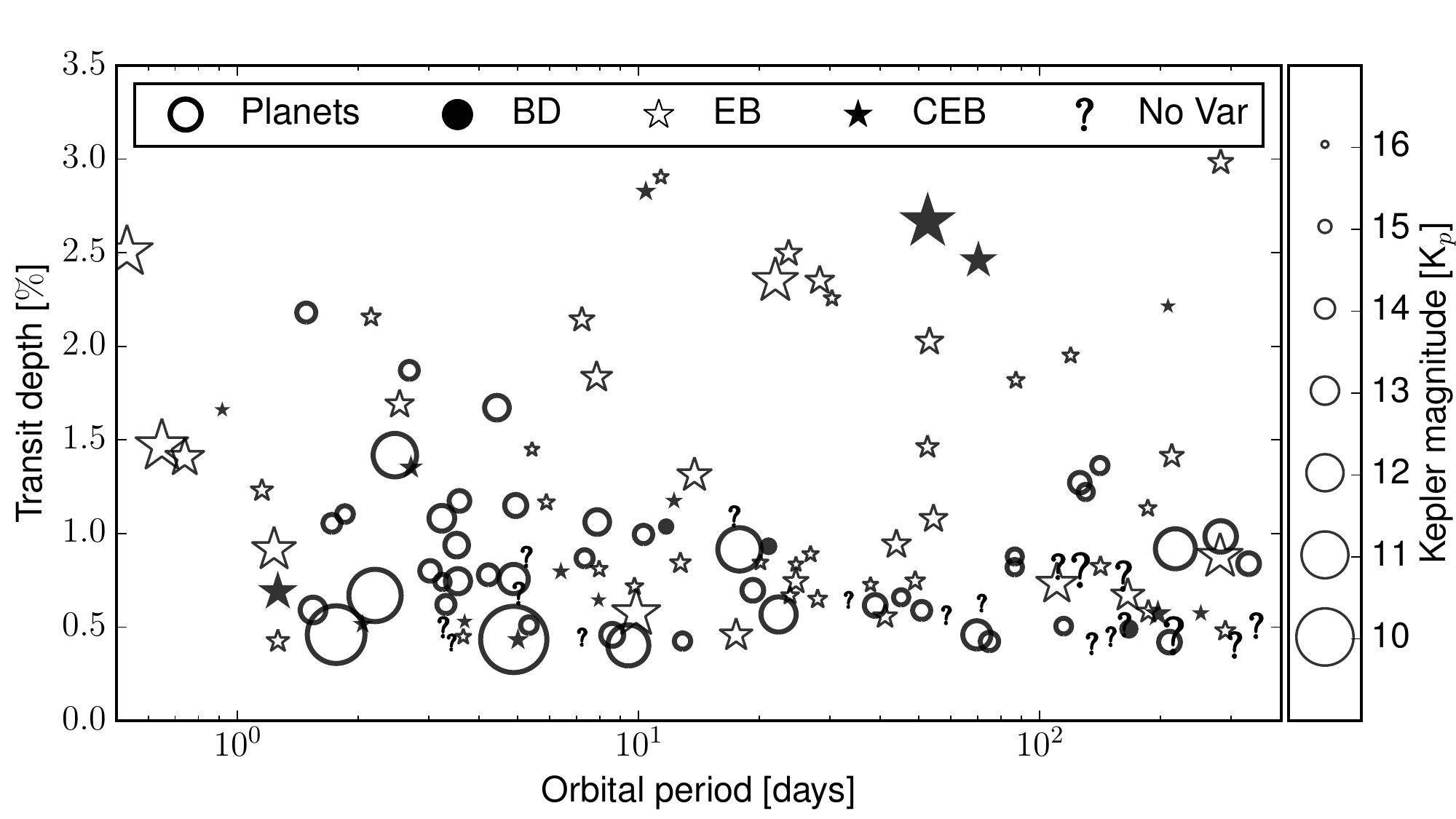}
\caption{Same as lower panel of Fig. \ref{KOIsSelection} but the marks indicate the nature of the candidates: BD stands for brown dwarfs, EB for eclipsing binaries, CEB for contaminating binaries and No Var for the unsolved cases that show no significant variation in radial velocity.}
\label{NatKOIs}
\end{center}
\end{figure}

The nature of the candidates is unveiled in the case by case in the Appendix \ref{SpectroObs}, reported in Table \ref{BigTable} and displayed in Fig. \ref{NatKOIs}. We present below a summary of the different populations of candidates.

\subsubsection{Bona-fide planets and brown dwarfs}
\label{planets}

In the sample of 129 giant-planet candidates, 30 of them are bona-fide planets already established and characterised by other spectroscopic facilities \citep[e.g.][]{2010ApJ...713L.140L,2011ApJS..197...13E, 2013A&A...557A..74G}, by transit timing variation analyses \citep[e.g.][]{2014A&A...561A.103O}, and by the ``multiplicity-boost'' validation \citep{2014ApJ...784...45R}. Except for a few cases, we did not observe them with SOPHIE, relying on the candidate nature that has been secured in the respective papers. In this sample, our team established and characterised 18 EGPs and brown dwarfs that were already published in previous papers of this series.

The distinction between EGPs and brown dwarfs has been widely discussed \citep[e.g.][]{2014prpl.conf..619C}, and remains uncertain except if we would know the formation history of these objects. With a mass of $\sim$~18~\Mjup, Kepler-39~b \citep{2011A&A...533A..83B} is somewhat arbitrarily considered as a brown dwarf. Considering it as a planet would not change significantly the results of this paper, except for Section \ref{OccBDs}. The two other massive substellar companions in our sample \citep[KOI-205~b ($\sim$~40~\Mjup) and KOI-415~b ($\sim$~62~\Mjup):][]{2013A&A...551L...9D,2013A&A...558L...6M} are very likely to be brown dwarfs. Finally, the case of the 78~\Mjup-companion to KOI-189 has been classified as a very-low-mass star by \citet{2014A&A...572A.109D}. This leads to a total number of bona-fide EGPs in our candidate sample of 45 and 3 brown dwarfs. All the references are provided in Table \ref{BigTable}. 

\subsubsection{Eclipsing binaries and contaminating eclipsing binaries}
\label{EBs}

Among the 129 candidates, we detected 63 EBs showing up to three different sets of lines in the spectra. The spectroscopic observations, analyses and conclusions are described in the Appendix \ref{SpectroObs} or in \citet{2012AA...545A..76S}. When 2 or 3 sets of lines is detected in the spectra, we fitted the cross-correlation function with two or three Gaussian profiles. For these cases, we estimated the RV photon noise using the following equation:
\begin{equation}
\sigma_{RV}\, [\kms] = A_{i} \times \frac{\sqrt{FWHM\, [\kms]}}{CTRS [\%] \times S/N},
\end{equation}
with A$_{\rm HE}$ = 3.4 and A$_{\rm HR}$ = 1.7 for both instrumental modes of SOPHIE. The $S/N$ is the signal-to-noise ratio per pixel computed by averaging the flux in the 200 pixels at the center of the spectral order \#26 (i.e. at about 550nm) and CTRS is the contrast of the averaged line profile. This photon noise estimate has been calibrated on a set of standard stars, following the same procedure as described in \citet{2005A&A...431.1105B}.

Among those 63 EBs, 48 are spectroscopic binaries showing one or two set(s) of lines (hence an SB1 or SB2). In most cases, we observed them only two or three times, which is not enough to fully characterise the mass and eccentricity of these binaries. To estimate the companion mass of an SB1, we assumed a circular orbit at the transit ephemeris and no significant RV drift. Several caveats in our analyses might significantly change the reported companion masses. First, the circular orbit assumption is not reasonable for binaries with an orbital period longer than about 10 days \citep{2003A&A...397..159H, 2010ApJS..190....1R}. Second, the primary mass estimate from \citet{2014ApJS..211....2H} that we used might be affected by the presence of a stellar companion. Finally, if the orbital periods of these binary are twice the ones detected by \textit{Kepler}, the reported masses are also wrong.

For SB2 binaries, we used the slope of their RV correlation to measure the binary mass ratio \citep{1941ApJ....93...29W}. As for the SB1, we observed most of them only very few times which limits the possibility of determining their mass and eccentricity. 

These spectroscopic binaries are stars eclipsing the target. Their eclipse depth is likely not diluted by a substantial third light, otherwise, we would have detected it in our spectroscopic data. They are able to mimic a giant-planet candidate because they have a grazing eclipse with a depth compatible with the one of an EGP. A few binaries are stars with an EGP-like radius, which identification is impossible from the light curve only, unless they present a deep secondary eclipse \citep[as in][]{2013ApJ...776L..35Z} or a large beaming, ellipsoidal, or reflection effect.

In this sample of EBs, we detected 16 eccentric systems \citep[2 already characterised in][and 14 new ones described in the Appendix \ref{SpectroObs}]{2012AA...545A..76S} that present only a secondary eclipse, the primary eclipse invisible from Earth \citep{2013A&A...557A.139S}. Two other candidates are secondary-only EBs in more complex multiple stellar systems. These numbers are fully compatible with the predictions of \citet{2013A&A...557A.139S}.

We also found 15 stellar systems that either present three stellar components in the spectra, or SB2 with RV that are not anti-correlated, revealing the presence of a third, unseen star in the system. Those candidates, most likely triple systems, have an eclipse depth severely diluted by the target star. In these cases, even a relatively deep eclipse might mimic the transit depth of a planet. Moreover, if the EB is eccentric, only the primary or secondary eclipse could be visible. Triple systems might be difficult to identify by spectroscopy because the brightest star in the system is not the eclipsed star. Moreover, if the eclipsing system is physically bound with the target star, they are most likely blended in both photometry and spectroscopy. Using the variation of line-profile \citep[the bisector and the FWHM, see][]{2015MNRAS.451.2337S}, we identified some triple systems with relatively faint companions compared with the target star. However, if the eclipse host contributes to less than about 5\% of the total flux of the system (magnitude difference more than 3, or mass ratio smaller than $\sim$0.5), we would not be able to detect the second set of lines in the cross-correlation functions, nor its impact on the target line-profile shape. If such systems are present in our sample, we would not be able to identify them as false positives. Therefore, the actual number of diluted EBs might exceed what we found.

In the Table \ref{BigTable} and in the rest of the paper, we will refer to ``eclipsing binaries'' (or EB) the 48 systems with an undiluted eclipse depth. We will also refer to ``contaminating eclipsing binaries'' (or CEB) the 15 ones with a diluted eclipse depth, which are either triple systems or background EBs.

Note that, among the 63 EBs we detected, 54 are already included in the \textit{Kepler} EB catalog (v3)\footnote{The \textit{Kepler} EB catalog is available at: \url{http://keplerebs.villanova.edu}} of Kirk et al. (in prep). The other 9 are unveiled by our observations and were not previously identified as such based on the \textit{Kepler} light curve. In this catalog, we found two candidates listed as EB but our observations do not support this statement. In previous versions of the catalog, some bona-fide exoplanets were also listed \citep[as discussed already in][]{2012AA...545A..76S}.

\subsubsection{No variation cases}
\label{NoVar}

For 18 giant-planet candidates, we found no significant RV, bisector, nor FWHM variation. The nature of these candidates remains uncertain: they might be planets that have too low a mass for our RV precision or they might be diluted EBs with a large flux ratio between the eclipse host and the target star, which make them undetectable in our spectroscopic data.

Assuming these candidates are planets, we derived their upper-limits in mass. For that, we analysed the data with the MCMC algorithm of the \texttt{PASTIS} software \citep{2014MNRAS.441..983D}. We used an uniform prior for the RV amplitudes (between 0 and 100 \kms), for the systemic RV (between -100 \kms\, and +100 \kms), and for the argument of periastron (between 0 \degr\, and 360 \degr). For the eccentricity, we used a Beta distribution as prior as recommended by \citet{2013MNRAS.434L..51K}. We fixed the periods and epochs of transit to the ones found by \textit{Kepler}. When only two or three different observed epochs were available, we fixed the eccentricity to zero. When enough RV were available, we fitted a Keplerian orbit. If sub-giant planet candidates were detected in the light-curve of the same system, we also included them in the model, even if their RV contribution is expected to be negligible. The choice of the model (circular vs eccentric) as well as the number of planets are described in the Appendix \ref{SpectroObs} together with the derived upper-limits. We report these upper-limits on the mass of the candidates, assuming they are planets, in the Table \ref{upperlims}.

Among the 18 unsolved cases, one has a mass constraint which is still compatible with a brown dwarf (KOI-2679.01) and another one has a mass constraint compatible with a low-mass star (KOI-3783.01). Those two cases are giant-planet candidates transiting fast rotating stars for which precise RV measurements are difficult to obtain.

\subsubsection{Particular cases}
\label{PC}
 
Some candidates we observed have masses that were already constrained by spectroscopy or TTV analysis. Our mass constraints are fully compatible except in two cases. The first case is KOI-1353.01. Assuming a circular orbit, we find a planet mass of 1.55 $\pm$ 0.34 \Mjup\, while \citet{2014ApJ...795..167S} reported a mass of 0.42 $\pm$ 0.05 \Mjup\, for the same planet based on a TTV analysis. Our mass constrain is therefore significantly higher (at the 3.3-$\sigma$) than the one found by TTVs. At least three reasons could explain this discrepancy: first, we find a host which is also more massive at the 3.7-$\sigma$ level, second, the star is active which might have impacted significantly our RV or the transit times \citep{2013MNRAS.430.3032B, 2013A&A...556A..19O}, and finally this planet might be significantly eccentric even if a low eccentricity has been reported by \citet{2014ApJ...795..167S}. For this case, more data and a better precision are needed to firmly conclude.

The second case is KOI-372.01 for which the mass was recently reported in \citet{2015arXiv150404625M} based on RV observations with the CAFE spectrograph. They found a RV amplitude of 132 $\pm$ 6 \ms\, while our SOPHIE HR RV show no significant variation with a $rms$ of 24 \ms. The analysis of the SOPHIE spectra and their comparison with the CAFE observations will be presented in a forthcoming paper (Demangeon et al., in prep.). We considered this case as unsolved.

Lastly, KOI-3663~b / Kepler-86~b, previously validated statistically by \citet{2013ApJ...776...10W}, reveals some line-profile variations correlated with the RV data (See appendix \ref{3663}). More observations are needed to conclude on this case, and KOI-3663~b might not be a planet but a triple system. Without further evidence, we  consider it as a planet in the rest of this article.




\section{The false-positive rate}
\label{FPR}

Based on the results of our spectroscopic survey, we can measure the false-positive rate of the \textit{Kepler} giant-planet candidates, an extension of the previous rate 34.8 $\pm$ 6.5 \% measured by \citet{2012AA...545A..76S} for EGPs within 25 days orbital period.

\subsection{The giant-planet false-positive rate}

\begin{figure*}[h!]
\begin{center}
\sidecaption
\includegraphics[width=12cm]{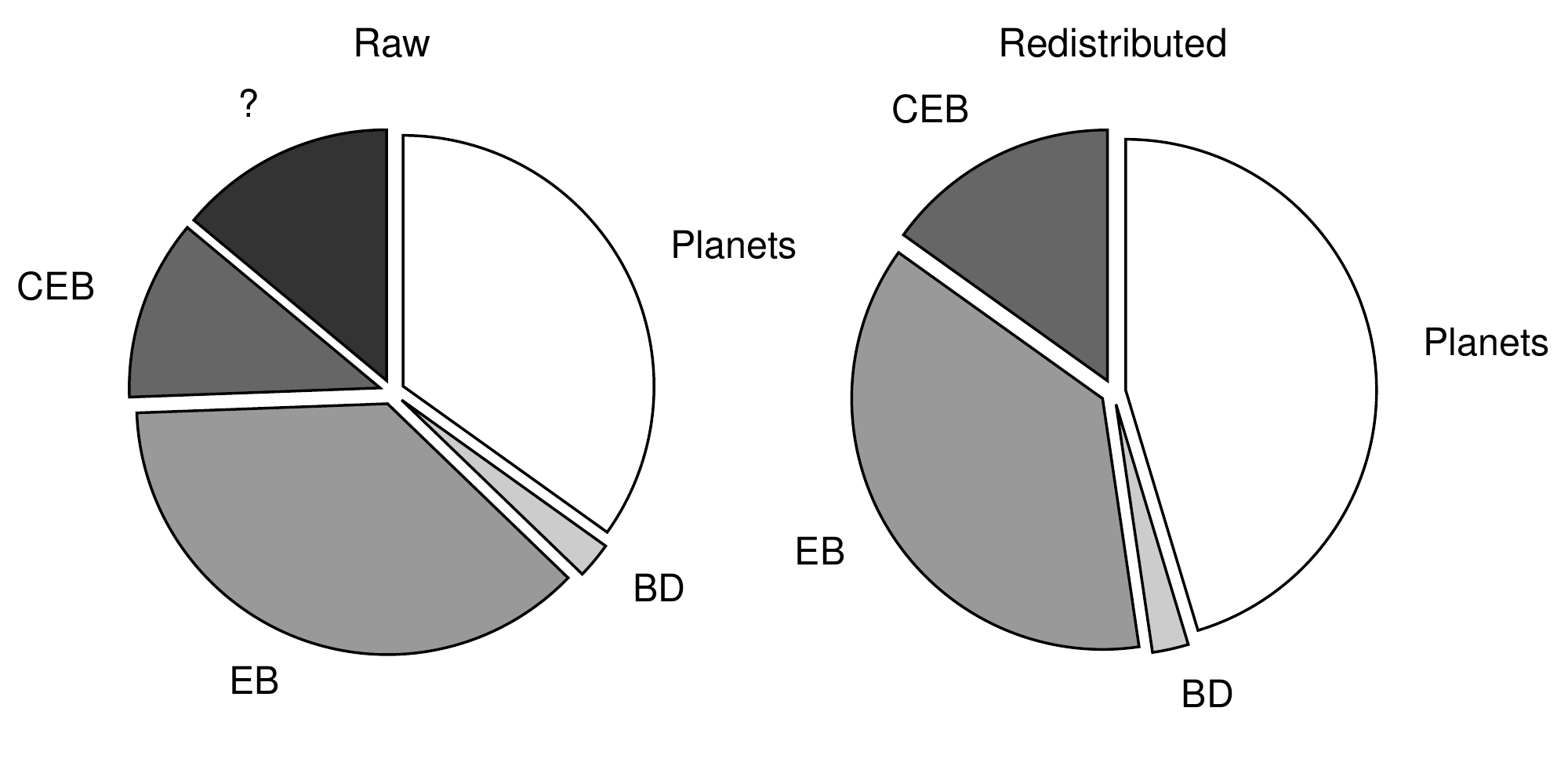}
\caption{Pie charts showing the distribution by nature of the giant-planet candidates: (left) the observed distribution of the candidates; (right) the underlying distribution of the candidates under the assumption that the unsolved cases are composed by 75\% of planets and 25\% of contaminating eclipsing binaries (see text). BD refers to brown dwarfs, EB to eclipsing binaries, CEB to contaminating eclipsing binaries and the question mark to the unsolved cases.}
\label{figFPP}
\end{center}
\end{figure*}

Among the 129 selected KOIs, we identified 34.9~$\pm$~5.2~\% of planets, 2.3~$\pm$~1.3~\% of brown dwarfs, 37.1~$\pm$~5.4~\% of EBs, 11.6~$\pm$~3.0~\% of CEBs, and 14.0~$\pm$~3.2~\% of unsolved cases, assuming a Poisson noise (see Fig. \ref{figFPP}).

The unsolved cases are not EBs nor brown dwarfs, otherwise a large RV variation would have been detected. They could be either planets with a mass lower than what can be detected with SOPHIE or a stellar or planetary companion eclipsing a different star than the target one. In this later case, if the flux ratio between the target and the eclipse host is low enough, it is not possible to detect its contribution on the spectra, either by detecting its impact on the target line-profile shape \citep{2015MNRAS.451.2337S}, or by detecting directly its line in the spectrum. This CEB could be either bound or chance-aligned with the target star. Following \citet{2012AA...545A..76S}, we assumed that the unsolved cases are composed by planets and CEBs with the same ratio as the observed one. This means that 75~$\pm$~11~\% of these unsolved cases are assumed to be planets, and 25~$\pm$~6~\% are likely faint CEBs.

We then find that the giant-planet candidates sample is composed by 45.3~$\pm$~5.9~\% of planets, 2.3~$\pm$~1.3~\% of brown dwarfs, 37.2~$\pm$~5.4~\% of EBs, and 15.1~$\pm$~3.4~\% of CEBs. This repartition of the nature of the giant-planet candidates is displayed in (Fig. \ref{figFPP}). This gives a giant-planet false-positive rate of 54.6~$\pm$~6.5~\%. Depending on the nature of the unsolved cases, the false-positive rate has a lower limit of 51.2~$\pm$~6.3~\% (if all unsolved cases are planets), and an upper limit of 65.1~$\pm$~7.1~\% (if they are false positives). Note that this value does not account for the false positives (about 50\% of the total number of EGP transit detection) already identified by \textit{Kepler} team.

If we repeat this analysis by dividing the sample in two, one for candidates with periods of less than 25 days \citep[i.e. an updated value for the sample of][]{2012AA...545A..76S} and one for the candidates with periods longer than 25 days, we find that the false-positive rate is 53.4~$\pm$~8.5~\% and 56.4~$\pm$~10.1~\%, respectively. Note that the value for the short-period sample is higher than the one reported in \citet{2012AA...545A..76S} for two reasons: (1) new candidates have been found on stars that were not observed by \textit{Kepler} in 2012 and (2) we included in this study the candidates that were flagged with a poor vetting flag in \citet{2011ApJ...736...19B} and rejected from the \citet{2012AA...545A..76S} sample.

The false-positive rate is however not uniform over orbital periods. If we split the sample in three for candidates with periods of less than 10 days, between 10 and 85 days, and between 85 and 400 days (see Section \ref{rates} for the reasons of this sub-samples selection), we find false-positive rates of 46.7~$\pm$~9.3~\%, 68.6~$\pm$~12.9~\%, and 50.2~$\pm$~12.1~\% (respectively). The false-positive rate is therefore the lowest for short-period candidates and the highest for intermediate period ones.

\subsection{Comparison with other false-positive rate estimates}

The false-positive rate of the \textit{Kepler} mission is a key element that describes the reliability of the \textit{Kepler} candidates catalog for statistical analyses. Together with the pipeline completeness \citep{2013ApJS..207...35C, 2015arXiv150705097C}, this information is needed to accurately assess the underlying occurrence of planets, down to Earth-size planets in the habitable zone. The latter is the main objective of the \textit{Kepler} prime mission \citep{2009Sci...325..709B, 2014PNAS..11112647B}.

By modelling the expected distribution of planets and binaries in the \textit{Kepler} field of view, \citet{2011ApJ...738..170M} found that the median false-positive probability among the \citet{2011ApJ...736...19B} candidates was as low as 5\%. This value was not supported by spectroscopic observations of a sample of 44 giant candidates which revealed a false-positive rate as high as 34.8 $\pm$ 6.5\% \citep{2012AA...545A..76S}, nor by the narrow-band GTC photometry of four small candidates in which two were found to be false positives \citep{2012MNRAS.426..342C}.

Later on, \citet{2013ApJ...766...81F} performed a new modelling of the expected population of planets and EBs in the \textit{Kepler} field of view, based on the \citet{2013ApJS..204...24B} candidate list. They found a median value of 9.4\%, with a higher rate (29.3~$\pm$~3.1~\% within 25 days) for the giant-planet candidates which is compatible with the measurement of 34.8 $\pm$ 6.5\% \citep{2012AA...545A..76S}. This median value was then revised by \citet{2013A&A...557A.139S} from 9.45\% to 11.3\% by accounting for secondary-only false positives.

Recently, \citet{2015ApJ...804...59D} found a false-positive rate as low as 1.3\% (upper-limit of 8.8\% at 3-$\sigma$) based on the \textit{Spitzer} near-infrared photometry of 51 candidates. However, this small set of candidates were selected to be representative of the KOI list from \citet{2011ApJ...736...19B}, and not a well defined sample. As pointed out by the authors, the extrapolation of the false-positive rate, from this small sample which represents 1.1\% only of the planet candidates known today, to the entire sample of candidates, should be done with caution. Note that among these 51 \textit{Spitzer} targets, 33 of them are orbiting in multiple systems that are known to have a very low a-priori probability of being false positives \citep{2012ApJ...750..112L,2014ApJ...784...44L}. The 18 remaining ones are relatively small planets, and only two are EGPs\footnote{These two EGPs are KOI-12~b and KOI-13~b. None of them is a false positive.}.


Using high-resolution spectroscopy and RV, we find that more than half of the giant-planet candidates are actually not planets. This value is significantly higher than all the other values reported so far. This value is however difficult to compare with the previous ones for two main reasons: (1) the list of candidates are different -- we used the Q1 -- Q17 candidate list from Coughlin et al., in prep., while most of the aforementioned studies used the Q1 -- Q6 candidate list from \citet{2013ApJS..204...24B}, with half as many candidates -- and (2) the selection criteria are also different. As an example, \citet{2013ApJ...766...81F} selected as giant-planet candidates all transit detections with an expected radius between 6 and 22~\Rearth, while our selection criteria is based on the observed transit depth (see Section \ref{sample}). Therefore, we will not compare directly the numbers, but qualitatively discuss the differences and similarities found. Since the work of \citet{2013ApJ...766...81F} is the most up-to-date simulation of the entire catalog of candidates, we will focus on the comparison between our observations and their results. 

\citet{2013ApJ...766...81F} predicted a false-positive rate of 17.7~$\pm$~2.9~\% for all the giant planet candidates within 418 days\footnote{Note that no giant-planet candidate has been found between 400 and 418 days.}. This value is significantly lower than our observational value. However, \citet{2013ApJ...766...81F} did not consider the fact that EBs might mimic the transit of an EGP. The underlying reason is that such false positives have a V-shaped transit (i.e. an impact parameter $b$ $\gtrsim$ 1) and can be easily rejected. Some grazing planets, like CoRoT-10~b \citep{2010A&A...520A..65B} or KOI-614~b \citep{2015A&A...575A..71A}, also present the same V-shaped transit. Since those V-shaped candidates actually are in the catalogs, this scenario of false positive should be considered, as it is done in \citet{2012ApJ...761....6M}. 

By not considering EBs as an important source of false positives, \citet{2013ApJ...766...81F} overestimated the occurrence rate of EGPs in the \textit{Kepler} field of view. Since they used this occurrence rate of EGPs to estimate the amount of planets transiting a physical companion to the target star, they also overestimated the abundance of this false-positive scenario. This scenario being the main source of false positives in the \textit{Kepler} list of candidates \citep[according to][]{2013ApJ...766...81F}, it has an impact on all the population of planets. 

Not all the EBs we identified are member of the \textit{Kepler} EB catalog (14\% are missing - Kirk et al., in prep) and two (KOI-1271.01 and KOI-6132.01) members of this catalog were not confirmed by our data to be EBs. The completeness of this catalog, used to estimate the fraction of false positives involving stellar systems, is thus lower than expected and the fraction of false positives composed by stellar systems is underestimated in the \citet{2013ApJ...766...81F} analysis.

As an illustration, \citet{2013ApJ...766...81F} predicted that among all target stars observed by \textit{Kepler}, there are 4.7 triple systems, 8.0 background EB, and about 24.5 planets transiting physical companion to the target star that mimic EGPs. By observing only 125 stars among the bright half of the candidates, we found 15 candidates that we considered as CEBs. They are likely bound with the target stars (hence triple systems) because they have a systemic RV similar to the target one\footnote{As discussed in \citet{2015arXiv150601668B}, this seems not to be the case of KOI-3783.01.}. This number of triple systems is three times larger than the one predicted by \citet{2013ApJ...766...81F}, for all the candidates. Hence, we predict roughly six times more triple systems than predicted\footnote{This assumes that there is the same rate of triple systems mimicking giant-planet candidates around targets brighter and fainter than K$_{p}=$14.7, and there are $\sim$equal numbers of stars in these categories.}. On the other hand, we found no clear evidence for planets transiting a physical companion to the target star. They might be however among the unsolved cases. Therefore we observe a higher rate of triple systems than false positives made of planets, which is the opposite of what \citet{2013ApJ...766...81F} predicted. The aforementioned reasons might explain this discrepancy. 
Note that only 66\% of the EGP candidates were released in \citet{2013ApJS..204...24B}, which might also explain this difference with \citet{2013ApJ...766...81F}.

\subsection{Extrapolation towards smaller planet candidates}

Even if our spectroscopic observations bring no constraints to the large sample of small-planet candidates detected by \textit{Kepler}, we can use the EGPs as a reference to qualitatively extrapolate the false-positive rate of small planets.

\subsubsection{Undiluted-depth eclipsing binaries}
\label{undilutedFPs}

The populations of small-planet candidates should be much less contaminated by EBs. Grazing eclipses (by stars or the rare brown dwarfs) can produce any transit depth but their occurrence rate is expected to decrease for transit depths below 1\% \citep{2013A&A...557A.139S}. Therefore, this source of false positives should completely disappear towards shallower candidates.

\subsubsection{Diluted-depth eclipsing binaries and transiting planet}
\label{dilutedFPs}

Shallower transits produce lower S/N events at a given stellar magnitude, and the false-positive diagnoses are expected to be less efficient for shallow transits, e.g., the duration of the transit ingress and egress or the presence of a secondary eclipse is poorly constrained if the primary transit S/N is low. Therefore, false positives that mimic small planets are more difficult to screen out compared to the large ones.

In addition, the dilution ratio also impacts the analysis. A planet candidate transiting a star of magnitude $m_{t}$ with an observed depth of $\delta_{t}$ can be mimicked by an eclipse of depth $\delta_{c}$ on a contaminating star of magnitude $m_{c}$ such as:
\begin{equation}
m_{t} - m_{c} = 2.5\log_{10}\left(\frac{\delta_{t}}{\delta_{c}}\right)
\end{equation}
Decreasing $\delta_{t}$ is achieved by decreasing $\delta_{c}$ and/or increasing $m_{c}$: i.e. fainter false-positive hosts and/or smaller companions. Those fainter hosts could be either smaller or farther away stars that are, in both cases, more common, as previously discussed by \citet{2003ApJ...593L.125B}.

\subsubsection{The false-positive rate of small planets}

By combining both effects discussed in Section \ref{dilutedFPs}, we thus expected that the total number of false positives invoking diluted-depth transits or eclipses increases towards smaller candidates and that we should be less efficient to rule them out. However, the false-positive rate is defined as the relative fraction of false positives against bona-fide planets among the candidates. Thus, decreasing the transit depth of the candidates corresponds to explore smaller planet populations that are more common according to planet-formation synthesis \citep[e.g.][]{2009A&A...501.1161M, 2014arXiv1411.5264N} and the results from radial velocity surveys \citep[e.g.][]{2010Sci...330..653H, 2011arXiv1109.2497M}. Therefore, even if the absolute number of false positive increases by decreasing the transit depths, the relative value (i.e the false-positive rate) might not necessarily increase. 

We discussed previously that the absolute number of false positives is expected to increase towards smaller planet candidates.  As pointed out by \citet{2011ApJ...732L..24L} and \citet{2011ApJS..197....8L}, however, about one third of the candidates smaller than Jupiter are found in multiple systems, in agreement with the first results of RV surveys \citep[e.g.][]{2009A&A...496..527B, 2011arXiv1109.2497M}. Furthermore, those multiple candidates have a very low $a priori$ probability of being false positives \citep{2012ApJ...750..112L, 2014ApJ...784...44L}. Therefore, our predicted increase of false positives towards small planet candidates should be mostly concentrated on the candidates that are not in multiple systems. Any physical interpretation on the nature of these small-and-single, unconfirmed candidates might thus lead to wrong conclusions. 

\subsection{Comparison between giant-planet and false-positive host properties}
\label{ComparHosts}

\begin{figure}[h]
\begin{center}
\includegraphics[width=\figw]{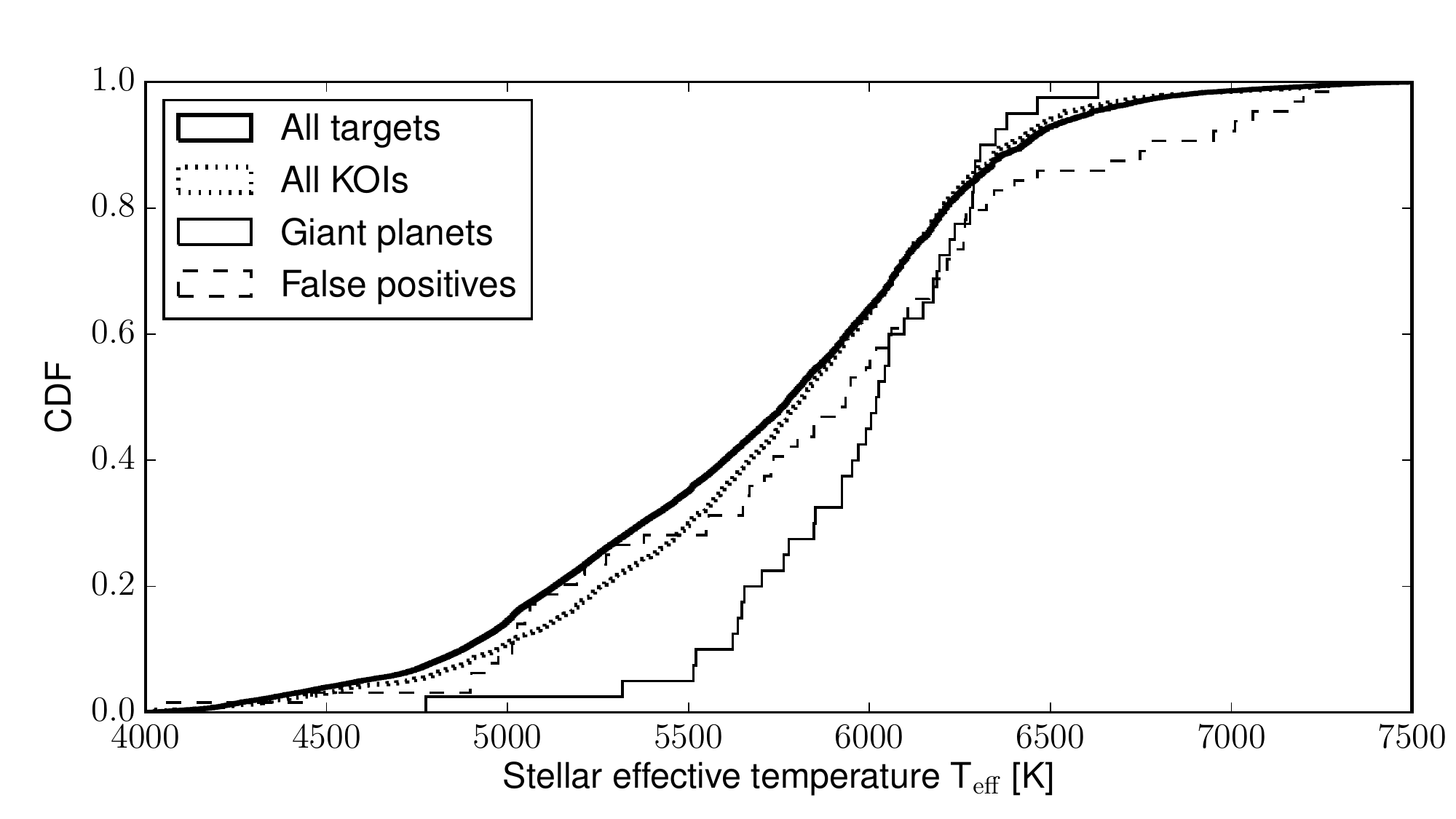}\\
\includegraphics[width=\figw]{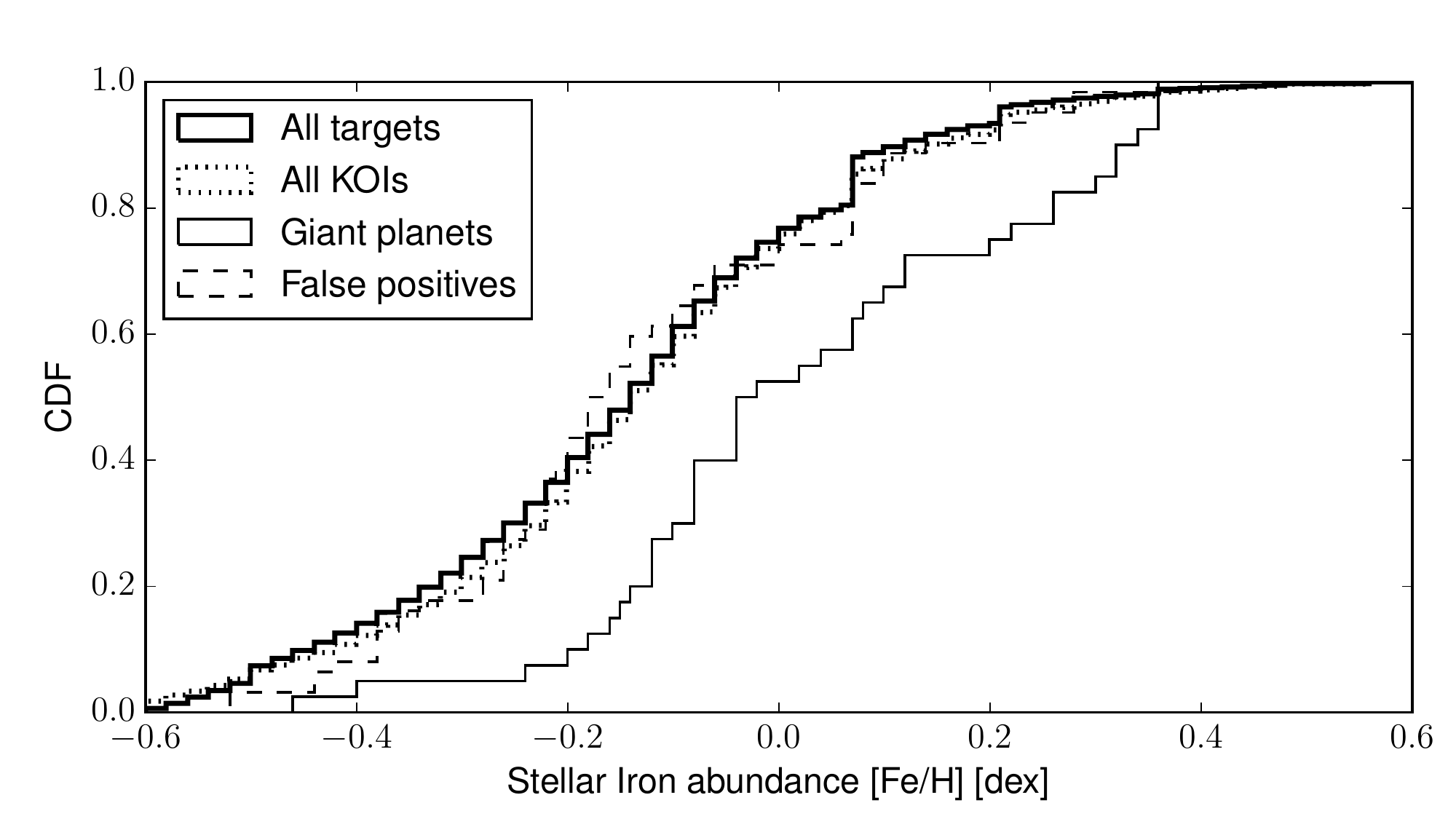}
\caption{Cumulative distribution functions (CDF) of the effective temperature (top panel) and the iron abundance (bottom panel) for all the \textit{Kepler} targets (solid and thick line), the secured EGPs in our sample (solid line), and the false positives identified by our spectroscopic surveys (dashed line). The stellar parameters are from \citet{2014ApJS..211....2H}. }
\label{DistribFPs}
\end{center}
\end{figure}

Several studies have tried to infer some planet -- star properties correlation, such as with the effective temperature of the host star \citep[e.g.][]{2012ApJS..201...15H}, or its metallicity \citep[e.g.][]{2014Natur.509..593B, 2015AJ....149...14W}, as they provide direct tests to planet-formation theories. However, if the false positives are not accounted for and have a different host parameters distribution, they might alter the underlying correlation. To test this, we display in Fig. \ref{DistribFPs} the cumulative distributions of the effective temperature and iron abundance of the planets and false positives hosts as well as all the \textit{Kepler} targets observed during Q1 -- Q16. Note that we considered here as planets only the 45 ones that have been well established in our sample. The stellar parameters are from \citet{2014ApJS..211....2H}.

We find that most stars hotter than 6500 K host false positives and very few host EGPs. This might be an observational bias since planets around hot, fast-rotating, and active stars are more difficult to find and characterise than those around Sun-like stars. We also find relatively few EGPs orbiting stars cooler than the Sun and that the giant planets preferentially orbit metal-rich stars, which confirms the RV results \citep[e.g.][see also Section \ref{hostprop}]{2011A&A...533A.141S}. 

We computed the Anderson -- Darling (AD) test\footnote{The Anderson -- Darling test is recommended by \citet{2009ApJ...702.1199H}
to estimate the probability that two random variables are drawn from the same underlying distribution. It is more sensitive to the differences in the wings of the distribution, whereas the Kolmogorov -- Smirnov (KS) test is mostly sensitive to its median. Both tests are non parametric and distribution free. The AD test is more computational expensive than the KS test.} between the distributions of the giant-planet and the false-positive hosts with the \textit{Kepler} targets and candidates host ($p$-values listed in Table \ref{ADtest}). 
We find that the distributions of stellar effective temperature and iron abundance between the target stars and candidate hosts (all KOIs) are different\footnote{We consider as significantly different all $p$-values smaller than 1\%.}. This is also the case between the giant-planet hosts and both the target stars and the candidate hosts. However, the distributions of \met\, and \teff\, are not significantly different between the false-positive hosts and the target stars. Those results are expected since the fraction of binaries is relatively constant in the regime of stars we are the most sensitive to \citep[\teff\, between 5000~K and 6500~K and \met\, between -0.4~dex and 0.4~dex ;][]{2010ApJS..190....1R}.

By comparing the distributions of stellar properties for different samples of stars hosting either EGPs, false positives, candidates or just field stars, we show that the presence of false positives have two main implications. First, the determination of the occurrence rate of EGPs as a function of the stellar properties based on the candidates list cannot be correct with $\sim$ 55\% false positives. Then, one may overestimate the occurrence of small planets orbiting metal-rich stars, if a significant fraction of the false positives are made of EGPs transiting stellar companions to the target star \citep[as claimed by][]{2013ApJ...766...81F}. On the other hand, if small planets are mostly mimicked by EBs, their metallicity distribution might not be significantly different than the field stars. Therefore, the determination of the planet occurrence rate as a function of the stellar host properties, without screening out the false positives, should be done with caution, as it might lead to wrong results.


\section{Giant-planet occurrence rates}
\label{rates}

In this Section, we analyse the secured and likely EGPs in our sample. The first information we can derive from this cleaned sample is the occurrence rate of EGPs. 

\subsection{The occurrence rate of giant planets within 400 days}

To measure the occurrence rate of planets, we need to determine: (1) a reference stellar sample, (2) the number of transiting planets in this reference stellar sample, and (3) the various corrections that should be applied, such as the number of non-transiting planets and the planets missed by incompleteness of the pipeline. We discuss these points below.

\subsubsection{The stellar reference sample}
\label{RefSample}

The \textit{Kepler} prime mission focused on solar-like stars \citep{2014ApJS..211....2H}, we thus define our stellar reference sample to match the properties of such stars. Our transit-candidate selection is biased toward dwarf hosts, and is quite insensitive to sub-giant and giant hosts around which jovian planets have transit depth shallower than 0.4\%. Thus, we need to determine how many FGK dwarfs \textit{Kepler} observed.

In previous works that attempt to measure occurrence rates of planets \citep[e.g.][]{2012ApJS..201...15H, 2013PNAS..11019273P}, the observed atmospheric parameters (\teff, \logg) were used to select solar-like dwarfs, to fit the historical Morgan -- Keenan classification of stars \citep{1973ARA&A..11...29M}. Using these selection criteria, it is however difficult to make the distinction between main-sequence and sub-giant stars in the regime of early G- and F-type stars. For example, a star with \teff\,= 5000 K and \logg\, = 4.1 \cmss\, (at solar metallicity) is a sub-giant while another star with the same surface gravity but with a \teff\, of 6500 K is still in the main-sequence. Because of their large radius, planets transiting sub-giants and giant stars are more difficult to detect. Thus, a stellar reference sample composed by a substantial fraction of evolved stars might lead to underestimation of the planet occurrence rates (unless this effect is taken into account). This problem does not occur for late G-, K-, and M-type stars because their lifetime in the main-sequence is longer than the age of the universe. Since both the \teff\, and \logg\, vary during the evolution of stars in the main-sequence and beyond, they are not the best parameters to select a stellar reference sample. 

To determine our stellar reference sample, we chose the stellar mass and radius as selection parameters, The mass of stars does not change significantly during their evolution, except at very late stages. The mass is also the fundamental parameter used in planet-formation synthesis \citep[e.g.][]{2009A&A...501.1139M} since it is expected to scale with the mass of the disk, for the mass range considered here \citep{2013ApJ...771..129A}. During the main sequence and sub-giant phases, the stellar radius increases in a strictly monotonic way. These reasons make the stellar mass and radius better parameters to select a stellar sample composed only by main-sequence stars. This requires to define the radius of stars at the end of their main-sequence life.

\begin{figure}[h]
\begin{center}
\includegraphics[width=\figw]{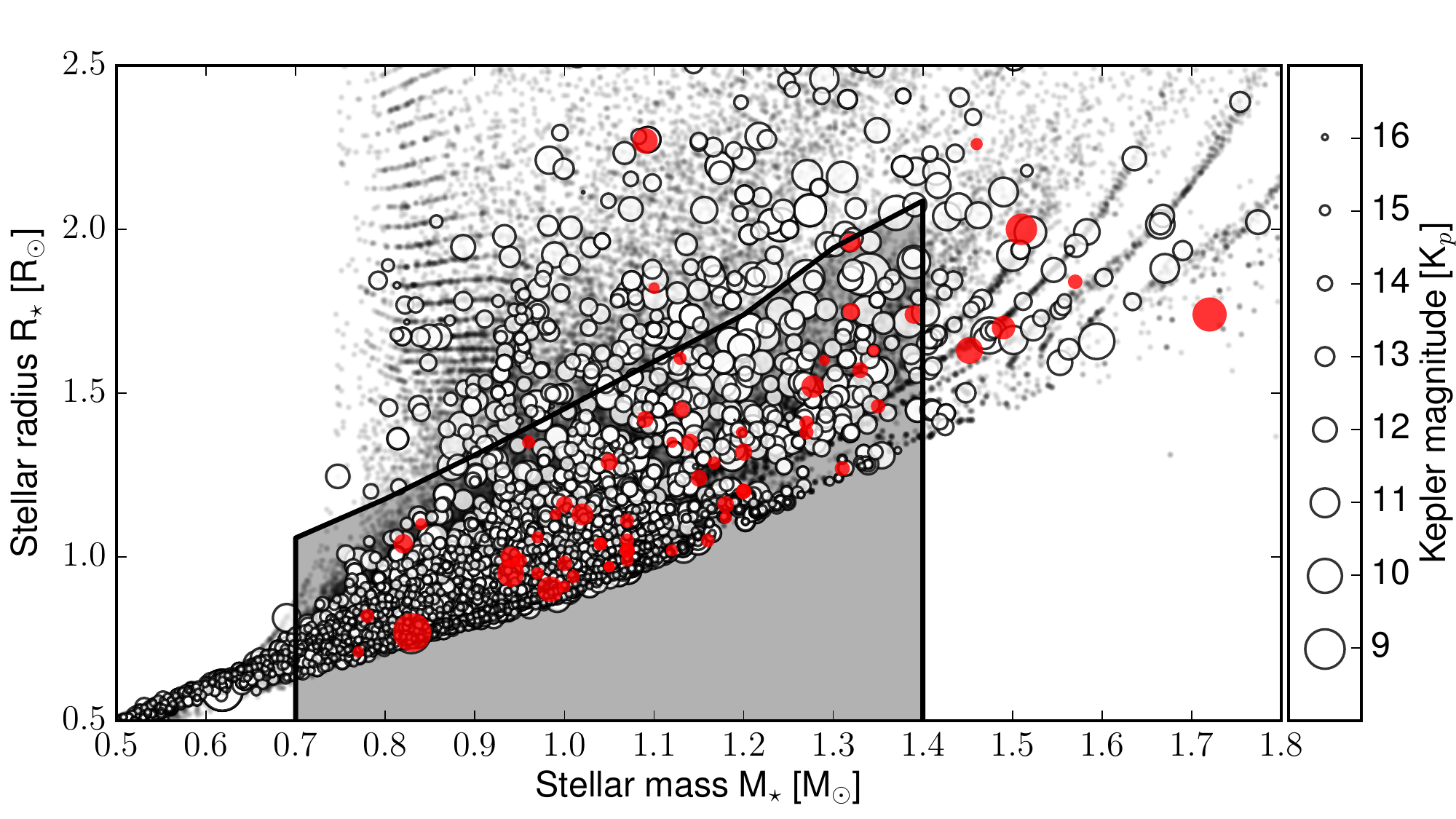}
\caption{Stellar mass -- radius diagramme of the targets observed by \textit{Kepler} during the quarters Q1 -- Q16 (black dots), the candidate hosts (white circles) and the bright giant-planet hosts (red circles). The data are from \citet{2014ApJS..211....2H} except for the giant-planet hosts which are taken from the Table \ref{PlanetParams}. The size of the mark for the two latter samples corresponds to the \textit{Kepler} magnitude. The grey region displays the selected dwarfs. Note that one secured EGP transits a star bigger than 2.5 \Rsun\, (KOI-680) and is not represented here.}
\label{KeplerTargets}
\end{center}
\end{figure}

We used the latest version of the STAREVOL stellar evolution code \citep[][Amard et al., in prep.]{2004IAUS..215..440C, 2012A&A...543A.108L}, with the solar composition following \citet{2009ARA&A..47..481A}. The metallicity is fixed to a solar value ($Z = 0.0134$) and the mixing length parameter calibrated to a solar model taken as $\alpha_{MLT} = 1.702$. We determined the end of the main-sequence when the Hydrogen abundance in the core is X(\ion{H}{i}) $ < 10^{-7}$. The main parameters of solar-type stars at the end of the main sequence are listed in Table \ref{EndMSparams}. We adopted the stellar radius listed in this table as the maximum value to select dwarf stars. Figure \ref{KeplerTargets} displays all the \textit{Kepler} targets, candidate host, and the bright giant-planet hosts in the M$_{\star}$ -- R$_{\star}$ space. The adopted maximum radius for the dwarf stars is also represented.

During the \textit{Kepler} prime mission, a relatively small fraction of dwarf stars were observed with M$_{\star} <$ 0.7 \Msun\, (10.0\%) and M$_{\star} >$ 1.4 \Msun\, (3.9\%). By selecting only the dwarfs that have a magnitude K$_{p} < 14.7$, only 6.2\% of them are smaller than 0.7 \Msun\, and 9.1\% are more massive than 1.4 \Msun. By selecting the bright dwarfs in the range 0.7 -- 1.4 \Msun, which corresponds to a spectral type F5 -- K5 \citep{2000asqu.book.....C}, we selected 84.8\% of the observed bright dwarfs. Because a relatively small amount of bright low-mass or massive dwarfs have been observed by \textit{Kepler}, measuring the occurrence rate of EGPs around those stars will be strongly limited by small-number statistics.

To determine our stellar reference sample, we selected the \textit{Kepler} targets that have a magnitude K$_{p} <$ 14.7, a mass in the range M$_{\star} \in [0.7 ; 1.4]$ \Msun, and a radius smaller than the ones listed in the Table \ref{EndMSparams}. We used the stellar parameters of \citet{2014ApJS..211....2H} and found a total number of bright, solar-type dwarfs observed by \textit{Kepler} of 58831. In spite of the large uncertainties on the stellar masses ($\sim$~20\%) and radii ($\sim$~40\%) in the \citeauthor{2014ApJS..211....2H}'s catalog, the total number of dwarfs is expected to be statistically accurate. 

Note that if we select the stellar reference sample based on \teff\, $\in [4410 ; 6650]$ K and \logg\, $\in [4.0 ; 4.9]$ \cmss, we find a number of 59873 bright dwarfs observed by \textit{Kepler}\footnote{This corresponds to F5 -- K5 dwarfs according to \citet{2000asqu.book.....C}.}. Thus, selecting the stellar reference sample based on \teff\, and \logg\, or stellar mass and radius does not change significantly our results.

\subsubsection{Sample of transiting planets}

Once the reference sample has been well defined, we need to determine how many EGPs in total are transiting those stars. In our giant-planet candidate sample, there are 45 secured transiting planets and 18 candidates that could be either planets or false positives. Since we expected the majority of the latter to be planets, we consider them as ''likely planets''. We report in Table \ref{PlanetParams} the transit, planet and stellar parameters of these 63 objects, from literature values. When planets have been analysed in different papers, we kept as adopted values the most updated or complete analysis of the systems. Similarly, we adopted the stellar parameters from, e.g.  \citet{2012ApJ...757..161T} or \citet{2013A&A...556A.150S}, when available, because those studies used higher-S/N data than in the discovery papers, leading to more reliable results. By default, when no detailed analysis of the photometric or spectroscopic data have been reported, we used the transit and planet parameters provided in NASA exoplanet archive\footnote{\url{http://exoplanetarchive.ipac.caltech.edu}} and the stellar parameters from \citet{2014ApJS..211....2H}. The parameters in Table \ref{PlanetParams} are thus heterogeneous.
\subsubsection{Survey corrections}
\label{Scorr}

\begin{figure}[h]
\begin{center}
\includegraphics[width=\figw]{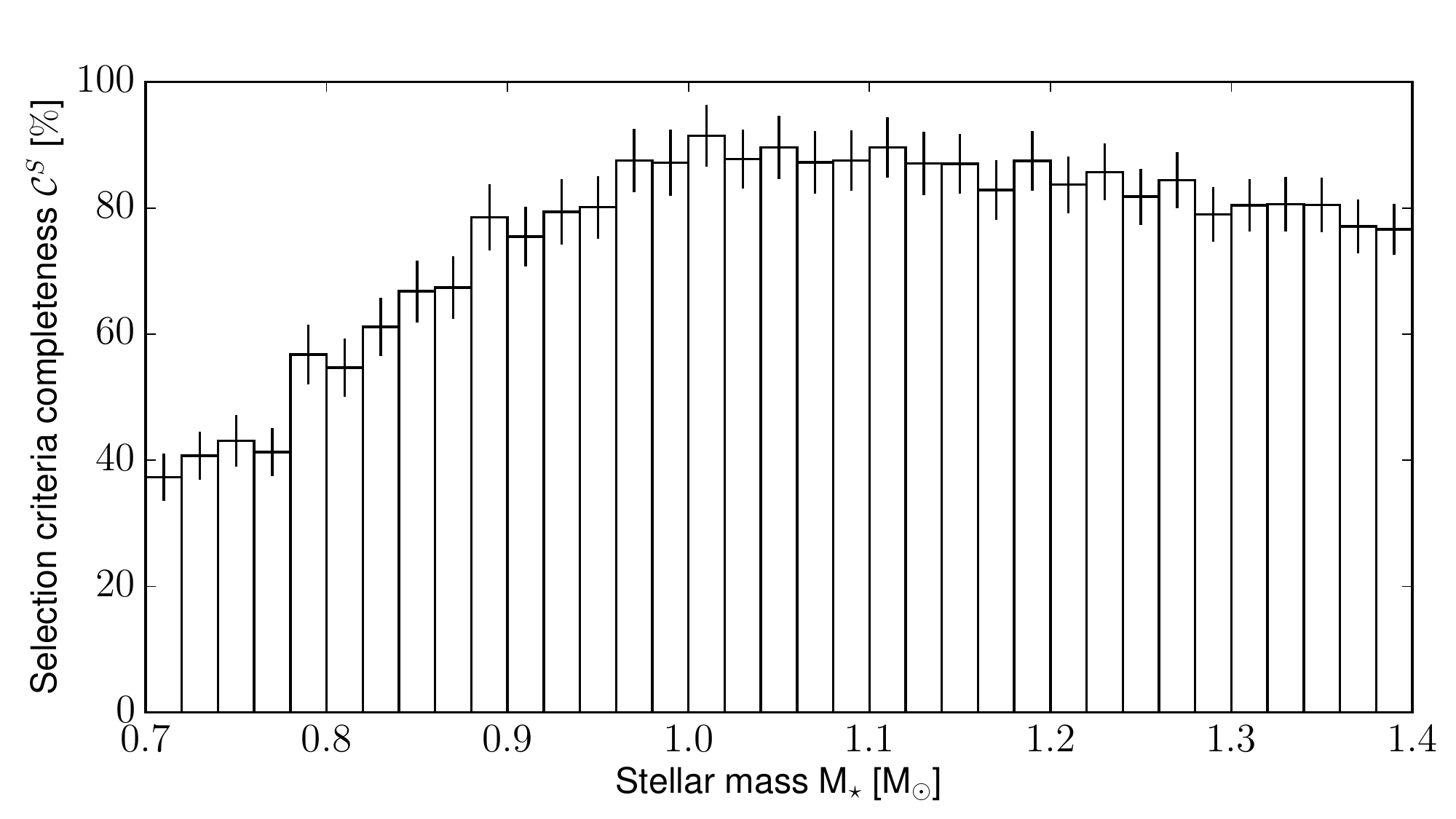}\\
\includegraphics[width=\figw]{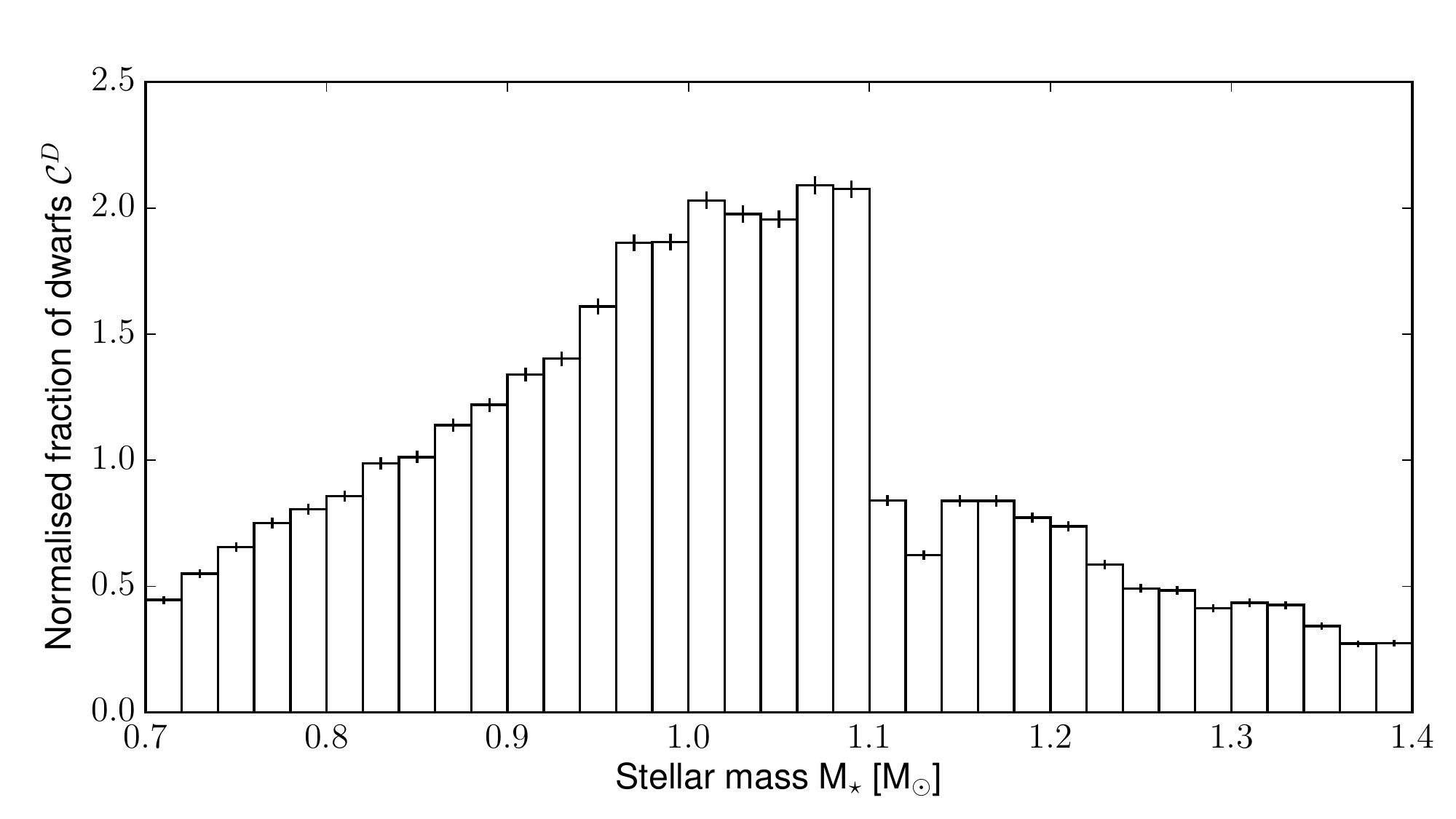}
\caption{Correction factors to compensate for the candidates selection ($\mathcal{C}^{S}$ -- upper panel) and the non-unformity distribution of the dwarfs stars ($\mathcal{C}^{D}$ -- lower panel).}
\label{Complet}
\end{center}
\end{figure}

We identified six corrections that have to be accounted for to derive the occurrence rates of EGPs within 400 days, based on our data. We call them $\mathcal{C}^{T}$, $\mathcal{C}^{R}$, $\mathcal{C}^{L}$, $\mathcal{C}^{S}$, $\mathcal{C}^{D}$, and $\mathcal{C}^{C}$. We describe and discuss them below.
\begin{itemize}
\item $\mathcal{C}^{T}$: correction for the geometric transit probability (the upper script $T$ refers to transit probability).
 Following \citet{2012ApJS..201...15H}, for each planet transiting a star of radius $R_{\star}$ with a semi-major axis $a$, there are $a/R_{\star}$ times more planets (both transiting and non-transiting). Therefore, we defined the correction for the transit probability such as $\mathcal{C}^{T} = a/R_{\star}$. This parameter is directly measured on the light curve and does not rely on the stellar parameters.

\item$\mathcal{C}^{R}$: correction for the probability that the planet host belongs to the stellar reference sample or not (the upper script $R$ refers to the reference sample). 
To estimate this correction, we bootstrapped 1000 times the planet-host mass and radius within their uncertainty, assuming they follow a Gaussian distribution. Then, we apply our reference sample criteria, defined in Section \ref{RefSample}, and determined $\mathcal{C}^{R}$ as the fraction of hosts that satisfy the reference sample cirteria. The values of $\mathcal{C}^{R}$ ranges from 0\% for evolved stars like KOI-680 \citep{2015A&A...575A..71A}, to 100\% for well-characterised solar-like dwarfs such as KOI-1 \citep[aka TrES-2;][]{2014ApJS..211....2H}. 

\item $\mathcal{C}^{L}$ : correction for the likelihood of the object to be a planet or not (the upper script $L$ refers to planet likelihood). 
The majority of the EGPs considered here have been well established using various techniques ($\mathcal{C}^{L} = 1$). For the candidates for which we detected no significant RV variation, we failed in ruling out all false-positive scenarios, as already discussed in Sections \ref{NoVar} and \ref{FPR}. However, we estimated in Section \ref{FPR} that about 75\% of them should be planets, so $\mathcal{C}^{L} = 0.75$.

\item $\mathcal{C}^{S}$: correction to the selection criteria used to define the transiting EGP sample (the upper script $S$ refers to the selection criteria). 
Depending on the stellar and planetary radii, the transit depth of EGPs might be lower than 0.4\% or larger than 3\%. These selection criteria were defined to include the majority of planets, but a fraction might have been missed, mostly grazing planets. To estimate this correction, we simulated 10$^{5}$ mock planetary systems. The stellar mass and radius were chosen uniformly within our definition of solar-type dwarfs (see Section \ref{RefSample})\footnote{While the maximum stellar radius is defined based on the evolution tracks (see Table \ref{EndMSparams}), we determined the minimum stellar radius as the lower envelope of the \textit{Kepler} targets.}. The orbital inclination was drawn from a Sine distribution, which corresponds to a uniform distribution of both the inclination and longitude of ascending node. The period was fixed to 10 days which is close to the median of the giant-planet periods and we considered circular orbits. Then, we assumed a radius distribution of EGPs that corresponds to the observed one (based on EGP radius listed in the NASA exoplanet archive). This radius distribution is an asymmetric Gaussian such as R$_{\rm p} = 1.19^{_{+0.18}}_{^{-0.21}}$ \Rjup. We used the \texttt{JKTEBOP} code \citep{2008MNRAS.386.1644S} assuming the limb darkening coefficients from \citet{2011A&A...529A..75C} to simulate the transit light curve and determine the transit depth. Finally, $\mathcal{C}^{S}$ is defined as the fraction of planets with transit depth in the range $[0.4\% ; 3\%]$ over the total number of transiting planets. We determined this correction per bin on stellar mass that we displayed in Fig. \ref{Complet}. The mean value of $\mathcal{C}^{S}$ over all stellar masses is 77\%, with values ranging from 37\% for the lowest mass stars in our reference sample to 91\% for Sun-like stars.  

\item $\mathcal{C}^{D}$: correction to the non-uniform distribution dwarf stars (the upper script $D$ refers to dwarfs distribution). 
We have more chances a priori of finding a transiting planet among the most abundant population of stars, i.e., stars with mass in the range 1 -- 1.1 \Msun). However, the distribution of giant-planet hosts might be (and actually is) different. To completely account for this effect, we would need to explore all the stellar parameters (at least the mass and \met) simultaneously, but we do not have a stellar reference sample large enough for that. So, for a first-order correction, we considered only the distribution of stellar masses. We defined $\mathcal{C}^{D}$ as the normalised distribution of dwarf masses in the stellar reference sample. This distribution is displayed in Fig. \ref{Complet}. The values of $\mathcal{C}^{D}$ range from 0.25 to 2.14.

\item $\mathcal{C}^{C}$: correction to account for the detection pipeline completeness, i.e. the number of transiting planets missed by the detection pipeline (the upper script $C$ refers to the completeness). This has been thoroughly studied in \citet{2013ApJS..207...35C, 2015arXiv150705097C}. In particular, they found that the detection efficiency of EGPs transiting bright, solar-type dwarfs is better than 95\% over orbital periods up to 400 days (Christiansen et al., private communication) based on the Q1 -- Q17 data. Therefore, we assigned a value of $\mathcal{C}^{C} = 0.95$ for all EGPs in our sample.
\end{itemize}

The values of the all correction factors, but $\mathcal{C}^{C}$ which is constant, are provided in Table \ref{CorrecFactor}.

\subsubsection{The occurrence rates and their uncertainties}

The occurrence rate is defined as the ratio between the number of transiting planets, $n_{t}$, corrected by the six aforementioned effects over the total number of dwarfs in the reference sample, $N_{\star}$: 
\begin{equation}
\label{occeq}
\mathcal{O}= \frac{1}{N_{\star}}\sum_{i=1}^{n_{t}}\frac{\mathcal{C}_{i}^{T}\;\mathcal{C}_{i}^{R}\;\mathcal{C}_{i}^{L}}{\mathcal{C}_{i}^{S}\;\mathcal{C}_{i}^{D}\;\mathcal{C}_{i}^{C}}\,.
\end{equation}

The main uncertainty is based on the fact that we are dealing with relatively small number statistics. This occurrence rate is based only on 63 transiting planets (secured and likely). Thus, we consider that our uncertainty is dominated by a Poisson noise that scales with the number of detected transiting planets ($n_{t}$), and we define the occurrence rate uncertainty, $\sigma_{\mathcal{O}}$, such as:
\begin{equation}
\label{occeqerr}
\sigma_{\mathcal{O}} = \mathcal{O}\frac{\sqrt{n_{t}}}{n_{t}}\,.
\end{equation}

Applying the two latter equations in the entire sample, we find that the occurrence rate of EGPs within 400 days is 4.6 $\pm$ 0.6 \%. 

\begin{figure*}[h!]
\begin{center}
\includegraphics[width=17cm]{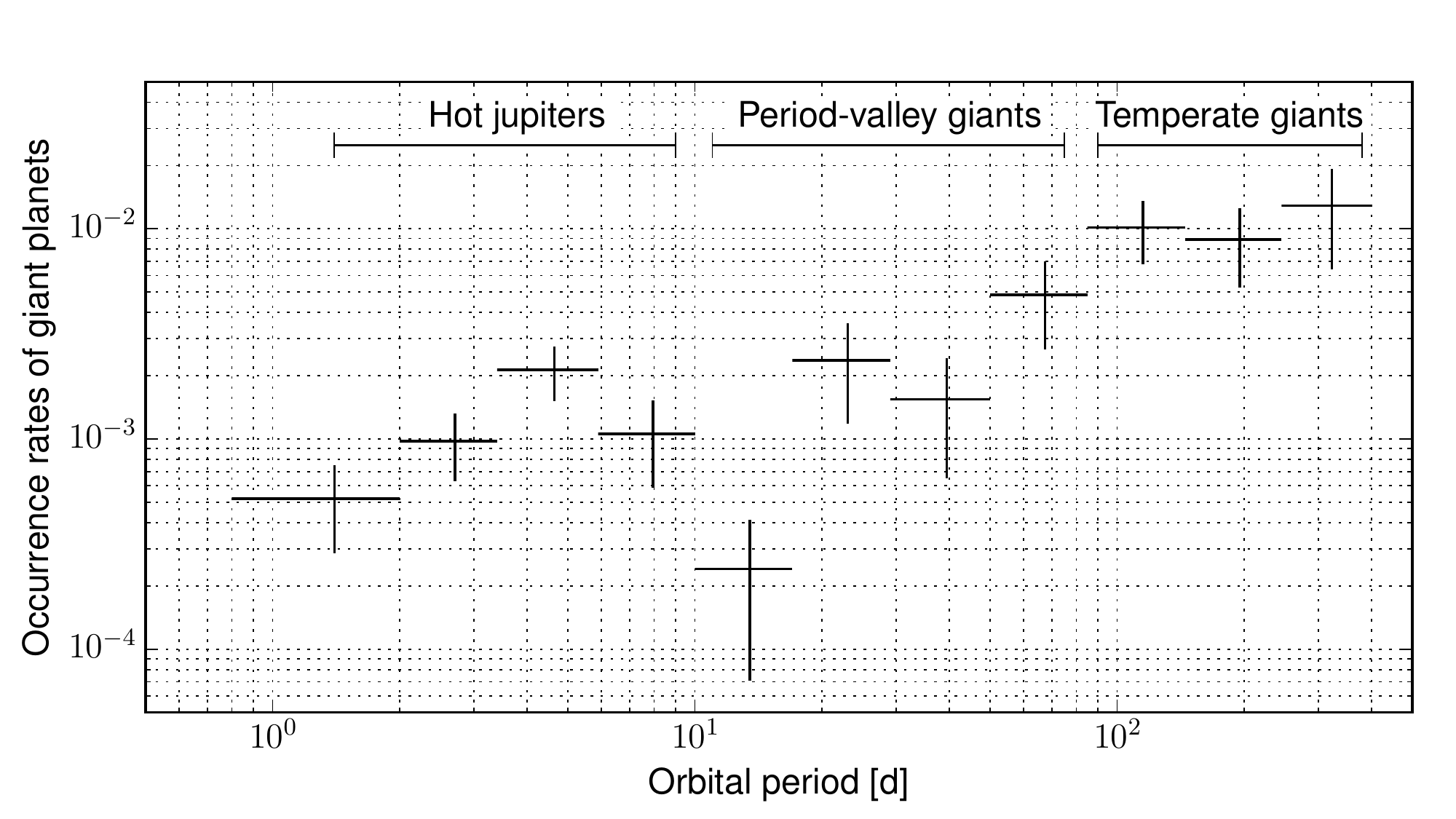}
\caption{Occurrence rates of giant planets as a function of the orbital period. The horizontal bars indicate the range of periods used in a given bin.}
\label{OccGP}
\end{center}
\end{figure*}

We computed the occurrence rates in different bins of orbital periods, as in \citet{2013ApJ...766...81F}. Tables \ref{EndMSparams}, \ref{PlanetParams}, and \ref{CorrecFactor} provide all the values needed to derive the occurrence rates of EGPs in different ranges of orbital periods. Our derived values are displayed in Fig. \ref{OccGP} and listed in Table \ref{OccurRateComp}. This shows that the overall occurrence rate of EGPs increases towards longer orbital periods. However, this increase is not monotonic. We can clearly see the pile-up of hot jupiters at about 5 days followed by a sharp decrease in the occurrence rate for planets with orbital periods in the range 10 -- 17 days. The occurrence rate in this period range is one order of magnitude lower than the one at 5 days. Then, the occurrence rate increases up to about 85 days before reaching a plateau up to 400 days.

These variations of the occurrence rate highlight the underlying populations of hot jupiters, period-valley giants and temperate giants that were already pointed out by RV surveys more than a decade ago by, e.g. \citet{2003A&A...407..369U}. These populations of giant planets, especially the pile-up of hot jupiters, were however not confirmed by previous analyses of the \textit{Kepler} detections \citep{2012ApJS..201...15H, 2013ApJ...766...81F}. The reason for that is the presence of false positives that have a different period distribution and dilute the underlying distribution of planets.

\subsection{Comparison with other yields}
\label{CompYield}

We compare now our results with the two major estimates of the giant-planet occurrence rates: the one of \citet{2013ApJ...766...81F} that is also based on \textit{Kepler} photometry (using only the Q1 -- Q6 results), and the one of \citet{2011arXiv1109.2497M} that is based on HARPS and CORALIE RV. All analyses were performed on similar stellar populations (FGK dwarfs) located in different regions: a few hundreds of parsec above the galactic plane for the \textit{Kepler} field of view (FOV) and in the solar neighborhood for HARPS and CORALIE. However, the selection of the EGPs is slightly different between the analyses: while we selected EGPs based on their deep transit, \citet{2013ApJ...766...81F} selected all planets with an expected radius in the range 6 -- 22~\Rjup, and \citet{2011arXiv1109.2497M} considered the limit for the runaway accretion of 50~\Mearth\, to select EGPs. These differences in the definition of what is an EGP is clearly a limitation for this comparison, and thus, it should be interpreted with caution.

To compare our results with the ones of \citet{2011arXiv1109.2497M}, we re-compute the occurrence rates of EGPs in their period ranges and masses above 50~\Mearth. We used their detection limits to correct for the missing planets and derive the occurrence rates. We assumed an uncertainty which follows a Poisson noise on the number of detected planets, as in equation \ref{occeqerr}. Our determination of the \citet{2011arXiv1109.2497M} occurrence rates for the different ranges of periods is reported in Table \ref{OccurRateComp}. We also report in this table the values from \citet{2013ApJ...766...81F}. In this table, we report both the values in each bin of periods and the cumulative values.

We find no significant difference (within less than 1-$\sigma$) between our estimation of the EGP occurrence rates and the one using \citeauthor{2011arXiv1109.2497M}'s data, in all the bins. The values of the occurrence rate integrated within 400 days are also compatible between the three analyses. 

\begin{figure*}[h!]
\begin{center}
\includegraphics[width=17cm]{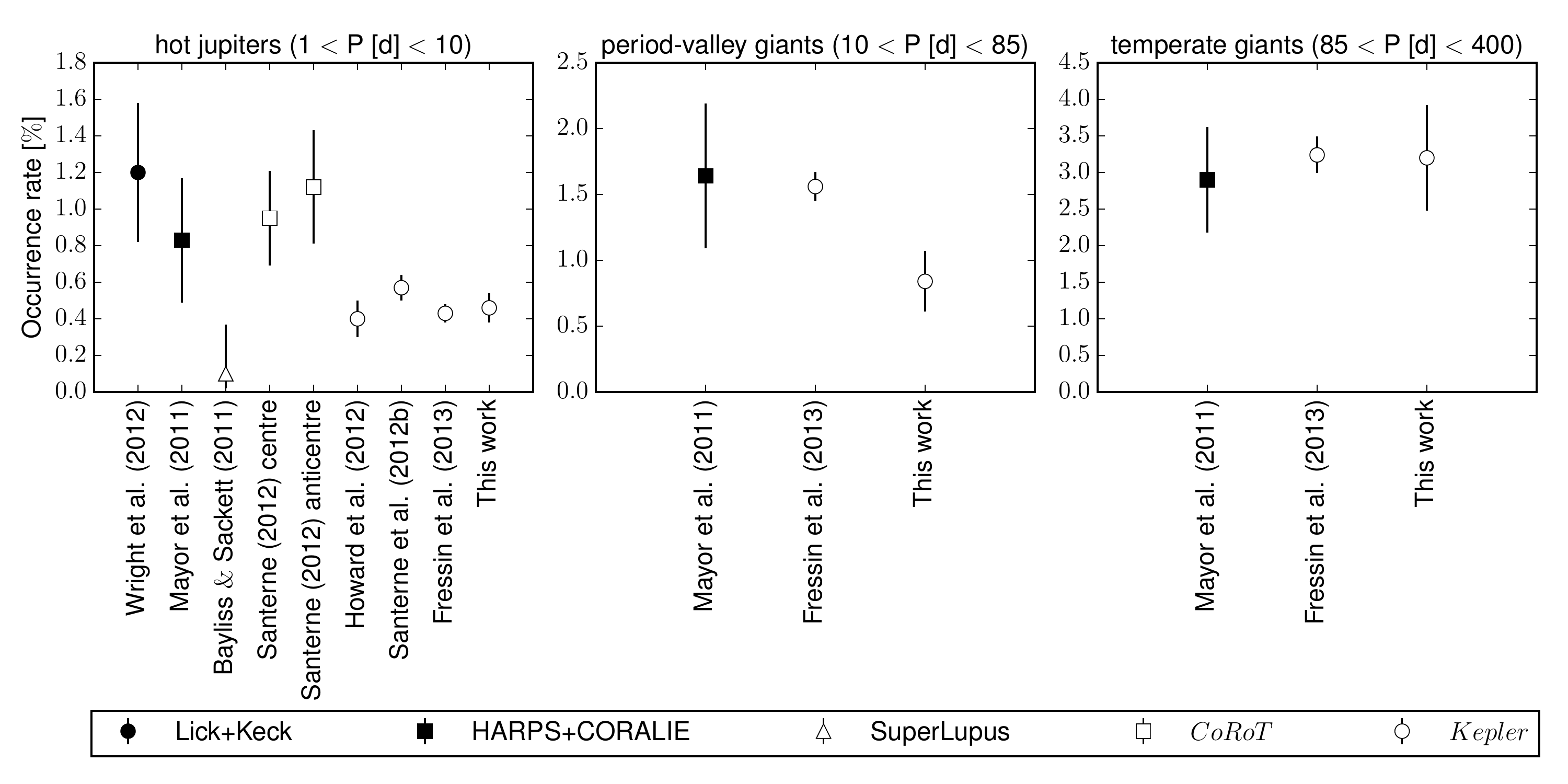}
\caption{Comparison of the occurrence rates of the three populations of giant planets. The black marks represent values based on RV surveys and the white ones are from photometric surveys.}
\label{CompOR}
\end{center}
\end{figure*}

To further compare the results between the three studies, we computed the occurrence rates for each population of EGPs: the hot jupiters with orbital periods of less than 10 days, the period-valley giants with orbital periods between 10 and 85 days, and finally the temperate giants with orbital periods between 85 and 400 days. We also compare the values found in the literature for the occurrence rate of hot jupiters. All these values are listed in Table \ref{OccurRateComp}, and plotted in Fig. \ref{CompOR}.

The four occurrence rates of hot jupiters based on the \textit{Kepler} data \citep[i.e. ][and this work]{2012ApJS..201...15H, 2012AA...545A..76S, 2013ApJ...766...81F} are fully compatible, in spite of differences in the candidate or planet selections, and stellar reference sample. The reported values are in the range 0.4 -- 0.5~\% for FGK dwarfs. However, this value seems to be systematically different with the values measured independently by RV in the California Planet Survey \citep{2012ApJ...753..160W} and the Swiss-led planet survey \citep{2011arXiv1109.2497M}. The latter values are also fully compatible with the estimates from the \textit{CoRoT} space mission, in both galactic directions \citep[towards the center and anti-center of the galaxy][]{SanternePhDThesis, 2013Icar..226.1625M}. In spite of their large uncertainties, these four estimates reported an occurrence rate of $\sim$ 1 \%, hence about twice more hot jupiters than in the \textit{Kepler} field (see Fig. \ref{CompOR}). \citet{2011ApJ...743..103B} reported a hot jupiter occurrence rate as low as 0.10$^{_{+0.27}}_{^{-0.08}}$ \% from the ground-based SuperLupus survey. This is however based on very small statistics, since only one hot jupiter has been established in this survey, with two other candidates. Therefore, this result is difficult to interpret and to compare with other transit surveys, which detected several tens of planets.

This difference, if real, might be explained by various effects. First, it might be an overestimation of the \textit{Kepler} pipeline completeness. This is quite unlikely since hot jupiters present high-S/N transits and thus are easily detected. Even though they would have been missed by the \textit{Kepler} detection pipeline, they would have been found by the Planet Hunters community \citep{2012MNRAS.419.2900F}. Then, it might be an overestimation of the numbers of dwarfs in the reference sample. If the \logg\, of the \textit{Kepler} targets are systematically overestimated, there would be a large fraction of giant and subgiant stars in our reference sample. As discussed in section \ref{CompHuber}, we have no evidence for this systematic bias in the \logg. Finally, the discrepancy with RV results could come from an overestimation of the hot Jupiter population in RV surveys, due to the minimum mass parameter rather than the true mass. Some low-mass stars with low inclination \citep[as in][]{2012A&A...538A.113D, 2013ApJ...770..119W} could contaminate the sample. This is however quite unlikely since they would produce line-profile variations \citep{2015MNRAS.451.2337S} that were monitored by \citet{2011arXiv1109.2497M}.

If this difference between the occurrence rates of hot jupiters found by \textit{Kepler} and other instruments is real, it should have a physical origin. The metallicity of the host star is well known to drive the formation rate of EGPs \citep[e.g.][]{2001A&A...373.1019S}. Therefore, if the different stellar populations probed by these surveys have significantly different metallicities, it should have an impact on the number of EGPs found. The median metallicity of dwarfs in the solar neighborhood has been found to be of about -0.08 dex \citep{2008A&A...487..373S} and $\sim$ 0 dex for both \textit{CoRoT} pointing directions \citep{2010A&A...523A..91G, 2015arXiv150602956C}. The median metallicity of the \textit{Kepler} dwarfs in our reference sample is -0.18 dex using the values from \citet{2014ApJS..211....2H} or -0.03 dex from LAMOST \citep{2014ApJ...789L...3D}.
The difference of metallicity of about 0.15 -- 0.2 dex between the \textit{Kepler} dwarfs \citep[using the metallicities from][]{2014ApJS..211....2H} and the ones from the solar neighborhood and the \textit{CoRoT} fields could well explain a factor of two in the occurrence rates of EGPs, as predicted by e.g. \citet{2005ApJ...622.1102F}. 

Recently, \citet{2015ApJ...799..229W} suggested that the difference of hot jupiters between the solar neighbourhood and the \textit{Kepler} FOV might be explained by the difference of stellar multiplicity rate, hence affecting their formation rate. To test that, we can use the fraction of detached EBs as a proxy of the stellar multiplicity rate. Using the results of \citet{2010ApJS..190....1R}, \citet{2013A&A...557A.139S} estimated a fraction of EBs with transit depth deeper than 3\% in the solar neighbourhood to be 0.53 $\pm$ 0.14\%. In the \textit{Kepler} FOV, it has been estimated to be 0.79 $\pm$ 0.02 \% using the second version of the \textit{Kepler} EB catalog \citep{2011AJ....142..160S}. In the \textit{CoRoT} fields, the value is 0.94 $\pm$ 0.02\% (Deleuil et al., in prep.). Therefore, if the stellar multiplicity rate was the reason of the difference in the occurrence rate of hot jupiters, there should be even less of those planets in the \textit{CoRoT} fields, which seems excluded. If the multiplicity affects the formation rate of EGPs, it is probably a second-order effect compared with the stellar metallicity.

The occurrence rates of period-valley giants also show about a factor of two between our value and the ones of \citet{2011arXiv1109.2497M} and \citet{2013ApJ...766...81F}. With only one independent estimation of the occurrence rate outside the \textit{Kepler} FOV, this difference might be only the results of small numbers statistics and thus is not significant. 

In the population of the temperate EGPs, the three values are fully compatible. The occurrence rate in the \textit{Kepler} FOV is not lower by a factor of two compared with the solar neighborhood. Note however that this population of transiting planets is the most difficult one to establish by RV, since the expected amplitudes are lower than for shorter periods planets. As a result, only half of the \textit{Kepler} objects used to compute the occurrence rate in this period range are well established\footnote{For comparison, in the population of hot jupiters more than 80\% of the objects are secured.}. 

\subsection{Occurrence rate of brown dwarfs in the brown-dwarf desert}
\label{OccBDs}

We identified three transiting candidates that have a mass in the brown dwarf regime. These objects have orbital periods of less than a 400 days and thus are rare members of the so called brown-dwarf desert \citep{2002MNRAS.330L..11A}. The orbital periods probed by \textit{Kepler} and our observations correspond to the ''driest region'' of this desert \citep{2014MNRAS.439.2781M, 2015A&A...580A.125R}. In spite of their very low number, we can derive a first measurement of their occurrence rate. We followed the same procedure as for EGPs, with the same stellar reference sample (see Section \ref{RefSample}). We list in Table \ref{BDParams} the adopted parameters of these brown dwarfs and in Table \ref{CorrecFactorBD} their occurrence correction factors, similar to the ones described in section \ref{Scorr}. We derived an overall occurrence rate of brown dwarfs within 400 days of orbital period to be as low as 0.29~$\pm$~0.17~\%\footnote{Since no brown dwarfs were found with orbital period longer than 200 days, the same value would be found for brown dwarfs within 200 days of orbital period}. Therefore brown dwarfs are about 15 times less common than EGPs in the considered range of periods. 

This value is fully compatible with the one derived by \citet{2015arXiv150805763C} based on the \textit{CoRoT} data. Note however that the brown dwarfs detected by \textit{CoRoT} have orbital periods of less than 10 days, while in our sample, they have periods between 10 and 170 days. This difference is most likely due to small number statistics.


\section{The physical properties of giant planets and their hosts}
\label{physics}

In this section we analyse in more details the physical properties of the 63 EGPs (secured and likely) in our sample and discuss them in the context of the other planets characterised so far, using the physical properties listed in Table \ref{PlanetParams}.

\subsection{Mass and density of giant planets}
\label{Mrho}

\begin{figure*}[h!]
\begin{center}
\sidecaption
\includegraphics[width=12cm]{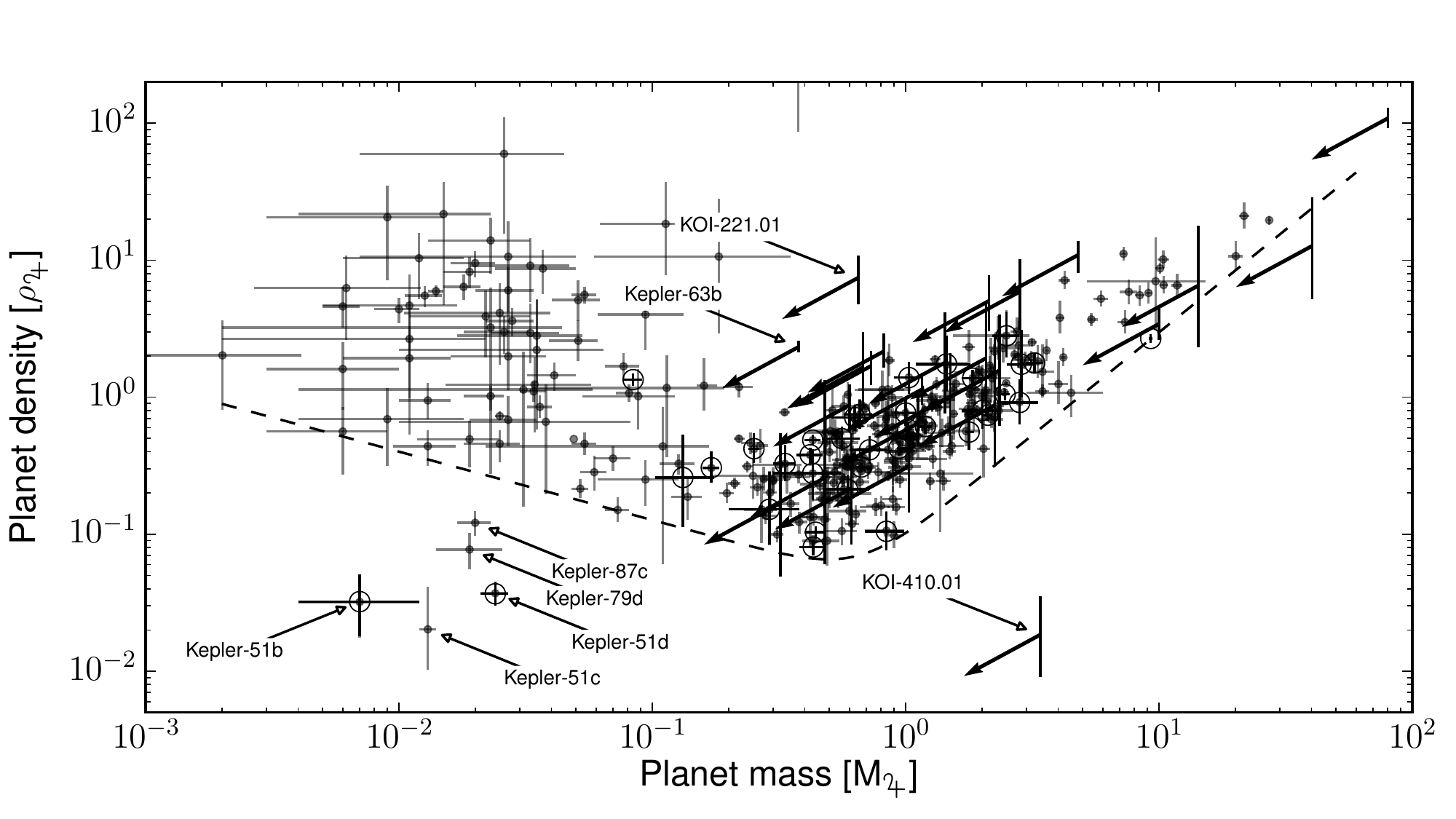}
\caption{\citeauthor{2015arXiv150605097H} diagramme of exoplanets, showing the bulk density as function of their mass. The open circles are the giant planets defined in our sample and the dots are other planets. Note that we kept as planets only those with a density characterised at a level better than 1-$\sigma$. The arrows represent the candidates in our sample for which it was only possible to derive an upper-limit on their mass (see Table \ref{upperlims}). The vertical bars attached to the arrow are the uncertainties on the densities imposed by the uncertainties on the radii. The dashed line represents the empirical lower envelope density for giant and low-mass planets (see eq. \ref{lowerenv}).}
\label{MDensGPs}
\end{center}
\end{figure*}

Our spectroscopic survey of \textit{Kepler} giant-planet candidates provides mass constraints to 40 giant exoplanets (15 well characterised and 25 upper limits). The 23 remaining planets were characterised by other means (e.g. other spectroscopic facilities or TTVs analysis) published in the literature. Combined with the radius measured by \textit{Kepler}, this allows us to derive the bulk density of these exoplanets. We display their bulk density as a function of their mass in Fig. \ref{MDensGPs}, similarly as in \citet{2015arXiv150605097H}. 

As expected, the large majority of the EGPs in our sample follow the same trend as pointed out in \citet{2015arXiv150605097H}: the bulk density of EGPs strongly correlate with their mass. This is a direct consequence of the fact that the radius of EGPs and brown dwarfs is nearly constant (within $\sim30\%$) over two decades in mass, provided they are mostly made of hydrogen and helium \citep{2005AREPS..33..493G}.

However, three objects that we considered as EGPs in our sample are clearly outliers in this diagram. We annotate their name in Fig. \ref{MDensGPs}. First, there is the case of KOI-410.01 which has a radius \citep[according to][]{2015arXiv150400707R} of 4.9~\Rjup\, for a mass upper-limit of 3.4~\Mjup\, \citep{2011A&A...533A..83B}. This gives to this candidate an extremely low density of less than 0.02~g.cm$^{-3}$. Therefore, KOI-410.01 is either a unique case of extreme inflation for a hot jupiter or, most likely, it is not a planet but a CEB. Kepler-63~b \citep{2013ApJ...775...54S} and KOI-221.01 (this work) are well above the giant-planet branch. These two objects are most likely low-mass planets contaminating our giant-planet sample. These three outliers with no mass determination are probably not EGPs and will not be considered in the rest of the discussion.

Using all the planets well characterised so far, we find an empirical lower envelope from the density -- mass diagram of planets (see Fig. \ref{MDensGPs}). This lower envelope is of the functional form:
\begin{equation}
\label{lowerenv}
\rho_{\rm lower}\, [\rho_{\jupiter}] = \sqrt{\left(\frac{M_{p}^{1.5}}{2.2^{3}}\right)^{2} + \left(\frac{0.04}{M_{p}^{0.5}}\right)^{2}}\, ,
\end{equation}
with $M_{p}$ the planet mass expressed in \Mjup. This form was defined using all exoplanets with a mass constrained at better than 1-$\sigma$. Considering only those constrained at better than 3-$\sigma$ does not change the form of this lower enveloppe.

Assuming this lower density envelope for exoplanets, we find that Kepler-63~b and KOI-221.01 have lower limits in mass of 0.035~\Mjup\,(11~\Mearth) and 0.019~\Mjup\,(6.2~\Mearth), respectively. For a given mass, we expect to find close to this limit objects which have the highest fraction of hydrogen and helium (i.e. lowest fraction of heavy elements), highest irradiation level and youngest age of the sample \citep[see][]{2005AREPS..33..493G}. Evaporation also plays a role and could explain the functional form of Eq.~\eqref{MDensGPs} at least in the small mass domain \citep[see][]{2012ApJ...761...59L, 2013ApJ...775..105O}.

Applying the same lower density envelope to all the EGPs in our sample for which we only have a upper-limit constraint, we can estimate their minimum mass and thus their minimum RV amplitude. This can be used then to determine the precision needed by future follow-up observations to characterise these objects. We list in Table \ref{NoVarMinMass} the minimum mass we find for these candidates and planets together with their minimum RV amplitude assuming a circular orbit. 

Finally, from the population of giant and low-mass planets in this M -- $\rho$ diagram, five objects seem not to follow the global trend: Kepler-51~b, c, d \citep{2014ApJ...783...53M}, Kepler-79~d \citep{2014ApJ...785...15J}, and Kepler-87~c \citep{2014A&A...561A.103O}. These planets are very low-mass low-density planets. No modelling of their internal structure has been reported in the literature so far. We believe that determining the internal structure of such low-density planets might be challenging for current models but this would allow us to better understand the nature of these particular objects. We note, however, that these five planets have been characterised thanks to TTV analyses, which might be biased in the presence of an unseen (i.e. non-transiting) companion or in the presence of stellar activity \citep{2013A&A...556A..19O, 2013MNRAS.430.3032B}. Some cases of TTV-mass determination were revealed to be systematically lower than RV mass determination as pointed out by \citet{2014ApJ...783L...6W}. To date, very few objects have been characterised independently by both techniques \citep{2013ApJ...768...14W, 2014A&A...561L...1B, 2015A&A...573A.124B}. A RV follow-up of these very low-mass low-density planets might reveal a completely different nature for these objects.

\subsection{Radius vs. irradiation}
\label{Inflation}

The radii of EGPs largely depends on the irradiation level that they receive which regulates the rate at which they cool down and contract \citep{1996ApJ...459L..35G}. This is controlled both by their atmosphere and interior radiative zone, with a higher irradiation implying a warmer atmosphere and a slower contraction. However, some EGPs, like KOI-680 \citep{2015A&A...575A..71A}, exhibit a radius that cannot be explained by conventional models, up to about 2~\Rjup. The reason for this inflation of EGPs is yet not completely understood. Different physical processes are proposed such as mechanisms driven by stellar flux heating, tidal heating, or Ohmic dissipation \citep[see][for a review]{2014prpl.conf..763B}. To identify the inflation mechanism, but also to further understand how the atmosphere controls the contraction, we need to characterise EGPs over a large range of physical properties. By exploring a sample of giant transiting exoplanets up to orbital periods of 400 days, we probed planets receiving a large range of stellar insolation flux. This insolation flux is defined as:
\begin{equation}
S_{\rm eff} = \sigma_{sb}\,a^{-2}\,\mathrm{T_{eff}}^{4}\,R_{\star}^{2}\, ,
\end{equation}
with $\sigma_{sb} = 5.6704\times10^{-5}$ the Stefan -- Boltzmann constant, $a$ the semi-major axis of the planet, and \teff\, and $R_{\star}$, the effective temperature and radius of the host star.

\begin{figure}
\begin{center}
\includegraphics[width=\figw]{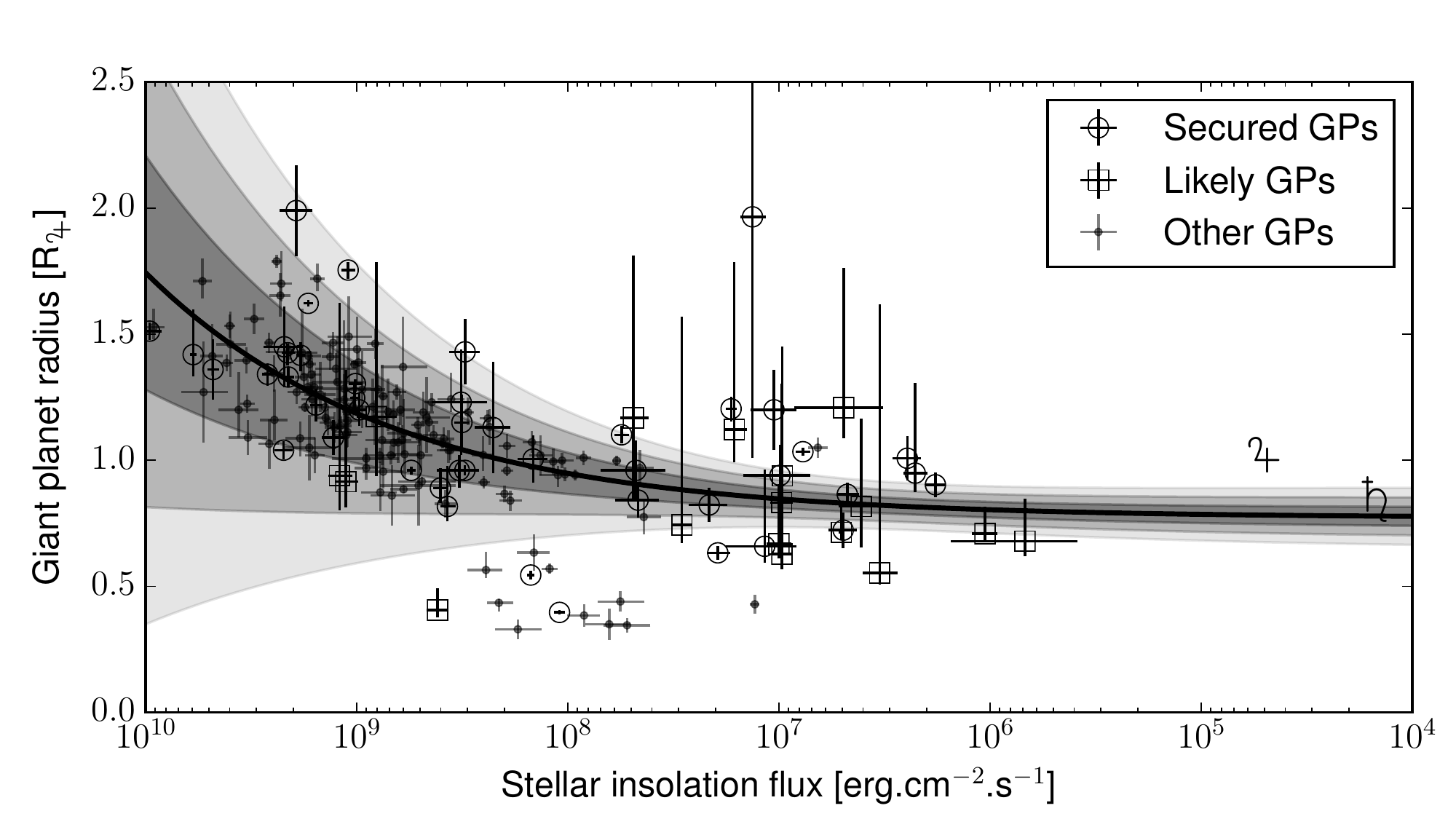}\\
\includegraphics[width=\figw]{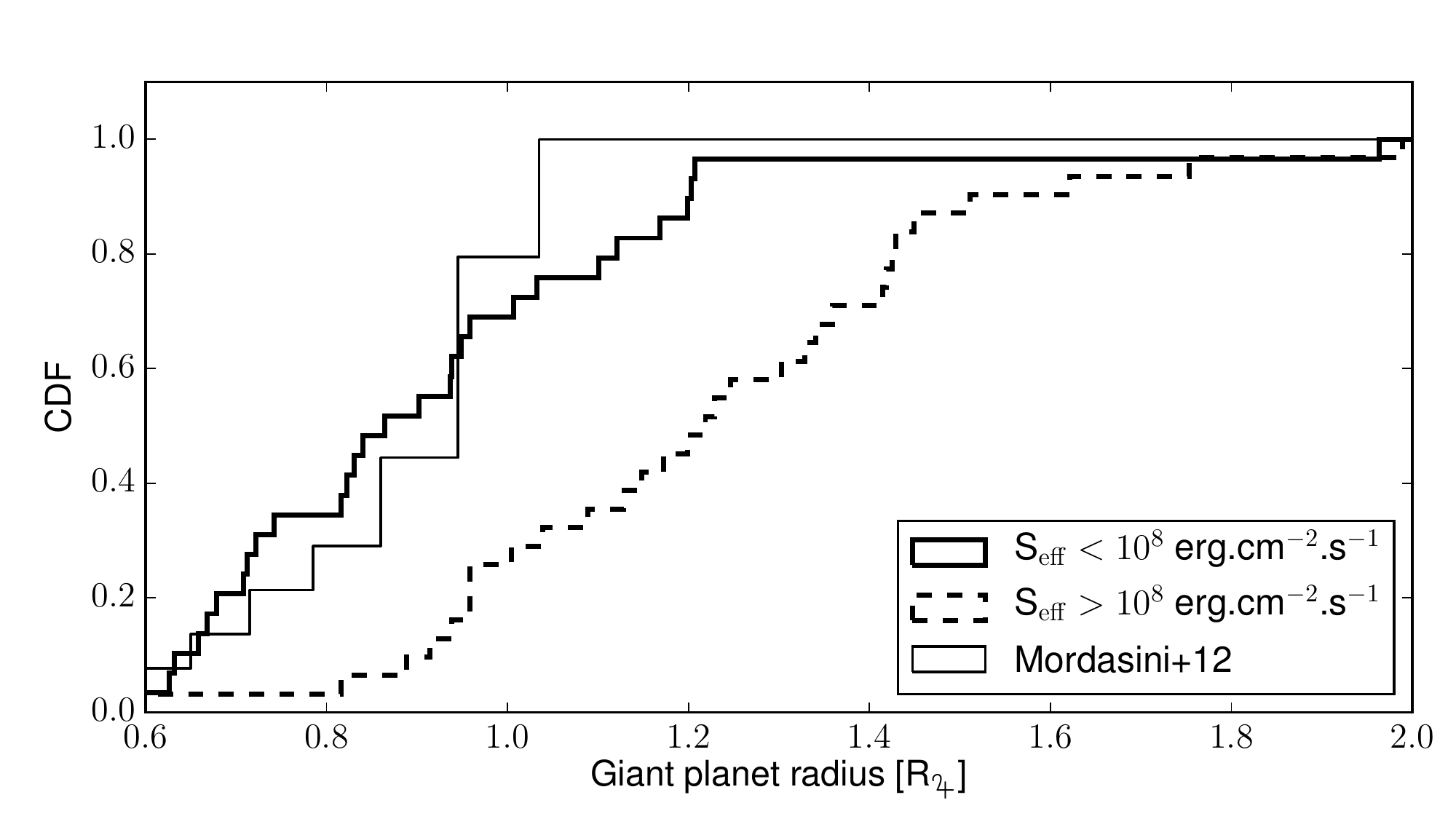}
\caption{Top panel: radius of giant planets as a function of their stellar insolation flux $S_{\rm eff}$. The open circles are the secured planets, the open squares are the likely ones and the dots are the non-\textit{Kepler} objects. The solid line represent the best model of eq. \ref{Infleq} and the greys regions represent the 1-, 2-, 3-$\sigma$ (from dark to light grey) confidence interval for this best model, as described by the covariance matrix provided in eq. \ref{Inflcov}. Bottom panel: normalised cumulative distribution (CDF) of the radius of giant planets listed in Table \ref{PlanetParams} for objects receiving more (dashed thick line) or less (solid thick line) insolation flux than $10^{8}$~erg.cm$^{-2}$.s$^{-1}$. The prediction from \citet{2012A&A...547A.112M} is shown for comparison (age of 5~Gyr - solid thin line). This prediction is for planet with orbit $>0.1$AU, which correspond to $1.4\times10^{8}$~erg.cm$^{-2}$.s$^{-1}$ for a Sun-like star.}
\label{InflationGPs}
\end{center}
\end{figure}

We display in Fig. \ref{InflationGPs} the measured radius of EGPs from our sample as a function of the stellar insolation flux they received. In this figure, we make the distinction between the secured and likely exoplanets and show the other EGPs\footnote{We selected as EGPs all objects listed in NASA exoplanet archive with a radius larger than 0.3~\Rjup.} detected and characterised by ground-based observatories like Super-WASP \citep{2006PASP..118.1407P}, HAT \citep{2004PASP..116..266B}, and from the \textit{CoRoT} space telescope \citep{2006cosp...36.3749B}.

First, we can see that thanks to $\sim$ 4.5 years of observation, \textit{Kepler} was able to explore EGPs receiving about 100 times less flux from their stars than the ones found by ground-based observatories, hence paving the way between hot jupiters and the solar system giants. The least irradiated object in our sample is KOI-1411.01 (likely a planet) with an insolation of $S_{\rm eff} \approx 6.8\times10^{5}$~erg.cm$^{-2}$.s$^{-1}$. This is only 13 times more than received by Jupiter. Among the non-Kepler detections, only CoRoT-9~b \citep{2010Natur.464..384D} is an EGP with an insolation below $10^{7}$~erg.cm$^{-2}$.s$^{-1}$. 

Then, as already pointed out by \citet{2011ApJS..197...12D}, there is a clear lack of inflated EGPs receiving a moderate irradiation (see Fig. \ref{InflationGPs}). Only KOI-3681.01 shows a radius of about 2~\Rjup\, for an insolation of $\sim10^{7}$~erg.cm$^{-2}$.s$^{-1}$, but has of a large uncertainty. However, our preliminary results show that the stellar host is not a F-IV star as reported by \citet{2014ApJS..211....2H} but a dwarf with a radius of about half the value reported in Table \ref{PlanetParams}. Thus this planet is much smaller and less irradiated than it appears here. All  planets with an insolation of $S_{\rm eff} < 10^{8}$~erg.cm$^{-2}$.s$^{-1}$ have a radius smaller than $\sim$1.2~\Rjup\,  (see Fig. \ref{InflationGPs}). A few objects show a radius of $\sim$0.5~\Rjup\, for an insolation of $S_{\rm eff} \approx 10^{8}$~erg.cm$^{-2}$.s$^{-1}$, but they are likely of different composition than H -- He gas giants.

In Fig. \ref{InflationGPs} (lower panel), we display the cumulative histograms of the EGP radius (both likely and secured planets) for the ones receiving more or less insolation than $10^{8}$~erg.cm$^{-2}$.s$^{-1}$. The Anderson -- Darling test gives a $p$-value at the level of 2$\times10^{-4}$ that they have the same distribution. This clearly shows that the atmosphere of EGPs receiving a high insolation are dominated by different physical processes than the low-insolation ones. 

Fitting all the planets displayed in Fig. \ref{InflationGPs}, we find that the distribution of EGP radius might be modelled as a function of the insolation flux with the following relation:
\begin{equation}
\label{Infleq}
R_{p} = a_{s} \times\left(S_{\rm eff}\right)^{b_{s}} + c_{s}\, ,
\end{equation}
with $a_{s} = 1.895\times10^{-4}$, $b_{s} = 0.371$, $c_{s} = 0.772$, and the following covariance matrix, derived by bootstrapping the planets 10$^{5}$ times\footnote{Note that Jupiter and Saturn were also included in the fit. They do not change significantly the results}:
\begin{dmath}
\label{Inflcov}
cov(a_{s}, b_{s}, c_{s}) = \left\|\begin{array}{ccc}3.59\times10^{-8} & -5.46\times10^{-6} & -6.31\times10^{-6} \\-5.46\times10^{-6} & 1.08\times10^{-3} & 1.14\times10^{-3} \\-6.31\times10^{-6} & 1.14\times10^{-3} & 1.67\times10^{-3}\end{array}\right\| .
\end{dmath}

This shows that EGPs in the \textit{Kepler} field with moderate and low irradiation ($S_{\rm eff} < 10^{8}$~erg.cm$^{-2}$.s$^{-1}$) tend to have a radius of $c_{s} = $0.77~$\pm$~0.04~\Rjup. This value is smaller than the one predicted by \citet{2012A&A...547A.112M} of $\sim$1~\Rjup. Computing the Anderson -- Darling test between the two distributions\footnote{This was done by drawing a statistically-large number (in this case 10$^{5}$) of planets from the interpolated distribution of \citet{2012A&A...547A.112M} at 5 Gyr.}, we find a $p$-value that they are similar at the level of $5\times10^{-3}$. This means the two distributions are significantly different. The main difference resides in the fact that \citet{2012A&A...547A.112M} predicted a pile-up of EGPs at about 1~\Rjup, while the observed distribution is nearly uniform between 0.6~\Rjup\, and 1.2~\Rjup. Note however that the radius uncertainty is relatively high and might explain this discrepancy. 

One might argue that our selection criteria might have biased this value, but given the absence of inflated planets in the regime of moderate irradiation, we might only have missed EGPs with transit depth shallower than 0.4\%. If they exist, they would be even smaller than our selected ones, hence increasing the discrepancy. The choice of the functional form might also bias this value. Computing the median radius for planets with $S_{\rm eff} < 10^{8}$~erg.cm$^{-2}$.s$^{-1}$, we find a value of 0.865~$\pm$~0.05~\Rjup(see Fig. \ref{InflationGPs}), still compatible with $c_{s}$. This would indicate that the mean radius of EGPs receiving a moderate or low irradiation is smaller than the one predicted in \citet{2012A&A...547A.112M}. The precise characterisation and modelling of these objects \citep[as done in e.g.][]{2011A&A...531A...3H} receiving a low irradiation should allow us to better understand the physics of the atmosphere of EGPs and provide new insights to planet formation. 

\subsection{Planet -- host properties}
\label{hostprop}

One of the main ingredients of EGP formation is the metallicity of the disk \citep[e.g.][]{2012A&A...541A..97M, 2015arXiv150207585N}. The correlation between the fraction of giant-planet hosts and their metallicity was revealed early in the solar neighborhood \citep{2001A&A...373.1019S}, and revised as the number of detections increases \citep{2005ApJ...622.1102F, 2010PASP..122..905J, 2011A&A...533A.141S, 2013A&A...551A.112M}. Using our sample of secured giant transiting exoplanets, we might test this correlation in the \textit{Kepler} field of view.

\begin{figure}[h]
\begin{center}
\includegraphics[width=\figw]{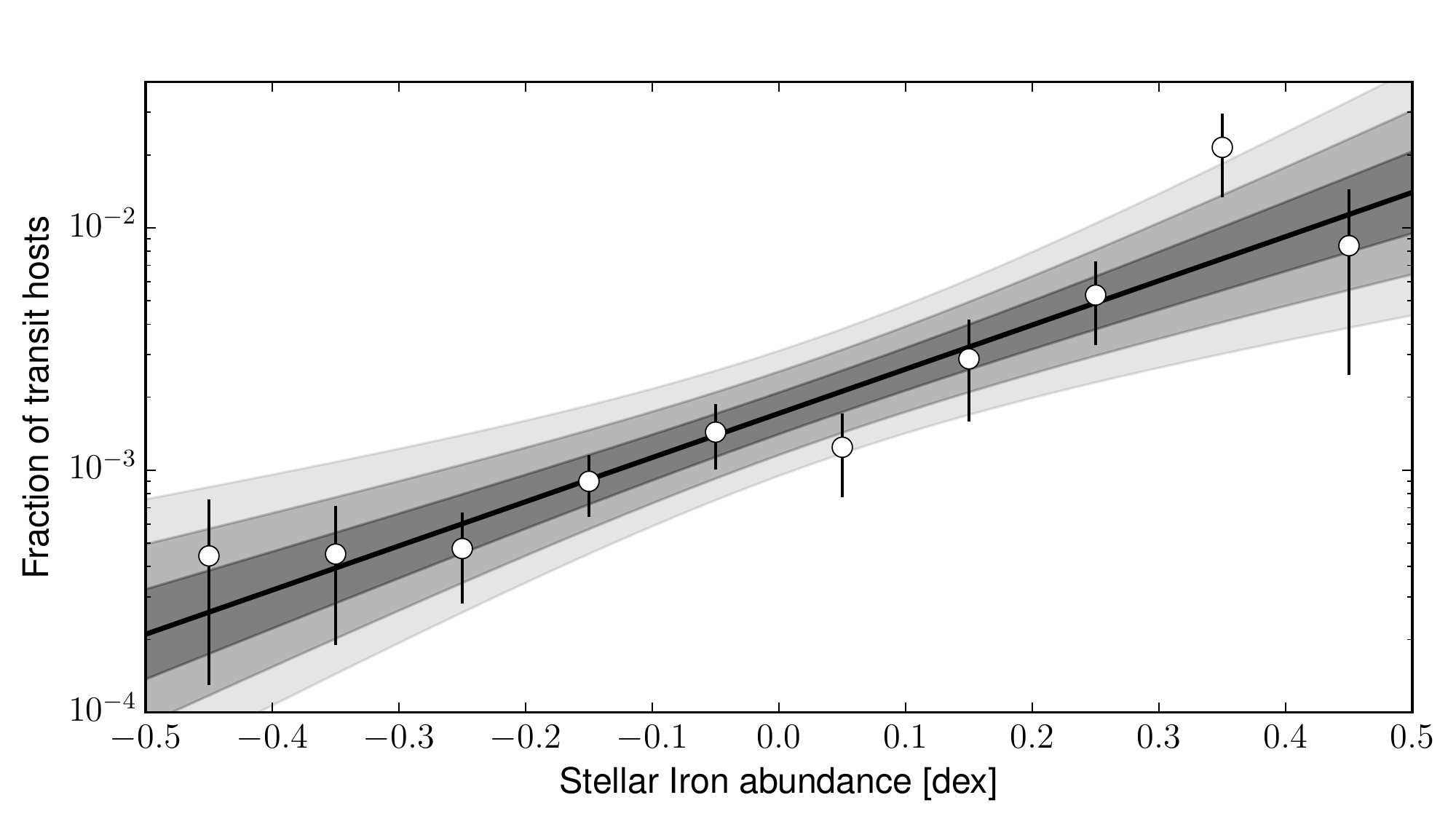}
\caption{Fraction of dwarf stars transited by a giant planet as a function of the stellar iron abundance. The solid line is the best model found and the grey regions represent the 1-, 2-, and 3-$\sigma$ confidence intervals (from dark to light grey).}
\label{PropHosts}
\end{center}
\end{figure}

For homogeneity, we used only the metallicity reported for all the \textit{Kepler} targets by \citet{2014ApJS..211....2H} for both exoplanet hosts and field stars, selecting only the dwarf stars that respect the criteria defined in Section \ref{RefSample}. We then compute the fraction of transiting giant-planet hosts as a function of iron abundance (Fig. \ref{PropHosts}). The planet-metallicity correlation is clearly visible in the \textit{Kepler} giant-planet sample. We fitted this correlation with a power law of the form:
\begin{equation}
\log_{10}\left(f({\rm [Fe/H]})\right) = a_{F}\times{\rm [Fe/H]} + b_{F}\, ,
\end{equation}
with $a_{F} = 1.82$, $b_{F} = -2.77$, and the following covariance matrix, obtained by bootstrapping 1000 times the values within their uncertainties:
\begin{equation}
cov(a_{F}, b_{F}) = \left\|\begin{array}{cc}0.097 & -0.003 \\ -0.003 & 0.007 \end{array}\right\|\, .
\end{equation}
Our value of $a_{F}$ is fully compatible with all the values reported for the solar neighborhood in the aforementioned papers. Since we are using here transit hosts, the value of $b_{F}$ cannot be compared directly with RV results. Indeed, our value of $b_{F}$ integrates the transit probability over the entire sample. 

We also explored the correlation between the occurrence of planets and the mass of the host. This correlation is well established in RV detections but we failed in confirming it. Assuming a correlation with a functional form of $f(M_{\star}) = a_{M}M_{\star}^{b_{M}}$ \citep[as in, e.g.][]{2010PASP..122..905J, 2013A&A...551A.112M}, we find that $b_{M} = 1.9 \pm 1.3$. We are limited here by a lack of precision and a too small sample.

\begin{figure}[h]
\begin{center}
\includegraphics[width=\figw]{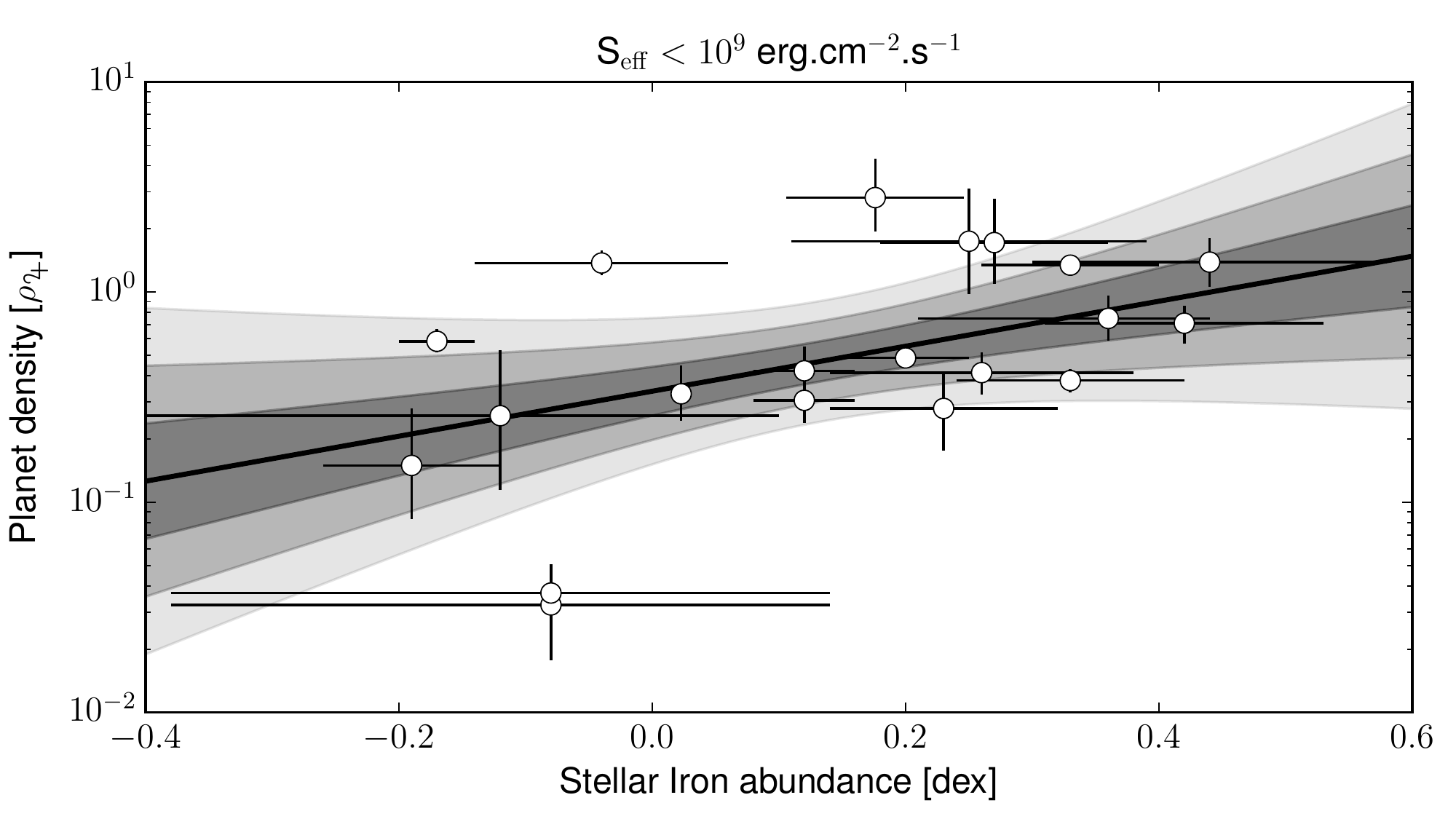}
\caption{Correlation between the bulk density of giant planets receiving a moderate irradiation as a function of the iron abundance of their host star. The solid line is the best model found and the grey regions represent the 1-, 2-, and 3-$\sigma$ confidence intervals (from dark to light grey).}
\label{PropDens}
\end{center}
\end{figure}

Finally, we search for a possible correlation between the density of EGPs and their host star metal content, directly from the observational data and independently of any model. A correlation between the core mass and the stellar metalicity has been proposed earlier on the basis of a comparison between theoretical interior models and observations of transiting planets \citep{2006A&A...453L..21G,2007ApJ...661..502B,2011ApJ...736L..29M,2013Icar..226.1625M}. Except for highly-irradiated planets, the bulk density of the planet might be used as a proxy of their core mass (for the same age, a planet with a massive core will have a higher density than a planet with no core).
We hence selected the objects in our giant-planet sample with a stellar insolation $S_{\rm eff} < 10^{9}$~erg.cm$^{-2}$.s$^{-1}$ (Fig. \ref{PropDens})\footnote{If we limit to $S_{\rm eff} < 10^{8}$~erg.cm$^{-2}$.s$^{-1}$, we don't have enough well characterised planets to search for a correlation. None of the planets in our sample with $S_{\rm eff} < 10^{9}$~erg.cm$^{-2}$.s$^{-1}$ exhibits a radius larger than $\sim$1.2~\Rjup. Thus we can consider that this sub-sample of planets are not significantly inflated.}.

We fitted this correlation with a model of the form:
\begin{equation}
\log_{10}\rho_{p} = a_{\rho}\times{\rm [Fe/H]} + b_{\rho}\, ,
\end{equation}
with $\rho_{p}$ the density of the planet expressed in Jupiter unit, $a_{\rho} = 1.07$, $b_{\rho} = -0.47$, and the following covariance matrix, obtained by bootstrapping 1000 times the values within their uncertainties:
\begin{equation}
cov(a_{\rho}, b_{\rho}) = \left\|\begin{array}{cc}0.24 & -0.03 \\ -0.03 & 0.01 \end{array}\right\|\, .
\end{equation}
This corresponds only to a hint of correlation at the 2.2-$\sigma$ level. Note that if we removed the two low-mass EGPs in the Kepler-51 system (see Section \ref{Mrho}), this significance drops to 1.3-$\sigma$. To confirm this possible correlation, it is important to characterise more EGPs in this regime of low irradiation.


\section{Summary and conclusion}
\label{conclusion}

In this paper, we studied the physical properties of giant exoplanets orbiting within 400 days of period. For that, we used the latest catalog of \textit{Kepler} transit candidates (Coughlin et al., in prep.) in which we defined  sample of EGP candidates (see Section \ref{sample}) as the 129 candidates having a transit depth between 0.4\% and 3\%, a period up to 400 days, and a host star brighter than K$_{p}=14.7$. These 129 objects orbit 125 different stars. We performed an extensive RV follow-up of these candidates using the SOPHIE spectrograph on the 1.93-m telescope of the Observatoire de Haute-Provence during 6 observing seasons. This allowed us to unveil the nature of the candidates and we found that 45 bona-fide planets (30 already known and 15 new ones characterised by our team), 3 sub-stellar companions that are likely brown dwarfs, 63 multiple stellar systems (SB1, SB2, and SB3) out of which 48 are eclipsing binaries and 15 are more complex stellar systems. Finally, for 18 objects, we rejected that they are grazing EBs or brown dwarfs, but we could not establish their planetary nature. For these, we were able to put an upper-limit on their mass. 

We then derived a false-positive rate of 54.6~$\pm$~6.5~\% for the EGPs, with a value ranging from 51.2~$\pm$~6.3~\% to 65.1~$\pm$~7.1\% depending on the true nature of the unsolved cases (see Section \ref{FPR}). This value is significantly higher than all the previously derived values \citep{2011ApJ...738..170M, 2012AA...545A..76S, 2013ApJ...766...81F}. We argued that this higher rate of false positives could have a significant and non-uniform impact on the various planet populations derived by \citet{2013ApJ...766...81F}. We also showed that the absolute number of false positives is expected to increase towards candidates with a smaller radius. 

In section \ref{ComparHosts}, we compared the properties of the false-positive, giant-planet, candidates hosts with the ones of the target stars. We found no statistical difference between the metallicity distribution of false-positive, candidate hosts and the target stars, while there is a difference between giant-planet host properties and other categories. This implies that either the fraction of candidate host (in which a majority of candidates are smaller than Neptune) does not depend on metallicity or -- more speculatively -- there is a substantial fraction of false positives among the catalog of planet candidates. Therefore, the nature of the candidates should be carefully scrutinised before inferring exoplanet properties, that can then be used to constrain planet formation models.

Thanks to our spectroscopic survey of the giant-planet candidates detected by \textit{Kepler}, we cleaned this sample from false positives. This allowed us to derive an occurrence rate of EGPs orbiting F5 -- K5 dwarfs within 400-day periods of 4.6~$\pm$~0.6\%. By computing this occurrence rate as a function of orbital periods, we recovered the three populations of giant planets already identified by RV surveys in the solar neighborhood \citep[e.g.][]{2003A&A...407..369U}: the hot jupiters orbiting with periods of up to 10 days, the period-valley giants with periods between 10 and $\sim$ 85 days, and the population of temperate giants with periods longer than $\sim$ 100 days. We note that these populations of giant planets, in particular the pile-up of hot jupiters, were not recovered in previous studies of the \textit{Kepler} candidates \citep{2012ApJS..201...15H, 2013ApJ...766...81F}. This was only possible because we rejected from our sample more than half of the candidates as false positives, which have a different period distribution.

The occurrence rate of hot jupiters in the \textit{Kepler} field seems systematically lower by a factor of $\sim$2 compared with other surveys. Even if this difference is not statistically significant, it could result from the EGP -- host metallicity correlation \citep[e.g.][]{2001A&A...373.1019S, 2005ApJ...622.1102F}. This lower occurrence rate of hot jupiters compared with, e.g. the HARPS and CORALIE RV survey \citep{2011arXiv1109.2497M}, has no counterparts for EGPs with orbital period longer than $\sim$ 85 days. The reasons for these differences, if real, might be caused by the mechanisms forming only hot jupiters. 

In the section \ref{OccBDs}, we provided an estimate of the formation rate of brown dwarfs in the brown-dwarf desert, at the level of 0.29~$\pm$~0.17~\% for orbital periods of less than 400 days.

Finally, in section \ref{physics}, we studied the physical properties of the EGPs in our sample and the ones of their host stars. We find that EGPs receiving an insolation lower than $S_{\rm eff} < 10^{8}$~erg.cm$^{-2}$.s$^{-1}$ are not inflated and exhibit a median radius of $\sim$ 0.8~\Rjup. This confirms the results of \citet{2011ApJS..197...12D}, with more objects receiving a moderate irradiation and that were all filtered out by spectroscopic means. Interestingly, we find that the radius distribution of EGPs with a moderate irradiation is significantly different from the one predicted by \citet{2012A&A...547A.112M}. These planets are found to be, in average, smaller than predicted. The detailed characterisation of the internal structure of these moderate-irradiation planets should provide new constraints to planet formation and evolution theories. In particular, we find a hint of correlation between the bulk density of these planets and the metallicity of the host stars. This correlation needs however to be confirmed with more planets in this regime. We also confirm that the EGP -- host metallicity correlation previously found by RV surveys \citep[e.g.][]{2011A&A...533A.141S, 2012AJ....143..111J, 2013A&A...551A.112M} holds for the transit exoplanet population in the \textit{Kepler} field. This suggests that similar formation processes are at work in both fields, at least for EGPs.

To probe any relation between the occurrence of EGPs and the properties of the stellar field, it is mandatory to explore more stellar populations. Measuring the abundance of $\alpha$-elements of the hosts could also provide information about the galactic populations these planet hosts belong to \citep{2012A&A...547A..36A}. Different stellar populations are currently observed by the  \textit{K2} mission \citep{2014PASP..126..398H} and will be explored by the upcoming space missions \textit{TESS} and \textit{PLATO}. This will give us a unique opportunity to further probe the difference of properties of planet population in different stellar environments. This would, however, require a deep characterisation of the stellar fields, which will be possible thanks to the \textit{GAIA} mission. Similar studies using the ground-based photometric surveys like SuperWASP \citep{2006PASP..118.1407P} and HAT-Net \citep{2004PASP..116..266B} could also provide us unprecedented constraints on EGP formation and migration, but the calibration of the detection limits might be challenging because of the lack of uniformity in the data. Note that \textit{TESS} \citep{2015JATIS...1a4003R}  and \textit{PLATO} \citep{2014ExA....38..249R} will be observing much brighter stars than the \textit{Kepler} targets. Hence we will have access to better RV precision with the SOPHIE spectrograph, allowing us to do a similar work on the populations of smaller planets, down to hot super-Earths and warm neptunes \citep{2015arXiv150607144C}.

\begin{acknowledgements}
We thank the referee for his/her time reviewing this quite long paper and for the fruitful comments and suggestion that improved the quality of this manuscript. AS thanks Jon Jenkins, Natalie Batalha, and Jessie Christiansen for their help concerning the \textit{Kepler} data, catalog of KOIs and pipeline completeness. AS also thanks Gibor Basri for his suggestion to derive the occurrence rate of brown dwarfs, Maxime Marmier for the updated detection list of the HARPS and CORALIE survey, and Pedro Figueira for fruitful discussions and his valuable comments on the manuscript. We are all grateful to the staff at the Observatoire de Haute-Provence maintaining the SOPHIE spectrograph and the 1.93-m telescope. In particular, we acknowledge the difficult but essential work of the telescope operators: Jean-Pierre Troncin, St\'ephane Favard, Didier Gravallon (a.k.a ``Le Didou de l'OHP''), and Florent Zapillon. 

The Porto group acknowledges the support by the European Research Council/European Community under the FP7 through Starting Grant agreement number 239953 and support from Funda\c{c}\~ao para a Ci\^encia e a Tecnologia (FCT, Portugal) in the form of grants reference SFRH/BPD/70574/2010 and PTDC/FIS-AST/1526/2014. NCS also acknowledges the support from FCT in the form of grant reference PTDC/CTE-AST/098528/2008 and through Investigador FCT contract of reference IF/00169/2012 as well as POPH/FSE (EC) by FEDER funding through the program ``Programa Operacional de Factores de Competitividade - COMPETE''. AS is supported by the European Union under a Marie Curie Intra-European Fellowship for Career Development with reference FP7-PEOPLE-2013-IEF, number 627202. ASB acknowledges funding from the European Union Seventh Framework Programme (FP7/2007 -- 2013) under Grant agreement number 313014 (ETAEARTH). SCCB acknowledges support by grants 98761 by CNES and the FCT through the Investigador FCT Contract No. IF/01312/2014. Part of this work was supported by FCT through the research grant UID/FIS/04434/2013. JMA acknowledges funding from the European Research Council under the ERC Grant Agreement n. 337591-ExTrA. 

This research has made use of the VizieR catalogue access tool, CDS, Strasbourg, France. The original description of the VizieR service was published in A\&AS 143, 23. This research has made use intensively of the NASA Exoplanet Archive, which is operated by the California Institute of Technology, under contract with the National Aeronautics and Space Administration under the Exoplanet Exploration Program.
\end{acknowledgements}


\Online

\onecolumn
\begin{landscape}
\begin{longtable}{cccccccccccl}
\caption{The sample of \textit{Kepler} candidates from the Q1 -- Q17 dataset that respect all the criteria defined in Section \ref{sample}.}\\
\hline
\hline
KIC & KOI & Kepler & Period & Depth & a/R$_{\star}$ & K$_{p}$ & \teff & \logg & \met & nature & reference\\
ID & ID & ID & [d] &  [\%] & & & [K] & [\cmss]  &  [dex]  & & \\
\hline
\endfirsthead
\caption{Continued.} \\
\hline
KIC & KOI & Kepler & Period & Depth & a/R$_{\star}$ & K$_{p}$ & \teff & \logg & \met & nature & reference\\
\hline
\endhead
\hline
\endfoot
\hline
\hline
\endlastfoot
11446443  &  1.01  &  1b  &  2.471  &  1.42  &  8.39$^{_{+0.03}}_{^{-0.03}}$  &  11.3  &  5850$^{_{+50}}_{^{-50}}$  &  4.46$^{_{+0.03}}_{^{-0.03}}$  &  -0.15$^{_{+0.10}}_{^{-0.10}}$  &  planet  &  OD06\\10666592  &  2.01  &  2b  &  2.205  &  0.67  &  4.71$^{_{+0.00}}_{^{-0.00}}$  &  10.5  &  6350$^{_{+80}}_{^{-80}}$  &  4.02$^{_{+0.01}}_{^{-0.01}}$  &  0.26$^{_{+0.08}}_{^{-0.08}}$  &  planet  &  P\'a08\\10748390  &  3.01  &  3b  &  4.888  &  0.43  &  16.72$^{_{+0.05}}_{^{-0.05}}$  &  9.2  &  4777$^{_{+93}}_{^{-91}}$  &  4.59$^{_{+0.01}}_{^{-0.04}}$  &  0.32$^{_{+0.10}}_{^{-0.14}}$  &  planet  &  Ba10\\6922244  &  10.01  &  8b  &  3.522  &  0.94  &  7.67$^{_{+0.20}}_{^{-0.20}}$  &  13.6  &  6225$^{_{+114}}_{^{-158}}$  &  4.17$^{_{+0.06}}_{^{-0.05}}$  &  -0.04$^{_{+0.12}}_{^{-0.16}}$  &  planet  &  Je10\\5812701  &  12.01  &  --  &  17.855  &  0.92  &  19.99$^{_{+0.02}}_{^{-0.02}}$  &  11.4  &  6635$^{_{+142}}_{^{-214}}$  &  4.19$^{_{+0.14}}_{^{-0.33}}$  &  -0.04$^{_{+0.25}}_{^{-0.36}}$  &  planet  &  Bou15\\9941662  &  13.01  &  13b  &  1.764  &  0.46  &  4.34$^{_{+0.04}}_{^{-0.04}}$  &  10.0  &  9107$^{_{+257}}_{^{-425}}$  &  3.87$^{_{+0.23}}_{^{-0.15}}$  &  0.07$^{_{+0.14}}_{^{-0.65}}$  &  planet  &  Sh11\\10874614  &  17.01  &  6b  &  3.235  &  1.08  &  7.54$^{_{+0.00}}_{^{-0.00}}$  &  13.3  &  5640$^{_{+99}}_{^{-110}}$  &  4.24$^{_{+0.05}}_{^{-0.05}}$  &  0.34$^{_{+0.10}}_{^{-0.16}}$  &  planet  &  Du10\\8191672  &  18.01  &  5b  &  3.548  &  0.75  &  6.43$^{_{+0.01}}_{^{-0.01}}$  &  13.4  &  6290$^{_{+105}}_{^{-120}}$  &  4.07$^{_{+0.04}}_{^{-0.05}}$  &  0.04$^{_{+0.10}}_{^{-0.16}}$  &  planet  &  Ko10\\
\end{longtable}%
\tablebib{Al15: \citet{2015A&A...575A..71A}; Ba10: \citet{2010ApJ...710.1724B}; Bo11: \citet{2011A&A...533A..83B}; Bo12: \citet{2012A&A...538A..96B}; Bog15: \citet{2015arXiv150601668B}; Bou15: \citet{2015arXiv150404130B}; Br15: \citet{2015A&A...573A.124B}; Ca14: \citet{2014ApJ...781...18C}; Da14: \citet{2014ApJ...791...89D}; D\'e11: \citet{2011ApJS..197...14D}; De14: \citet{2014A&A...564A..56D}; De+: Demangeon et al. (in prep.); D\'i13: \citet{2013A&A...551L...9D}; D\'i14: \citet{2014A&A...572A.109D}; Du10: \citet{2010ApJ...713L.136D}; En11: \citet{2011ApJS..197...13E}; En14: \citet{2014ApJ...795..151E}; Fa13: \citet{2013ApJ...771...26F}; Fo11: \citet{2011ApJS..197....9F}; Ga13: \citet{2013A&A...557A..74G}; H\'e13: \citet{2013A&A...554A.114H}; H\'e14: \citet{2014A&A...572A..93H}; H\'e+: H\'ebrard et al. (in prep.); Ho10: \citet{2010Sci...330...51H}; Je10: \citet{2010ApJ...724.1108J}; Ko10: \citet{2010ApJ...713L.131K}; La10: \citet{2010ApJ...713L.140L}; Mo13: \citet{2013A&A...558L...6M}; OD06: \citet{2006ApJ...651L..61O}; Of14: \citet{2014A&A...561A.103O}; P\'a08: \citet{2008ApJ...680.1450P}; Ro14: \citet{2014ApJ...784...45R}; Sa11: \citet{2011A&A...536A..70S}; Sa12: \citet{2012AA...545A..76S}; Sa14: \citet{2014A&A...571A..37S}; Sc14: \citet{2014ApJ...795..167S}; Sh11: \citet{2011AJ....142..195S}; SO13: \citet{2013ApJ...775...54S}; St13: \citet{2013MNRAS.428.1077S}; Ti14: \citet{2014A&A...567A..14T}; Wa13: \citet{2013ApJ...776...10W}; We13: \citet{2013ApJ...768...14W}; Z\&H13: \citet{2013ApJ...776L..35Z}.}
\tablefoot{The orbital period, transit depth, system scale a/R$_{\star}$ are from the Q1 -- Q17 data (Coughlin et al., in prep.) as provided by the NASA exoplanet archive. The Kepler magnitude, K$_{p}$, the host effective temperature, \teff, the host surface gravity, \logg, and the host iron abondance, \met, are from \citet{2014ApJS..211....2H}, except for KOI-51 which is not available. BD means brown dwarf, EB means eclipsing binary, and CEB means contaminating eclipsing binary (chance-aligned or physically bound). The question marks stand for candidates for which the nature is still uncertain because of no significant RV variation.}
\end{landscape}

\begin{table}[h]
\caption{Compilation of the results from the spectroscopic analyses performed in the context of the spectroscopic follow-up of \textit{Kepler} giant-planet candidates.}
\begin{center}
\begin{tabular}{llccccll}
\hline
\hline
KIC &KOI & \teff & \logg & \met & \vsini &  Method & References \\
  & &  [K]  &  [\cmss] &  [dex]  & [\kms] & & \\
\hline
 5812701  &  12  &  6800  $\pm$  120  &  4.34  $\pm$  0.15  &  0.09  $\pm$  0.15  &  60.0 $\pm$ 1.5 & VWA  &  Bou15\\
 9818381  &  135  &  6050  $\pm$  100  &  4.40  $\pm$  0.10  &  0.40  $\pm$  0.10  & 5.5 $\pm$ 1.5 &  MOOG  & Bon15\\
 5446285  &  142  &  5460  $\pm$  70  &  4.60  $\pm$  0.20  &  0.25  $\pm$  0.09  & 2 $\pm$ 1 &  VWA  &  Ba14\\
 5357901  &  188  &  5170  $\pm$  70  &  4.50  $\pm$  0.15  &  0.24  $\pm$  0.11  & 3 $\pm$ 1 &  VWA  &  H\'e14\\
 11391018 &  189  &  4850  $\pm$  100  &  4.60  $\pm$  0.12  &  -0.07  $\pm$  0.12 &  2.5 $\pm$ 1.5 &  VWA  & D\'i14\\
 7950644  &  192  &  5800  $\pm$  70  &  4.15  $\pm$  0.15  &  -0.19  $\pm$  0.07  &  3 $\pm$ 1 & VWA  &  H\'e14\\
 11502867 &  195  &  5725  $\pm$  90  &  4.50  $\pm$  0.15  &  -0.21  $\pm$  0.08 &  3 $\pm$ 1 &  VWA  &  H\'e14\\
 9410930  &  196  &  5750  $\pm$  100  &  4.20  $\pm$  0.10  &  0.38  $\pm$  0.11  & 6 $\pm$ 2  &  MOOG  & Bon15\\
 2987027  &  197  &  4995  $\pm$  126  &  4.77  $\pm$  0.30  &  -0.11  $\pm$  0.06  &  11 $\pm$ 1 &  MOOG  &  Sa12\\
 6046540  &  200  &  6000  $\pm$  100  &  4.50  $\pm$  0.10  &  0.42  $\pm$  0.11  & 5 $\pm$ 1 &  MOOG  & Bon15\\
 6849046  &  201  &  5526  $\pm$  231  &  4.52  $\pm$  0.40  &  0.28  $\pm$  0.15  &  9 $\pm$ 1 &  MOOG  &  Sa12\\
 7877496  &  202  &  5750  $\pm$  90  &  4.30  $\pm$  0.07  &  0.27  $\pm$  0.12 &  5 $\pm$ 1 &  VWA  & De14\\
 9305831  &  204  &  5800  $\pm$  100  &  4.10  $\pm$  0.10  &  0.15  $\pm$  0.10 & 4 $\pm$ 2  &  MOOG  & Bon15\\
 7046804  &  205  &  5400  $\pm$  75  &  4.70  $\pm$  0.10  &  0.18  $\pm$  0.12  & 2 $\pm$ 1  &  MOOG  & Bon15\\
 5728139  &  206  &  6340  $\pm$  140  &  4.00  $\pm$  0.30  &  0.06  $\pm$  0.19 &  11 $\pm$ 1 &  VWA  & Al15\\
 10723750 &  209  &  6260  $\pm$  80  &  4.40  $\pm$  0.11  &  0.10  $\pm$  0.13 &  6.8 $\pm$ 1 &  VWA  & Br15\\
 6471021  &  372  &  5776  $\pm$  46  &  4.12  $\pm$  0.12  &  -0.06  $\pm$  0.04 &  4.1 $\pm$ 1.2 &  MOOG  & This work\\
 6289650  &  415  &  5810  $\pm$  80  &  4.50  $\pm$  0.20  &  -0.24  $\pm$  0.11  &  1.1 $\pm$ 1 &  VWA  & Mo13\\
 9478990  &  423  &  6360  $\pm$  100  &  4.40  $\pm$  0.15  &  0.10  $\pm$  0.14  & 16.0 $\pm$ 0.3 &  MOOG  & Bon15\\
 10418224 &  428  &  6510  $\pm$ 100  &  4.10  $\pm$  0.20  &  0.10  $\pm$  0.15  &  9 $\pm$ 2 &  VWA  & Sa11\\
 7368664  &  614  &  5970  $\pm$  100  &  4.22  $\pm$  0.10  &  0.35  $\pm$  0.15  &  3 $\pm$ 1 &  VWA  & Al15\\
 7529266  &  680  &  6090  $\pm$  110  &  3.50  $\pm$  0.10  &  -0.17  $\pm$  0.10 &  6 $\pm$ 1 &  VWA  & Al15\\
 7906882  &  686  &  5750  $\pm$  120  &  4.50  $\pm$  0.15  &  0.02  $\pm$  0.12 &  3.5 $\pm$ 1.0 &  VWA  & D\'i14\\
 5358624  &  830  &  5150  $\pm$  100  &  5.00  $\pm$  0.40  &  0.09  $\pm$  0.17  & 2 $\pm$ 2 &  VWA  &  H\'e14\\
 757450   &  889  &  5200  $\pm$  100  &  4.60  $\pm$  0.15  &  0.30  $\pm$  0.12  & 3.5 $\pm$ 1.5 &  MOOG  & Bon15\\
 3247268  &  1089  &  6027  $\pm$  169  &  4.23  $\pm$  0.29  &  0.31  $\pm$  0.13 &  3.5 $\pm$ 1.2 &  MOOG  & This work\\
 8751933  &  1257  &  5520  $\pm$  80  &  4.32  $\pm$  0.10  &  0.27  $\pm$  0.09 &  4 $\pm$ 2 &  VWA  &  Sa14\\
 8631160  &  1271  &  6600  $\pm$  122  &  4.28  $\pm$  0.23  &  -0.06  $\pm$  0.09  &  4.6 $\pm$ 1.2 &  MOOG  & This work\\
 7303287  &  1353  &  6326  $\pm$  126  &  4.50  $\pm$  0.25  &  0.33  $\pm$  0.09 &  5.3 $\pm$ 1.2 &  MOOG  & This work\\
 9425139  &  1411  &  5687  $\pm$  146  &  4.32  $\pm$  0.29  &  0.47  $\pm$  0.12 &  4.6 $\pm$ 1.2 &  MOOG  & This work\\
 11075279 &  1431  &  5507  $\pm$  74  &  4.36  $\pm$  0.25  &  0.27  $\pm$  0.05  &  4.6 $\pm$ 1.2 &  MOOG  & This work\\
 10005758 &  1783  &  6298  $\pm$  150  &  4.30  $\pm$  0.30  &  0.49  $\pm$  0.12  &  3.0 $\pm$ 1.2 &  MOOG  & This work\\
 2975770  &  1788  &  4890  $\pm$  232  &  4.58  $\pm$  0.59  &  0.05  $\pm$  0.39  &  6.2 $\pm$ 1.2 &  MOOG  & This work\\
 12735740 &  3663  &  5649  $\pm$  162  &  4.27  $\pm$  0.34  &  0.14  $\pm$  0.12  &  3.4 $\pm$ 1.2 &  MOOG  & This work\\
 10795103 &  3683  &  6666  $\pm$  203  &  4.39  $\pm$  0.30  &  0.24  $\pm$  0.16 &  8.9 $\pm$ 1.2 &  MOOG  & This work\\
 7017372  &  3689  &  6154  $\pm$  253  &  4.51  $\pm$  0.36  &  0.31  $\pm$  0.21 &  2.4 $\pm$ 1.2 &  MOOG  & This work\\
 12645761 &  5976  &  4753  $\pm$  90  &  2.87  $\pm$  0.29  &  0.01  $\pm$  0.06  & 4.8 $\pm$ 1.2 & MOOG  & This work\\
\hline
\hline
\end{tabular}
\end{center}
\label{SpectroResults}
\tablefoot{The instrumental resolution of SOPHIE corresponds to a \vsini\, of $\sim$ 4 \kms\, for a solar-like star. Therefore, \vsini\, vales lower than 4 \kms\, should be considered as upper limits.}
\tablebib{Al15: \citet{2015A&A...575A..71A}; Ba14: \citet{2014A&A...561L...1B}; Bon15: \citet{2015A&A...575A..85B}; Bou15: \citet{2015arXiv150404130B}; Br15: \citet{2015A&A...573A.124B}; De14: \citet{2014A&A...564A..56D}; D\'i14: \citet{2014A&A...572A.109D}; H\'e14: \citet{2014A&A...572A..93H}; Mo13: \citet{2013A&A...558L...6M}; Sa11: \citet{2011A&A...528A..63S}; Sa12: \citet{2012AA...545A..76S}; Sa14: \citet{2014A&A...571A..37S}.}
\end{table}%

\twocolumn

\begin{table}[h]
\caption{Measured \vsini\, based on the SOPHIE CCFs.}
\begin{center}
\begin{tabular}{llccl}
\hline
\hline
KIC & KOI & \vsini & $\sigma$\vsini & Method \\
  & &  [\kms]  &  [\kms]  & \\
\hline
 11974540  &  129  & 18.98 & 0.03 & RotPro\\
  8506766  &  138  & 22.1 & 0.1 & RotPro\\
  10666242  &  198 & 4.1  & 1.2 & Bo10\\
  3937519  &  221 & 4.4 & 1.2 & Bo10\\
  11442793  &  351 & 4.1 & 1.2 & Bo10\\
  6603043  &  368 & 86.5 & 0.6 & RotPro\\
  6471021  &  372 & 4.1 & 1.2 & Bo10\\
  5779852  &  449 & 7.4 & 1.2 & Bo10\\
  8890783  &  464 & 4.8 & 1.2 & Bo10\\
  10395543  & 531 & 6.2 & 1.2 & Bo10\\
  11773022  &  620 & 5.7 & 1.2 & Bo10\\
  3247268  &  1089 & 3.5 & 1.2 & Bo10\\
  6470149  &  1230 & 4.7 & 1.2 & Bo10\\
  6665223  &  1232 & 4.1 & 1.2 & Bo10\\
  8631160  &  1271 & 4.6 & 1.2 & Bo10\\
  7303287  &  1353 & 5.3 & 1.2 & Bo10\\
  8958035  &  1391 & 5.1 & 1.2 & Bo10\\
  9425139  &  1411 & 4.6 & 1.2 & Bo10\\
  11122894  &  1426 & 2.8 & 1.2 & Bo10\\
  11075279  &  1431 & 4.6 & 1.2 & Bo10\\
  7449844  &  1452 & 36.2 & 0.2 & RotPro\\
  11702948  &  1465 & 5.4 & 1.2 & Bo10\\
  11909686  &  1483 & 3.6 & 1.2 & Bo10\\
  5475431  &  1546 & 5.9 & 1.2 & Bo10\\
  10028792  &  1574 & 3.3 & 1.2 & Bo10\\
  10005758  &  1783 & 3.0 & 1.2 & Bo10\\
  2975770  &  1788 & 6.2 & 1.2 & Bo10\\
  6716021  &  2679 & 29.8 & 0.2 & RotPro\\
  12735740  &  3663 & 3.4 & 1.2 & Bo10\\
  4150804  &  3678 & 4.5 & 1.2 & Bo10\\
  9025971  &  3680 & 3.9 & 1.2 & Bo10\\
  2581316  &  3681 & 5.9 & 1.2 & Bo10\\
  10795103  &  3683 & 8.9 & 1.2 & Bo10\\
  7017372  &  3689 & 2.4 & 1.2 & Bo10\\
  10735331  &  3720 & 24.7 & 0.1 & RotPro\\
  6775985  &  3780 & 5.5 & 1.2 & Bo10\\
  9533489  &  3783 & 71.7 & 0.1 & RotPro\\
  7813039  &  3787 & 3.7 & 1.2 & Bo10\\
  4769799  &  5086 & 4.6 & 1.2 & Bo10\\
  5179609  &  5132 & 4.9 & 1.2 & Bo10\\
  7377343  &  5384 & 43.6 & 0.3 & RotPro\\
  9724993  &  5708 & 7.3 & 1.2 & Bo10\\
  12645761  &  5976 & 4.8 & 1.2 & Bo10\\
  9221398  &  6066 & 3.9 & 1.2 & Bo10\\
  5034333	 &  6124 & 51.3 & 0.8 & RotPro\\
  5629353  &  6132 & 9.3 & 1.2 & Bo10\\
  8197761  &  6175 & 4.8 & 1.2 & Bo10\\
  1147460  &  6235 & 5.1 & 1.2 & Bo10\\
  12470041 & 6251 & 19.1 & 0.3 & RotPro\\
\hline
\hline
\end{tabular}
\end{center}
\label{vsini}
\tablefoot{The instrumental resolution of SOPHIE corresponds to a \vsini\, of $\sim$ 4 \kms\, for a solar-like star. Therefore, \vsini\, vales lower than 4 \kms\, should be considered as upper limits.}
\tablefoot{Bo10 refers to the method described in \citet{2010A&A...523A..88B} to measure the \vsini\, from the Gaussian width of the SOPHIE CCF. RotPro means that the \vsini\, was measured on the CCF using a rotation profile, as described in \citet{2012A&A...544L..12S}.}
\end{table}%

\begin{table}[h]
\caption{Derived upper-limits on the mass of the candidates for which we detected no significant RV variation. The eccentricity flag indicates if the eccentricity was a fixed (0) or a free (1) parameter in the analysis.}
\begin{center} 
\begin{tabular}{lccc}
 \hline
 \hline
Candidate & 99\% mass constraints & Eccentricity \\
 & [\Mjup] & flag \\
 \hline
 KOI-221.01 & $<$ 0.65 & 0 \\
 KOI-221.02 & $<$ 1.16 & 0 \\
 KOI-351.01 & $<$ 1.16 & 0 \\
 KOI-351.02 & $<$ 0.82 & 0 \\
 KOI-351.03 & $<$ 1.17 & 0 \\
 KOI-351.04 & $<$ 1.76 & 0 \\
 KOI-351.05 & $<$ 0.29 & 0 \\
 KOI-351.06 & $<$ 1.04 & 0 \\
 KOI-351.07 & $<$ 0.78 & 0 \\
 KOI-368.01 & $<$ 225 & 0\\
KOI-464.01 & $<$ 0.68 & 0\\
KOI-464.02 & $<$ 0.29 & 0\\
KOI-620.01 & $<$ 0.85 & 0\\
KOI-620.02 &$<$  2.45 & 0\\
KOI-620.03 &$<$  1.01 & 0\\
KOI-1089.01 & $<$ 1.12 & 1\\
KOI-1089.02 & $<$ 0.46 & 1\\
KOI-1271.01 & $<$ 2.35 & 1\\
KOI-1353.01 & [0.68, 2.43] & 0\\
KOI-1353.02 & $<$ 0.52 & 0\\
KOI-1353.03 & $<$ 1.41 & 0\\
KOI-1411.01 & $<$ 2.13 & 1\\
KOI-1426.01 & $<$ 0.69 & 0\\
KOI-1426.02 & $<$ 0.45 & 0\\
KOI-1426.03 & $<$ 1.03 & 0\\
KOI-1431.01 & $<$ 0.73 & 1\\
KOI-1574.01 & $<$ 2.25 & 0\\
KOI-1574.02 & $<$ 1.66 & 0\\
KOI-1574.03 & $<$ 0.61 & 0\\
KOI-1574.04 & $<$ 0.68 & 0\\
KOI-1783.01 & $<$ 2.83 & 0\\
KOI-1788.01 & $<$ 0.48 & 0\\
KOI-1788.02 & $<$ 3.0 & 0\\
KOI-2679.01 & $<$ 40.3 & 0\\
KOI-3678.01 & $<$ 1.43 & 0 \\
KOI-3683.01 & $<$ 2.08 & 0\\
KOI-3689.01 & $<$ 0.61 & 0\\
KOI-6132.01 & $<$ 2.25 & 0\\
KOI-6132.02 & $<$ 1.50 & 0\\
  \hline
  \hline
\end{tabular}
\end{center}
\label{upperlims}
\end{table}%

\begin{table}[h]
\caption{Estimated $p$-values from the Anderson -- Darling (AD) tests for the distributions of \teff\, and \met\, between all the \textit{Kepler} targets, all the KOIs, the giant-planet hosts, and false-positive hosts. Note that for speeding up the computation of the AD test, we took only one \textit{Kepler} target over 10. We tested that the conclusions are unchanged using different (but large enough) sub-sample of the \textit{Kepler} targets.}
\begin{center}
\begin{tabular}{lccc}
& \multicolumn{3}{c}{Stellar effective temperature}\\
\hline
\hline
& All targets & All KOIs & Giant planets\\
\hline
All KOIs & 2.4$\times10^{-5}$ & -- & -- \\
Giant planets & 7.6$\times10^{-4}$ & 1.5$\times10^{-3}$ & -- \\
False positives & 0.088 & 0.061 & 0.062 \\
\hline
\hline
&&&\\
& \multicolumn{3}{c}{Stellar iron abundance}\\
\hline
\hline
& All targets & All KOIs & Giant planets\\
\hline
All KOIs & 9.7$\times10^{-6}$ & -- & --\\
Giant planets & 1.4$\times10^{-5}$ & 4.6$\times10^{-5}$ & --\\
False positives & 0.61 & 0.50 & 2.2$\times10^{-4}$ \\
\hline
\hline
&&&\\
\end{tabular}
\end{center}
\label{ADtest}
\end{table}%

\begin{table}[h]
\caption{Stellar parameters at the end of the main sequence from the STAREVOL evolution tracks.}
\begin{center}
\begin{tabular}{cccc}
\hline
\hline
Mass & Radius & \teff & age \\
$\left[\right.$\Msun] & [\Rsun] & [K] & [Gyr]\\
\hline
0.7  &  1.058  &  4978  &  43 \\
0.8  &  1.177  &  5260  &  26 \\
0.9  &  1.311  &  5501  &  16\\
1.0  &  1.452  &  5721  &  10.7\\
1.1  &  1.596  &  5929  &  7.1\\
1.2  &  1.739  &  6125  &  4.9\\
1.3  &  1.943  &  6298  &  3.7\\
1.4  &  2.087  &  6538  &  2.8\\
\hline
\hline
\end{tabular}
\end{center}
\label{EndMSparams}
\end{table}%


\onecolumn

\begin{landscape}
\begin{center}
\setlength{\tabcolsep}{0.5mm}
\begin{longtable}{ccc|ccccc|ccccc|c|l}
\caption{Adopted parameters for the giant planets and their host.}\\
\hline
\hline
KIC & KOI & Kepler & Period & a/R$_{\star}$ & R$_{\rm p}$/R$_{\star}$ & R$_{\rm p}$ & M$_{\rm p}$ & \teff & \met & R$_{\star}$ & M$_{\star}$ & K$_{p}$  &  Status & References\\
ID & ID & ID & [d] &  & & [\Rjup] & [\Mjup] & [K] & [dex] & [\Rsun] & [\Msun] & & & \\ 
\hline
\endfirsthead
\caption{Continued.} \\
\hline
KIC & KOI & Kepler & Period & a/R$_{\star}$ & R$_{\rm p}$/R$_{\star}$ & R$_{\rm p}$ & M$_{\rm p}$ & \teff & \met & R$_{\star}$ & M$_{\star}$ & K$_{p}$  &  Status & References\\
\hline
\endhead
\hline
\endfoot
\hline
\hline
\endlastfoot
11446443  &  1.01  &    1b    &  2.471  &  7.903$^{_{+0.02}}_{^{-0.02}}$  &  0.12539$^{_{+4.9\times10^{-4}}}_{^{-3.5\times10^{-4}}}$  &  1.25$^{_{+0.05}}_{^{-0.04}}$  &  1.20$\pm$0.07  &  5795$\pm$73  &  0.06$\pm$0.08  &  0.95$\pm$0.02  &  0.94$\pm$0.05  &  11.3  &  S  &    Es15, Sa13, Hu14\\ 
10666592  &  2.01  &    2b    &  2.205  &  4.154$^{_{+0.00}}_{^{-0.00}}$  &  0.07752$^{_{+1.7\times10^{-5}}}_{^{-2.2\times10^{-5}}}$  &  1.42$^{_{+0.18}}_{^{-0.09}}$  &  1.78$^{_{+0.08}}_{^{-0.06}}$  &  6525$\pm$61  &  0.31$^{_{+0.07}}_{^{-0.11}}$  &  2.00$^{_{+0.01}}_{^{-0.02}}$  &  1.51$^{_{+0.04}}_{^{-0.05}}$  &  10.5  &  S  &    Es15, Sa13, Lu14\\ 
10748390  &  3.01  &    3b    &  4.888  &  16.510$^{_{+0.18}}_{^{-0.17}}$  &  0.05891$^{_{+2.1\times10^{-4}}}_{^{-2.5\times10^{-4}}}$  &  0.40$\pm$0.01  &  0.08$\pm$0.01  &  4792$\pm$69  &  0.33$\pm$0.07  &  0.77$^{_{+0.03}}_{^{-0.02}}$  &  0.83$^{_{+0.02}}_{^{-0.05}}$  &  9.2  &  S  &    M\"u13, So11, To12, Hu14\\ 
6922244  &  10.01  &    8b    &  3.522  &  6.854$^{_{+0.02}}_{^{-0.02}}$  &  0.09575$^{_{+1.9\times10^{-4}}}_{^{-2.3\times10^{-4}}}$  &  1.42$^{_{+0.05}}_{^{-0.06}}$  &  0.59$^{_{+0.13}}_{^{-0.12}}$  &  6251$\pm$75  &  0.05$\pm$0.09  &  1.45$^{_{+0.12}}_{^{-0.13}}$  &  1.13$^{_{+0.09}}_{^{-0.10}}$  &  13.5  &  S  &    Es15, To12, Hu14\\ 
5812701  &  12.01  &    --    &  17.855  &  20.000$\pm$1.50  &  0.09049$^{_{+8.0\times10^{-5}}}_{^{-8.0\times10^{-5}}}$  &  1.43$\pm$0.13  &  $< 10.0$  &  6820$\pm$120  &  0.09$\pm$0.15  &  1.63$\pm$0.15  &  1.45$\pm$0.09  &  11.4  &  S  &    Bou15\\ 
9941662  &  13.01  &    13b    &  1.764  &  4.501$^{_{+0.00}}_{^{-0.00}}$  &  0.08737$^{_{+2.3\times10^{-5}}}_{^{-2.4\times10^{-5}}}$  &  1.51$\pm$0.04  &  9.28$\pm$0.16  &  7650$\pm$250  &  0.20$\pm$0.20  &  1.74$\pm$0.04  &  1.72$\pm$0.10  &  10.0  &  S  &    Es15, Sh14\\ 
10874614  &  17.01  &    6b    &  3.235  &  7.503$\pm$0.02  &  0.09424$^{_{+1.2\times10^{-4}}}_{^{-1.1\times10^{-4}}}$  &  1.30$^{_{+0.02}}_{^{-0.03}}$  &  0.67$^{_{+0.04}}_{^{-0.04}}$  &  5640$^{_{+99}}_{^{-110}}$  &  0.34$^{_{+0.10}}_{^{-0.16}}$  &  1.29$^{_{+0.09}}_{^{-0.10}}$  &  1.05$^{_{+0.08}}_{^{-0.07}}$  &  13.3  &  S  &    Es15, Hu14\\ 
8191672  &  18.01  &    5b    &  3.548  &  6.450$^{_{+0.02}}_{^{-0.03}}$  &  0.07996$^{_{+8.7\times10^{-5}}}_{^{-7.1\times10^{-5}}}$  &  1.43$^{_{+0.04}}_{^{-0.05}}$  &  2.11$^{_{+0.07}}_{^{-0.09}}$  &  6290$^{_{+105}}_{^{-120}}$  &  0.04$^{_{+0.10}}_{^{-0.16}}$  &  1.75$^{_{+0.14}}_{^{-0.15}}$  &  1.32$^{_{+0.09}}_{^{-0.14}}$  &  13.4  &  S  &    Es15, Hu14\\ 
11804465  &  20.01  &    12b    &  4.438  &  8.019$^{_{+0.01}}_{^{-0.01}}$  &  0.11887$^{_{+8.5\times10^{-5}}}_{^{-9.4\times10^{-5}}}$  &  1.75$^{_{+0.03}}_{^{-0.04}}$  &  0.43$^{_{+0.05}}_{^{-0.05}}$  &  5953$^{_{+105}}_{^{-123}}$  &  0.08$^{_{+0.13}}_{^{-0.14}}$  &  1.42$^{_{+0.30}}_{^{-0.24}}$  &  1.09$^{_{+0.13}}_{^{-0.09}}$  &  13.4  &  S  &    Es15, Hu14\\ 
9631995  &  22.01  &    422b    &  7.891  &  15.078$\pm$0.49  &  0.09570$^{_{+4.8\times10^{-4}}}_{^{-5.5\times10^{-4}}}$  &  1.15$\pm$0.11  &  0.43$\pm$0.13  &  5972$\pm$84  &  0.23$\pm$0.09  &  1.24$\pm$0.12  &  1.15$\pm$0.06  &  13.4  &  S  &    En14\\ 
11554435  &  63.01  &    63b    &  9.434  &  19.120$\pm$0.08  &  0.06220$^{_{+1.0\times10^{-3}}}_{^{-1.0\times10^{-3}}}$  &  0.55$\pm$0.02  &  $< 0.4$  &  5576$\pm$50  &  0.05$\pm$0.08  &  0.90$^{_{+0.03}}_{^{-0.02}}$  &  0.98$^{_{+0.04}}_{^{-0.04}}$  &  11.6  &  S  &    SO13\\ 
6462863  &  94.01  &    89d    &  22.343  &  23.800$\pm$1.90  &  0.06802$^{_{+8.0\times10^{-5}}}_{^{-8.0\times10^{-5}}}$  &  1.00$\pm$0.10  &  0.33$\pm$0.04  &  6182$\pm$58  &  0.02$\pm$0.00  &  1.52$\pm$0.14  &  1.28$\pm$0.05  &  12.2  &  S  &    We13\\ 
5780885  &  97.01  &    7b    &  4.885  &  6.637$\pm$0.02  &  0.08294$^{_{+1.1\times10^{-4}}}_{^{-1.1\times10^{-4}}}$  &  1.62$\pm$0.01  &  0.44$^{_{+0.04}}_{^{-0.04}}$  &  6027$\pm$75  &  0.10$\pm$0.10  &  1.96$\pm$0.07  &  1.32$\pm$0.09  &  12.9  &  S  &    Es15, Hu14\\ 
8359498  &  127.01  &    77b    &  3.579  &  9.764$\pm$0.06  &  0.09924$^{_{+2.6\times10^{-4}}}_{^{-2.6\times10^{-4}}}$  &  0.96$\pm$0.02  &  0.43$\pm$0.03  &  5520$\pm$60  &  0.20$\pm$0.05  &  0.99$\pm$0.02  &  0.95$\pm$0.04  &  13.9  &  S  &    Ga13\\ 
11359879  &  128.01  &    15b    &  4.943  &  12.800$^{_{+1.20}}_{^{-1.50}}$  &  0.09960$^{_{+5.5\times10^{-4}}}_{^{-5.3\times10^{-4}}}$  &  0.96$^{_{+0.06}}_{^{-0.07}}$  &  0.66$^{_{+0.08}}_{^{-0.09}}$  &  5514$^{_{+89}}_{^{-109}}$  &  0.36$^{_{+0.08}}_{^{-0.15}}$  &  0.98$^{_{+0.16}}_{^{-0.06}}$  &  1.00$^{_{+0.03}}_{^{-0.06}}$  &  13.8  &  S  &    En11, Hu14\\ 
7778437  &  131.01  &    --    &  5.014  &  8.958$\pm$0.01  &  0.07600$^{_{+5.4\times10^{-5}}}_{^{-3.1\times10^{-5}}}$  &  0.94$^{_{+0.69}}_{^{-0.14}}$  &  $< 14.3$  &  6475$^{_{+169}}_{^{-250}}$  &  0.21$^{_{+0.15}}_{^{-0.43}}$  &  1.27$^{_{+0.93}}_{^{-0.19}}$  &  1.31$\pm$0.26  &  13.8  &  L  &    RT15, Sa12, Hu14\\ 
9818381  &  135.01  &    43b    &  3.024  &  6.975$^{_{+0.05}}_{^{-0.04}}$  &  0.08628$^{_{+3.6\times10^{-4}}}_{^{-3.3\times10^{-4}}}$  &  1.22$^{_{+0.07}}_{^{-0.06}}$  &  3.23$\pm$0.26  &  6050$\pm$100  &  0.40$\pm$0.10  &  1.38$^{_{+0.05}}_{^{-0.03}}$  &  1.27$\pm$0.04  &  14.0  &  S  &    Es15, Bon15\\ 
9651668  &  183.01  &    423b    &  2.684  &  8.601$\pm$0.14  &  0.12420$^{_{+8.9\times10^{-4}}}_{^{-3.7\times10^{-4}}}$  &  1.20$\pm$0.07  &  0.72$\pm$0.12  &  5970$\pm$116  &  0.26$\pm$0.12  &  0.99$\pm$0.05  &  1.07$\pm$0.05  &  14.3  &  S  &    En14\\ 
7950644  &  192.01  &    427b    &  10.291  &  14.200$\pm$2.10  &  0.09130$^{_{+3.0\times10^{-4}}}_{^{-3.0\times10^{-4}}}$  &  1.23$\pm$0.21  &  0.29$\pm$0.09  &  5800$\pm$70  &  -0.19$\pm$0.07  &  1.35$\pm$0.20  &  0.96$\pm$0.06  &  14.2  &  S  &    H\'e14\\ 
9410930  &  196.01  &    41b    &  1.856  &  5.053$\pm$0.02  &  0.10253$^{_{+4.3\times10^{-4}}}_{^{-3.5\times10^{-4}}}$  &  1.04$\pm$0.04  &  0.56$^{_{+0.10}}_{^{-0.09}}$  &  5620$\pm$140  &  0.29$\pm$0.16  &  1.02$\pm$0.03  &  1.12$\pm$0.07  &  14.5  &  S  &    Es15, Sa11\\ 
2987027  &  197.01  &    --    &  17.276  &  36.582$\pm$0.04  &  0.09300$^{_{+1.6\times10^{-4}}}_{^{-5.7\times10^{-5}}}$  &  0.74$^{_{+0.82}}_{^{-0.07}}$  &  $< 0.3$  &  5085$^{_{+165}}_{^{-131}}$  &  0.08$^{_{+0.31}}_{^{-0.28}}$  &  0.82$^{_{+0.91}}_{^{-0.08}}$  &  0.78$^{_{+0.14}}_{^{-0.06}}$  &  14.0  &  L  &    RT15, Sa12, Hu14\\ 
6046540  &  200.01  &    74b    &  7.341  &  15.470$\pm$0.18  &  0.09120$^{_{+9.0\times10^{-4}}}_{^{-9.0\times10^{-4}}}$  &  0.96$\pm$0.02  &  0.63$\pm$0.12  &  6000$\pm$100  &  0.42$\pm$0.11  &  1.12$\pm$0.04  &  1.18$\pm$0.04  &  14.4  &  S  &    Bon15\\ 
6849046  &  201.01  &    --    &  4.225  &  12.379$\pm$0.03  &  0.07900$^{_{+6.4\times10^{-5}}}_{^{-3.5\times10^{-5}}}$  &  0.82$^{_{+0.16}}_{^{-0.06}}$  &  $< 0.6$  &  5649$^{_{+98}}_{^{-129}}$  &  0.48$^{_{+0.08}}_{^{-0.20}}$  &  1.05$^{_{+0.21}}_{^{-0.08}}$  &  1.07$^{_{+0.06}}_{^{-0.08}}$  &  14.0  &  S  &    RT15, Sa12, Hu14\\ 
7877496  &  202.01  &    412b    &  1.721  &  4.841$^{_{+0.02}}_{^{-0.02}}$  &  0.10474$^{_{+5.4\times10^{-4}}}_{^{-7.7\times10^{-4}}}$  &  1.34$^{_{+0.04}}_{^{-0.05}}$  &  0.94$^{_{+0.12}}_{^{-0.02}}$  &  5750$\pm$90  &  0.27$\pm$0.12  &  1.29$\pm$0.04  &  1.17$\pm$0.09  &  14.3  &  S  &    Es15, De14\\ 
10619192  &  203.01  &    17b    &  1.486  &  5.480$\pm$0.02  &  0.13031$^{_{+2.2\times10^{-4}}}_{^{-1.8\times10^{-4}}}$  &  1.33$\pm$0.04  &  2.47$\pm$0.10  &  5781$\pm$85  &  0.26$\pm$0.10  &  1.05$\pm$0.03  &  1.16$\pm$0.06  &  14.1  &  S  &    Bo12, D\'e11\\ 
9305831  &  204.01  &    44b    &  3.247  &  7.070$^{_{+0.35}}_{^{-0.37}}$  &  0.08280$^{_{+8.0\times10^{-4}}}_{^{-8.0\times10^{-4}}}$  &  1.09$\pm$0.07  &  1.00$\pm$0.10  &  5800$\pm$100  &  0.42$\pm$0.11  &  1.35$\pm$0.08  &  1.12$\pm$0.08  &  14.7  &  S  &    Bon15\\ 
5728139  &  206.01  &    433b    &  5.334  &  6.440$\pm$0.62  &  0.06590$^{_{+1.5\times10^{-4}}}_{^{-1.5\times10^{-4}}}$  &  1.45$\pm$0.16  &  2.82$\pm$0.52  &  6340$\pm$140  &  0.06$\pm$0.19  &  2.26$\pm$0.25  &  1.46$\pm$0.17  &  14.5  &  S  &    Al15\\ 
10723750  &  209.01  &    117b    &  50.790  &  38.180$\pm$0.72  &  0.07052$^{_{+3.4\times10^{-4}}}_{^{-3.4\times10^{-4}}}$  &  1.10$\pm$0.04  &  1.84$\pm$0.18  &  6150$\pm$110  &  -0.04$\pm$0.10  &  1.61$\pm$0.05  &  1.13$^{_{+0.13}}_{^{-0.02}}$  &  14.3  &  S  &    Br15\\ 
11046458  &  214.01  &    424b    &  3.312  &  11.209$\pm$0.32  &  0.09610$^{_{+6.5\times10^{-3}}}_{^{-3.3\times10^{-3}}}$  &  0.89$^{_{+0.08}}_{^{-0.06}}$  &  1.03$\pm$0.13  &  5460$\pm$81  &  0.44$\pm$0.14  &  0.94$\pm$0.06  &  1.01$\pm$0.05  &  14.3  &  S  &    En14\\ 
3937519  &  221.01  &    --    &  3.413  &  10.600$\pm$0.08  &  0.05800$^{_{+2.7\times10^{-4}}}_{^{-8.2\times10^{-5}}}$  &  0.41$^{_{+0.09}}_{^{-0.03}}$  &  $< 0.7$  &  5332$^{_{+172}}_{^{-140}}$  &  -0.42$^{_{+0.38}}_{^{-0.28}}$  &  0.71$^{_{+0.16}}_{^{-0.05}}$  &  0.77$^{_{+0.11}}_{^{-0.07}}$  &  14.6  &  L  &    RT14, this work, Hu14\\ 
11442793  &  351.01  &    90h    &  331.601  &  180.700$\pm$4.70  &  0.08660$^{_{+7.0\times10^{-4}}}_{^{-7.0\times10^{-4}}}$  &  1.01$\pm$0.09  &  $< 1.2$  &  6080$^{_{+260}}_{^{-170}}$  &  -0.12$\pm$0.18  &  1.20$\pm$0.10  &  1.20$\pm$0.10  &  13.8  &  S  &    Ca14, this work\\ 
11442793  &  351.02  &    90g    &  210.603  &  127.300$\pm$4.10  &  0.06150$^{_{+1.1\times10^{-3}}}_{^{-1.1\times10^{-3}}}$  &  0.72$\pm$0.07  &  $< 0.8$  &  6080$^{_{+260}}_{^{-170}}$  &  -0.12$\pm$0.18  &  1.20$\pm$0.10  &  1.20$\pm$0.10  &  13.8  &  S  &    Ca14, this work\\ 
6471021  &  372.01  &    --    &  125.629  &  111.625$\pm$0.02  &  0.08200$^{_{+1.8\times10^{-4}}}_{^{-7.1\times10^{-5}}}$  &  0.71$^{_{+0.10}}_{^{-0.03}}$  &  $< 4.8$  &  5776$\pm$46  &  -0.06$\pm$0.04  &  1.13$^{_{+0.28}}_{^{-0.14}}$  &  1.02$\pm$0.04  &  12.4  &  L  &    RT15, De+, this work\\ 
3323887  &  377.01  &    9b    &  19.271  &  36.840$\pm$4.30  &  0.07885$^{_{+8.1\times10^{-4}}}_{^{-8.1\times10^{-4}}}$  &  0.84$\pm$0.07  &  0.25$\pm$0.01  &  5777$\pm$61  &  0.12$\pm$0.04  &  1.02$\pm$0.05  &  1.07$\pm$0.05  &  13.8  &  S  &    Ho10\\ 
3323887  &  377.02  &    9c    &  38.908  &  54.340$\pm$5.60  &  0.07708$^{_{+8.0\times10^{-4}}}_{^{-8.0\times10^{-4}}}$  &  0.82$\pm$0.07  &  0.17$\pm$0.01  &  5777$\pm$61  &  0.12$\pm$0.04  &  1.02$\pm$0.05  &  1.07$\pm$0.05  &  13.8  &  S  &    Ho10\\ 
5449777  &  410.01  &    --    &  7.217  &  13.540$\pm$3.30  &  0.47300$^{_{+4.7\times10^{-1}}}_{^{-1.3\times10^{-1}}}$  &  4.86$^{_{+1.88}}_{^{-0.62}}$  &  $< 3.4$  &  6266$^{_{+171}}_{^{-207}}$  &  -0.40$^{_{+0.26}}_{^{-0.30}}$  &  1.06$^{_{+0.41}}_{^{-0.14}}$  &  0.97$^{_{+0.14}}_{^{-0.11}}$  &  14.5  &  L  &    RT15, Bo11, Hu14\\ 
8890783  &  464.01  &    --    &  58.362  &  75.657$\pm$0.69  &  0.06800$^{_{+4.0\times10^{-4}}}_{^{-1.2\times10^{-4}}}$  &  0.63$^{_{+0.24}}_{^{-0.06}}$  &  $< 0.7$  &  5592$^{_{+154}}_{^{-153}}$  &  0.16$^{_{+0.20}}_{^{-0.26}}$  &  0.95$^{_{+0.36}}_{^{-0.09}}$  &  0.97$^{_{+0.10}}_{^{-0.08}}$  &  14.4  &  L  &    RT15, this work, Hu14\\ 
6309763  &  611.01  &    --    &  3.252  &  8.950$\pm$0.24  &  0.08400$^{_{+4.1\times10^{-3}}}_{^{-2.9\times10^{-3}}}$  &  0.92$^{_{+0.44}}_{^{-0.10}}$  &  $< 1.5$  &  6343$^{_{+161}}_{^{-206}}$  &  -0.16$^{_{+0.22}}_{^{-0.30}}$  &  1.11$^{_{+0.54}}_{^{-0.12}}$  &  1.07$^{_{+0.24}}_{^{-0.10}}$  &  14.0  &  L  &    RT15, Sa12, Hu14\\ 
7368664  &  614.01  &    434b    &  12.875  &  17.900$\pm$1.60  &  0.08350$^{_{+1.4\times10^{-2}}}_{^{-8.0\times10^{-3}}}$  &  1.13$^{_{+0.26}}_{^{-0.18}}$  &  2.86$\pm$0.35  &  5977$\pm$95  &  0.25$\pm$0.14  &  1.38$\pm$0.13  &  1.20$^{_{+0.09}}_{^{-0.93}}$  &  14.5  &  S  &    Al15\\ 
11773022  &  620.01  &    51b    &  45.155  &  61.500$^{_{+1.50}}_{^{-1.20}}$  &  0.07414$^{_{+5.9\times10^{-4}}}_{^{-6.1\times10^{-4}}}$  &  0.63$\pm$0.03  &  0.01$^{_{+0.01}}_{^{-0.00}}$  &  6046$^{_{+149}}_{^{-197}}$  &  -0.08$^{_{+0.22}}_{^{-0.30}}$  &  0.97$^{_{+0.39}}_{^{-0.09}}$  &  1.05$^{_{+0.17}}_{^{-0.14}}$  &  14.7  &  S  &    Ma14, Hu14\\ 
11773022  &  620.02  &    51d    &  130.177  &  124.700$^{_{+3.00}}_{^{-2.50}}$  &  0.10141$^{_{+8.4\times10^{-4}}}_{^{-8.5\times10^{-4}}}$  &  0.86$\pm$0.04  &  0.02$\pm$0.00  &  6046$^{_{+149}}_{^{-197}}$  &  -0.08$^{_{+0.22}}_{^{-0.30}}$  &  0.97$^{_{+0.39}}_{^{-0.09}}$  &  1.05$^{_{+0.17}}_{^{-0.14}}$  &  14.7  &  S  &    Ma14, Hu14\\ 
7529266  &  680.01  &    435b    &  8.600  &  6.350$\pm$0.51  &  0.06384$^{_{+2.0\times10^{-4}}}_{^{-2.0\times10^{-4}}}$  &  1.99$\pm$0.18  &  0.84$\pm$0.15  &  6090$\pm$110  &  -0.17$\pm$0.10  &  3.21$\pm$0.30  &  1.54$\pm$0.09  &  14.6  &  S  &    Al15\\ 
3247268  &  1089.01  &    418b    &  86.679  &  84.400$\pm$9.50  &  0.11040$^{_{+3.5\times10^{-3}}}_{^{-3.5\times10^{-3}}}$  &  1.20$\pm$0.16  &  $< 1.1$  &  6029$\pm$169  &  0.31$\pm$0.13  &  1.60$^{_{+0.68}}_{^{-0.36}}$  &  1.29$^{_{+0.21}}_{^{-0.10}}$  &  14.7  &  S  &    Ti14, this work\\ 
8751933  &  1257.01  &    420b    &  86.648  &  73.000$\pm$13.00  &  0.08220$^{_{+1.4\times10^{-3}}}_{^{-1.4\times10^{-3}}}$  &  0.94$\pm$0.12  &  1.45$\pm$0.35  &  5520$\pm$80  &  0.27$\pm$0.09  &  1.13$\pm$0.14  &  0.99$\pm$0.05  &  14.7  &  S  &    Sa14\\ 
8631160  &  1271.01  &    --    &  162.054  &  105.720$\pm$7.40  &  0.06900$^{_{+4.5\times10^{-4}}}_{^{-3.6\times10^{-4}}}$  &  0.94$^{_{+0.51}}_{^{-0.18}}$  &  $< 2.4$  &  6600$\pm$122  &  -0.06$\pm$0.09  &  1.57$^{_{+0.41}}_{^{-0.26}}$  &  1.33$^{_{+0.13}}_{^{-0.08}}$  &  13.6  &  L  &    RT15, this work\\ 
7303287  &  1353.01  &    289b    &  125.865  &  108.600$\pm$1.10  &  0.10620$^{_{+4.9\times10^{-4}}}_{^{-5.0\times10^{-4}}}$  &  1.03$\pm$0.02  &  0.42$\pm$0.05  &  6326$\pm$126  &  0.33$\pm$0.09  &  1.46$^{_{+0.27}}_{^{-0.15}}$  &  1.35$^{_{+0.11}}_{^{-0.07}}$  &  14.0  &  S  &    Sc14, this work\\ 
9425139  &  1411.01  &    --    &  305.076  &  295.600$\pm$97.10  &  0.06200$^{_{+4.7\times10^{-4}}}_{^{-3.4\times10^{-3}}}$  &  0.68$^{_{+0.17}}_{^{-0.06}}$  &  $< 2.1$  &  5687$\pm$146  &  0.47$\pm$0.12  &  1.35$^{_{+0.58}}_{^{-0.27}}$  &  1.14$^{_{+0.15}}_{^{-0.10}}$  &  13.4  &  L  &    RT15, this work\\ 
11122894  &  1426.02  &    297c    &  74.928  &  82.400$\pm$14.90  &  0.06600$^{_{+3.8\times10^{-4}}}_{^{-1.3\times10^{-3}}}$  &  0.66$^{_{+0.30}}_{^{-0.07}}$  &  0.13$^{_{+0.04}}_{^{-0.03}}$  &  6150$^{_{+151}}_{^{-193}}$  &  -0.12$^{_{+0.22}}_{^{-0.30}}$  &  1.04$^{_{+0.48}}_{^{-0.10}}$  &  1.04$^{_{+0.20}}_{^{-0.12}}$  &  14.2  &  S  &    RT15, HL14, Hu14\\ 
11122894  &  1426.03  &    --    &  150.019  &  127.200$\pm$29.50  &  0.11900$^{_{+4.8\times10^{-1}}}_{^{-2.6\times10^{-2}}}$  &  1.21$^{_{+0.56}}_{^{-0.12}}$  &  $< 1.0$  &  6150$^{_{+151}}_{^{-193}}$  &  -0.12$^{_{+0.22}}_{^{-0.30}}$  &  1.04$^{_{+0.48}}_{^{-0.10}}$  &  1.04$^{_{+0.20}}_{^{-0.12}}$  &  14.2  &  L  &    RT15, Hu14\\ 
11075279  &  1431.01  &    --    &  345.160  &  222.400$\pm$14.30  &  0.07600$^{_{+5.7\times10^{-4}}}_{^{-6.1\times10^{-4}}}$  &  0.71$^{_{+0.11}}_{^{-0.03}}$  &  $< 0.7$  &  5507$\pm$74  &  0.27$\pm$0.05  &  1.16$^{_{+0.38}}_{^{-0.18}}$  &  1.00$^{_{+0.07}}_{^{-0.04}}$  &  13.5  &  L  &    RT15, this work\\ 
12365184  &  1474.01  &    419b    &  69.727  &  45.170$\pm$7.80  &  0.06260$^{_{+2.0\times10^{-4}}}_{^{-2.0\times10^{-4}}}$  &  0.96$\pm$0.12  &  2.50$\pm$0.30  &  6430$\pm$79  &  0.18$\pm$0.07  &  1.74$\pm$0.07  &  1.39$^{_{+0.08}}_{^{-0.07}}$  &  13.0  &  S  &    Da14\\ 
10028792  &  1574.01  &    87b    &  114.736  &  57.400$^{_{+1.40}}_{^{-1.20}}$  &  0.06855$^{_{+2.6\times10^{-4}}}_{^{-2.8\times10^{-4}}}$  &  1.20$\pm$0.05  &  1.02$\pm$0.03  &  5600$\pm$50  &  -0.17$\pm$0.03  &  1.82$\pm$0.04  &  1.10$\pm$0.05  &  14.6  &  S  &    Of14\\ 
4570949  &  1658.01  &    76b    &  1.545  &  4.464$^{_{+0.05}}_{^{-0.04}}$  &  0.10330$^{_{+2.4\times10^{-3}}}_{^{-3.0\times10^{-3}}}$  &  1.36$\pm$0.12  &  2.01$^{_{+0.37}}_{^{-0.35}}$  &  6409$\pm$95  &  -0.10$\pm$0.20  &  1.32$\pm$0.08  &  1.20$\pm$0.20  &  13.3  &  S  &    Es15, Fa13\\ 
10005758  &  1783.01  &    --    &  134.479  &  94.470$\pm$4.40  &  0.07200$^{_{+1.2\times10^{-3}}}_{^{-9.4\times10^{-4}}}$  &  0.67$^{_{+0.24}}_{^{-0.06}}$  &  $< 2.8$  &  6300$\pm$150  &  0.49$\pm$0.12  &  1.84$^{_{+0.77}}_{^{-0.29}}$  &  1.57$^{_{+0.23}}_{^{-0.11}}$  &  13.9  &  L  &    RT15, this work\\ 
2975770  &  1788.01  &    --    &  71.525  &  98.609$\pm$0.36  &  0.07000$^{_{+9.7\times10^{-4}}}_{^{-4.4\times10^{-4}}}$  &  0.55$^{_{+1.06}}_{^{-0.05}}$  &  $< 0.5$  &  4890$\pm$233  &  0.05$\pm$0.39  &  1.10$^{_{+1.60}}_{^{-0.30}}$  &  0.84$^{_{+0.32}}_{^{-0.09}}$  &  14.5  &  L  &    RT15, this work\\ 
6716021  &  2679.01  &    --    &  110.756  &  78.250$\pm$1.30  &  0.10000$^{_{+4.9\times10^{-1}}}_{^{-4.8\times10^{-3}}}$  &  1.12$^{_{+0.66}}_{^{-0.13}}$  &  $< 40.3$  &  6528$^{_{+159}}_{^{-246}}$  &  -0.16$^{_{+0.22}}_{^{-0.32}}$  &  1.16$^{_{+0.68}}_{^{-0.13}}$  &  1.18$^{_{+0.27}}_{^{-0.17}}$  &  13.5  &  L  &    RT15, Hu14\\ 
12735740  &  3663.01  &    86b    &  282.525  &  176.700$^{_{+9.50}}_{^{-9.00}}$  &  0.09200$^{_{+2.0\times10^{-4}}}_{^{-4.0\times10^{-4}}}$  &  0.90$\pm$0.05  &  $< 80.0$  &  5629$^{_{+42}}_{^{-45}}$  &  -0.08$^{_{+0.03}}_{^{-0.03}}$  &  1.00$\pm$0.05  &  0.94$\pm$0.02  &  12.6  &  S  &    Wa13, this work\\ 
4150804  &  3678.01  &    --    &  160.885  &  120.000$\pm$12.30  &  0.08000$^{_{+1.4\times10^{-4}}}_{^{-5.1\times10^{-4}}}$  &  0.82$^{_{+0.35}}_{^{-0.16}}$  &  $< 1.4$  &  5650$^{_{+194}}_{^{-162}}$  &  -0.28$^{_{+0.34}}_{^{-0.24}}$  &  1.04$^{_{+0.45}}_{^{-0.21}}$  &  0.82$^{_{+0.15}}_{^{-0.06}}$  &  12.9  &  L  &    RT15, Hu14\\ 
9025971  &  3680.01  &    --    &  141.242  &  175.500$\pm$6.10  &  0.10700$^{_{+3.0\times10^{-4}}}_{^{-3.1\times10^{-4}}}$  &  0.95$^{_{+0.36}}_{^{-0.08}}$  &  2.20  &  5926$^{_{+161}}_{^{-172}}$  &  -0.12$^{_{+0.24}}_{^{-0.28}}$  &  0.91$^{_{+0.34}}_{^{-0.07}}$  &  1.00$^{_{+0.11}}_{^{-0.12}}$  &  14.5  &  S  &    RT15, this work, Hu14\\ 
2581316  &  3681.01  &    --    &  217.832  &  83.760$\pm$1.80  &  0.08900$^{_{+9.6\times10^{-5}}}_{^{-1.3\times10^{-4}}}$  &  1.97$^{_{+0.65}}_{^{-0.96}}$  &  4.20  &  6382$^{_{+204}}_{^{-215}}$  &  -0.84$\pm$0.30  &  2.27$^{_{+0.75}}_{^{-1.11}}$  &  1.09$^{_{+0.20}}_{^{-0.25}}$  &  11.7  &  S  &    RT15, this work, Hu14\\ 
10795103  &  3683.01  &    --    &  214.311  &  107.360$\pm$1.00  &  0.06200$^{_{+1.0\times10^{-4}}}_{^{-5.1\times10^{-5}}}$  &  0.83$^{_{+0.47}}_{^{-0.17}}$  &  $< 2.1$  &  6666$\pm$203  &  0.24$\pm$0.16  &  1.70$^{_{+0.57}}_{^{-0.26}}$  &  1.49$^{_{+0.19}}_{^{-0.13}}$  &  12.0  &  L  &    RT15, this work\\ 
7017372  &  3689.01  &    --    &  5.241  &  10.056$\pm$0.72  &  0.09100$^{_{+5.4\times10^{-4}}}_{^{-5.8\times10^{-4}}}$  &  1.17$^{_{+0.61}}_{^{-0.23}}$  &  $< 0.6$  &  6154$\pm$253  &  0.31$\pm$0.21  &  1.41$^{_{+0.59}}_{^{-0.24}}$  &  1.27$\pm$0.20  &  14.0  &  L  &    RT15, this work\\ 
5629353  &  6132.01  &    --    &  33.320  &  41.630$\pm$1.60  &  0.07362$^{_{+1.2\times10^{-3}}}_{^{-5.8\times10^{-4}}}$  &  1.17$^{_{+0.64}}_{^{-0.33}}$  &  $< 2.2$  &  6266$^{_{+168}}_{^{-247}}$  &  0.26$^{_{+0.14}}_{^{-0.32}}$  &  1.63$^{_{+0.90}}_{^{-0.46}}$  &  1.34$^{_{+0.25}}_{^{-0.27}}$  &  14.6  &  L  &    RT15, this work, Hu14
\label{PlanetParams}
\end{longtable}%
\tablebib{Al15: \citet{2015A&A...575A..71A}; Bo11: \citet{2011A&A...533A..83B}; Bo12: \citet{2012A&A...538A..96B}; Bon15: \citet{2015A&A...575A..85B}; Bou15: \citet{2015arXiv150404130B}; Br15: \citet{2015A&A...573A.124B}; Da14: \citet{2014ApJ...791...89D}; D\'e11: \citet{2011ApJS..197...14D}; De14: \citet{2014A&A...564A..56D}; De+: Demangeon et al. (in prep.); En11: \citet{2011ApJS..197...13E}; En14: \citet{2014ApJ...795..151E}; Es15: \citet{2015ApJ...804..150E}; Fa13: \citet{2013ApJ...771...26F}; Ga13: \citet{2013A&A...557A..74G}; HL14: \citet{2014ApJ...787...80H}; Ho10: \citet{2010Sci...330...51H}; Hu14: \citet{2014ApJS..211....2H}; Lu14: \citet{2014A&A...570A..54L}; Ma14: \citet{2014ApJ...783...53M}; M\"u13: \citet{2013A&A...560A.112M}; Of14: \citet{2014A&A...561A.103O}; RT15: \citet{2015arXiv150400707R}; Sa11: \citet{2011A&A...536A..70S}; Sa12: \citet{2012AA...545A..76S}; Sa13: \citet{2013A&A...556A.150S}; Sh14: \citet{2014ApJ...788...92S}; So11: \citet{2011MNRAS.417.2166S}; SO13: \citet{2013ApJ...775...54S}; To12: \citet{2012ApJ...757..161T}; We13: \citet{2013ApJ...768...14W}.
}
\tablefoot{Status of the giant planets: S for secured, L for likely.}
\end{center}
\end{landscape}

\begin{longtable}{cccccc}
\caption{Values of the correction factors used to derive the occurrence rates of giant planets in the \textit{Kepler} FOV.}\\
\hline
\hline
KOI & $\mathcal{C}^{T}$ & $\mathcal{C}^{R}$ & $\mathcal{C}^{L}$  & $\mathcal{C}^{S}$ & $\mathcal{C}^{D}$\\ 
ID &  & [\%] & [\%]  & [\%] &  \\ 
\hline
\endfirsthead
\caption{Continued.} \\
\hline
KOI & $\mathcal{C}^{T}$ & $\mathcal{C}^{R}$ & $\mathcal{C}^{L}$  & $\mathcal{C}^{S}$ & $\mathcal{C}^{D}$\\ 
\hline
\endhead
\hline
\endfoot
\hline
\hline
\endlastfoot
1.01  &  7.90  &  100.0  &  100  &  79.4  &  1.44  \\
2.01  &  4.15  &  1.9  &  100  &  76.6  &  0.28  \\
3.01  &  16.51  &  99.1  &  100  &  61.1  &  0.94  \\
10.01  &  6.85  &  84.5  &  100  &  87.1  &  0.62  \\
12.01  &  20.00  &  30.0  &  100  &  76.6  &  0.28  \\
13.01  &  4.50  &  0.1  &  100  &  76.6  &  0.28  \\
17.01  &  7.50  &  96.7  &  100  &  89.6  &  1.87  \\
18.01  &  6.45  &  58.3  &  100  &  80.6  &  0.43  \\
20.01  &  8.02  &  69.2  &  100  &  87.5  &  2.10  \\
22.01  &  15.08  &  100.0  &  100  &  87.0  &  0.83  \\
63.01  &  19.12  &  100.0  &  100  &  87.2  &  1.97  \\
94.01  &  23.80  &  98.0  &  100  &  84.4  &  0.49  \\
97.01  &  6.64  &  31.7  &  100  &  80.6  &  0.43  \\
127.01  &  9.76  &  100.0  &  100  &  80.1  &  1.66  \\
128.01  &  12.80  &  99.2  &  100  &  87.2  &  1.97  \\
131.01  &  8.96  &  29.2  &  75  &  80.4  &  0.43  \\
135.01  &  6.97  &  100.0  &  100  &  84.4  &  0.49  \\
183.01  &  8.60  &  100.0  &  100  &  87.3  &  2.06  \\
192.01  &  14.20  &  58.0  &  100  &  80.1  &  1.66  \\
196.01  &  5.05  &  100.0  &  100  &  87.1  &  0.62  \\
197.01  &  36.58  &  38.0  &  75  &  56.7  &  0.84  \\
200.01  &  15.47  &  100.0  &  100  &  82.9  &  0.83  \\
201.01  &  12.38  &  97.2  &  100  &  87.3  &  2.06  \\
202.01  &  4.84  &  99.3  &  100  &  82.9  &  0.83  \\
203.01  &  5.48  &  100.0  &  100  &  87.0  &  0.83  \\
204.01  &  7.07  &  96.9  &  100  &  87.1  &  0.62  \\
206.01  &  6.44  &  3.0  &  100  &  76.6  &  0.28  \\
209.01  &  38.18  &  85.2  &  100  &  87.1  &  0.62  \\
214.01  &  11.21  &  100.0  &  100  &  91.5  &  2.08  \\
221.01  &  10.60  &  86.4  &  75  &  41.3  &  0.73  \\
351.01  &  180.70  &  97.8  &  100  &  87.5  &  0.82  \\
351.02  &  127.30  &  97.6  &  100  &  87.5  &  0.82  \\
372.01  &  111.62  &  85.0  &  75  &  91.5  &  2.08  \\
377.01  &  36.84  &  100.0  &  100  &  87.3  &  2.06  \\
377.02  &  54.34  &  100.0  &  100  &  87.3  &  2.06  \\
410.01  &  13.54  &  69.5  &  75  &  87.5  &  1.83  \\
464.01  &  75.66  &  84.1  &  75  &  87.5  &  1.83  \\
611.01  &  8.95  &  63.6  &  75  &  87.3  &  2.06  \\
614.01  &  17.90  &  30.1  &  100  &  87.5  &  0.82  \\
620.01  &  61.50  &  84.4  &  100  &  89.6  &  1.87  \\
620.02  &  124.70  &  85.0  &  100  &  89.6  &  1.87  \\
680.01  &  6.35  &  0.0  &  100  &  76.6  &  0.28  \\
1089.01  &  84.40  &  31.6  &  100  &  79.0  &  0.41  \\
1257.01  &  73.00  &  96.8  &  100  &  87.2  &  1.97  \\
1271.01  &  105.72  &  48.9  &  75  &  80.6  &  0.43  \\
1353.01  &  108.60  &  57.6  &  100  &  80.5  &  0.33  \\
1411.01  &  295.60  &  57.9  &  75  &  87.1  &  0.62  \\
1426.02  &  82.40  &  71.3  &  100  &  87.8  &  2.07  \\
1426.03  &  127.20  &  71.6  &  75  &  87.8  &  2.07  \\
1431.01  &  222.40  &  71.1  &  75  &  87.2  &  1.97  \\
1474.01  &  45.17  &  52.3  &  100  &  76.6  &  0.28  \\
1574.01  &  57.40  &  0.6  &  100  &  87.5  &  2.10  \\
1658.01  &  4.46  &  76.2  &  100  &  87.5  &  0.82  \\
1783.01  &  94.47  &  1.4  &  75  &  76.6  &  0.28  \\
1788.01  &  98.61  &  30.0  &  75  &  61.1  &  0.94  \\
2679.01  &  78.25  &  43.8  &  75  &  82.9  &  0.83  \\
3663.01  &  176.70  &  100.0  &  100  &  79.4  &  1.44  \\
3678.01  &  120.00  &  60.2  &  75  &  54.7  &  0.89  \\
3680.01  &  175.50  &  85.7  &  100  &  87.2  &  1.97  \\
3681.01  &  83.76  &  28.7  &  100  &  87.5  &  2.10  \\
3683.01  &  107.36  &  11.3  &  75  &  76.6  &  0.28  \\
3689.01  &  10.06  &  45.5  &  75  &  84.4  &  0.49  \\
6132.01  &  41.63  &  23.6  &  75  &  80.5  &  0.33  \\
\label{CorrecFactor}
\end{longtable}%

\begin{table}[h]
\caption{Occurrence rates of giant planets for different ranges of orbital periods from different studies. All values are in percent.}
\begin{center}
\begin{tabular}{lccccccccccc}
\hline
\hline
\multirow{3}{*}{Orbital periods} & 0.8 d & 2.0 d & 3.4 d & 5.9 d & 10 d & 17 d & 29 d & 50 d & 85 d & 145 d & 245 d\\
 & -- & -- & -- & -- & -- & -- & -- & -- & -- & -- & -- \\
 & 2.0 d & 3.4 d & 5.9 d & 10 d & 17 d & 29 d & 50 d & 85 d & 145 d & 245 d & 400 d\\
\hline
\multirow{2}{*}{This work} & 0.051  &  0.10  &  0.21  &  0.11  &  0.025  &  0.23  &  0.16  &  0.49  &  1.05  &  0.88  &  1.27\\
& $\pm$0.023  &  $\pm$0.03  &  $\pm$0.06  &  $\pm$0.05  &  $\pm$0.017  &  $\pm$0.11  &  $\pm$0.09  &  $\pm$0.22  &  $\pm$0.35  &  $\pm$0.36  &  $\pm$0.63 \\
\hline
\multirow{2}{*}{\citet{2013ApJ...766...81F}} & 0.015  &  0.067  &  0.17  &  0.18  &  0.27  &  0.23  &  0.35  &  0.71  &  1.25  &  0.94  &  1.05\\
& $\pm$0.007  &  $\pm$0.018  &  $\pm$0.03  &  $\pm$0.04  &  $\pm$0.06  &  $\pm$0.06  &  $\pm$0.10  &  $\pm$0.17  &  $\pm$0.29  &  $\pm$0.28  &  $\pm$0.30\\
\hline
\multirow{2}{*}{\citet{2011arXiv1109.2497M}} & --  &  0.26  &  0.28  &  0.29  &  0.17  &  0.39 &  0.12  &  0.96  &  0.70  &  0.83  &  1.36\\
 & --  &  $\pm$0.19  &  $\pm$0.20  &  $\pm$0.20  &  $\pm$0.17  &  $\pm$0.28  &  $\pm$0.12  &  $\pm$0.43  &  $\pm$0.35  &  $\pm$0.42  &  $\pm$0.48\\
\hline
\hline
&&&&&&\\
&&&&&&\\
\hline
\hline
\multirow{3}{*}{Orbital periods} & 0.8 d & 0.8 d & 0.8 d & 0.8 d & 0.8 d & 0.8 d & 0.8 d & 0.8 d & 0.8 d & 0.8 d & 0.8 d\\
 & -- & -- & -- & -- & -- & -- & -- & -- & -- & -- & -- \\
 & 2.0 d & 3.4 d & 5.9 d & 10 d & 17 d & 29 d & 50 d & 85 d & 145 d & 245 d & 400 d\\
\hline
\multirow{2}{*}{This work} & 0.051  &  0.15  &  0.36  &  0.47  &  0.49  &  0.72  &  0.88  &  1.36  &  2.41  &  3.29  &  4.55\\
& $\pm$0.023  &  $\pm$0.04  &  $\pm$0.07  &  $\pm$0.08  &  $\pm$0.09  &  $\pm$0.12  &  $\pm$0.14  &  $\pm$0.21  &  $\pm$0.33  &  $\pm$0.43  &  $\pm$0.57 \\
\hline
\multirow{2}{*}{\citet{2013ApJ...766...81F}} & 0.015  &  0.08  &  0.25  &  0.43  &  0.70  &  0.93  &  1.29  &  2.00  &  3.24  &  4.19  &  5.24\\
& $\pm$0.007  &  $\pm$0.02  &  $\pm$0.04  &  $\pm$0.05  &  $\pm$0.08  &  $\pm$0.10  &  $\pm$0.14  &  $\pm$0.22  &  $\pm$0.37  &  $\pm$0.46  &  $\pm$0.55\\
\hline
\multirow{2}{*}{\citet{2011arXiv1109.2497M}} & --  &  0.26  &  0.54  &  0.83  &  1.00  &  1.39  &  1.51  &  2.47  &  3.18  &  4.01  &  5.37\\
 & --  &  $\pm$0.19  &  $\pm$0.27  &  $\pm$0.34  &  $\pm$0.38  &  $\pm$0.46  &  $\pm$0.48  &  $\pm$0.64  &  $\pm$0.73  &  $\pm$0.84  &  $\pm$0.96\\
\hline
\hline
&&&&&&\\
&&&&&&\\
\end{tabular}
\setlength{\tabcolsep}{1.1mm}
\begin{tabular}{l|ccc|ll}
\hline
\hline
\multirow{2}{*}{Reference} & Hot jupiters & Period-valley giants & Temperate giants & \multirow{2}{*}{Instrument} & \multirow{2}{*}{Stellar population$^\ast$}\\
 & $P<10$ d & $10<P<85$ d & $85<P<400$ d & & \\
\hline
\citet{2012ApJ...753..160W} & 1.20$\pm$0.38 & -- & -- & Keck+Lick & FGK V / SNH \\
\citet{2011arXiv1109.2497M} & 0.83$\pm$0.34$^{\dag}$  &  1.64$\pm$0.55  &  2.90$\pm$0.72 & HARPS+CORALIE & F5 -- K5 V / SNH\\
\hline
\citet{2011ApJ...743..103B} & 0.10$^{_{+0.27}}_{^{-0.08}}$ & -- & -- & SuperLupus & dwarfs / Lupus-FOV\\
\citet{SanternePhDThesis} & 0.95$\pm$0.26$^{\ddag}$ & -- & -- & \textit{CoRoT} & FGK V / center\\
\citet{SanternePhDThesis} & 1.12$\pm$0.31$^{\ddag}$ & -- & -- & \textit{CoRoT} & FGK V / anti-center\\
\citet{2012ApJS..201...15H} & 0.4$\pm$0.1 & -- & --  & \textit{Kepler} & GK V / K-FOV\\
\citet{2012AA...545A..76S} & 0.57$\pm$0.07 & -- & --  & \textit{Kepler} & FGK V / K-FOV\\
\citet{2013ApJ...766...81F} & 0.43$\pm$0.05  & 1.56$\pm$0.11  &  3.24$\pm$0.25  & \textit{Kepler} & FGK V / K-FOV \\
This work & 0.47$\pm$0.08  &  0.90$\pm$0.24  &  3.19$\pm$0.73 & \textit{Kepler} & F5 -- K5 V / K-FOV \\
\hline
\hline
\end{tabular}
\end{center}
\tablefoot{The horizontal line separates the values determined by RV surveys (above the line) from the ones determined by photometric transit surveys (below the line).\\
$^{\ast}$ SNH refers to solar neighborhood; center refers to the FOVs observed by \textit{CoRoT} toward the galactic center during the prime mission; anti-center refers to the FOVs observed by \textit{CoRoT} toward the galactic anti-center during the prime mission; IRa01 and LRa01 are two FOVs toward the galactic anti-center observed by \textit{CoRoT} K-FOV refers to the \textit{Kepler} prime mission FOV .\\
$^{\dag}$ This value slightly differs from the one provided in \citet{2011arXiv1109.2497M} for planets up to 11 days of period. The difference is the planet HD108147b (P=10.89 d) that we included in the population of the period-valley giants and not in the hot jupiters.\\
$^{\ddag}$ Preliminary results. For more robust values, see Deleuil et al. (in prep.).}
\label{OccurRateComp}
\end{table}%

\begin{landscape}
\begin{center}
\setlength{\tabcolsep}{0.5mm}
\begin{longtable}{ccc|ccccc|ccccc|l}
\caption{Adopted parameters for the brown dwarfs and their host.}\\
\hline
\hline
KIC & KOI & Kepler & Period & a/R$_{\star}$ & R$_{\rm c}$/R$_{\star}$ & R$_{\rm c}$ & M$_{\rm c}$ & \teff & \met & R$_{\star}$ & M$_{\star}$ & K$_{p}$  &  References\\
ID & ID & ID & [d] &  & & [\Rjup] & [\Mjup] & [K] & [dex] & [\Rsun] & [\Msun] & & \\ 
\hline
\endfirsthead
\caption{Continued.} \\
\hline
KIC & KOI & Kepler & Period & a/R$_{\star}$ & R$_{\rm c}$/R$_{\star}$ & R$_{\rm c}$ & M$_{\rm c}$ & \teff & \met & R$_{\star}$ & M$_{\star}$ & K$_{p}$  & References\\
\hline
\endhead
\hline
\endfoot
\hline
\hline
\endlastfoot
7046804  &  205.01  &    --    &  11.720  &  25.070$\pm$0.43  &  0.09906$^{_{+9.4\times10^{-4}}}_{^{-9.4\times10^{-4}}}$  &  0.82$\pm$0.02  &  40.80$^{_{+1.10}}_{^{-1.50}}$  &  5400$\pm$75  &  0.18$\pm$0.12  &  0.87$\pm$0.02  &  0.96$^{_{+0.03}}_{^{-0.04}}$  &  14.5  &    D\'i13, Bon15\\ 
6289650  &  415.01  &    --    &  166.788  &  100.000$^{_{+7.50}}_{^{-10.10}}$  &  0.06490$^{_{+1.7\times10^{-3}}}_{^{-1.3\times10^{-4}}}$  &  0.79$^{_{+0.12}}_{^{-0.07}}$  &  62.14$\pm$2.69  &  5810$\pm$80  &  -0.24$\pm$0.11  &  1.25$^{_{+0.15}}_{^{-0.10}}$  &  0.94$\pm$0.06  &  14.1  &    Mo13\\ 
9478990  &  423.01  &    39b    &  21.087  &  24.920$^{_{+1.90}}_{^{-1.50}}$  &  0.09100$^{_{+6.0\times10^{-4}}}_{^{-8.0\times10^{-4}}}$  &  1.24$^{_{+0.09}}_{^{-0.10}}$  &  20.10$^{_{+1.30}}_{^{-1.20}}$  &  6350$\pm$100  &  0.10$\pm$0.14  &  1.40$\pm$0.10  &  1.29$^{_{+0.06}}_{^{-0.07}}$  &  14.3  &    Bou11, Bon15
\label{BDParams}
\end{longtable}%
\tablebib{Bou11: \citet{2011A&A...533A..83B}; Bon15: \citet{2015A&A...575A..85B}; D\'i13: \citet{2013A&A...551L...9D}; Mo13: \citet{2013A&A...558L...6M}.
}
\end{center}
\end{landscape}

\twocolumn

\begin{table}[h]
\caption{Values of the correction factors used to derive the occurrence rates of brown dwarfs in the \textit{Kepler} FOV.}
\begin{center}
\begin{tabular}{ccccc}
\hline
\hline
KOI & $\mathcal{C}^{T}$ & $\mathcal{C}^{R}$ & $\mathcal{C}^{S}$ & $\mathcal{C}^{D}$\\ 
ID &  & [\%] & [\%] &  \\ 
\hline
205.01  &  25.07  &  100.0  &  80.1  &  1.57  \\
415.01  &  100.00  &  72.5  &  79.4  &  1.39  \\
423.01  &  24.92  &  96.0  &  79.0  &  0.40 \\
\hline
\hline
\end{tabular}
\end{center}
\label{CorrecFactorBD}
\end{table}%

\begin{table}[h]
\caption{Assumed lower limit in mass and RV amplitude (assuming a circular orbit) for the objects for which only an upper-limit in mass was determined.}
\begin{center}
\begin{tabular}{lccc}
\hline
\hline
KOI & min(M$_{p}$) & min(M$_{p}$) & min($K$)\\
ID & [\Mjup] & [\Mearth] & [\ms]\\
\hline
12.01  &  0.241  &  76.7  &  14.6  \\
63.01  &  0.035  &  11.1  &  3.4  \\
131.01  &  0.104  &  33.1  &  10.3  \\
197.01  &  0.065  &  20.7  &  6.0  \\
201.01  &  0.078  &  24.9  &  9.4  \\
221.01  &  0.019  &  6.2  &  3.1  \\
351.01  &  0.119  &  37.9  &  3.1  \\
351.02  &  0.061  &  19.5  &  1.9  \\
372.01  &  0.060  &  19.0  &  2.4  \\
464.01  &  0.046  &  14.7  &  2.5  \\
611.01  &  0.098  &  31.3  &  12.9  \\
1089.01  &  0.169  &  53.7  &  6.5  \\
1271.01  &  0.103  &  32.9  &  3.2  \\
1411.01  &  0.054  &  17.2  &  1.5  \\
1426.03  &  0.172  &  54.6  &  6.4  \\
1431.01  &  0.059  &  18.8  &  1.7  \\
1783.01  &  0.053  &  16.7  &  1.5  \\
1788.01  &  0.036  &  11.5  &  2.0  \\
2679.01  &  0.148  &  47.1  &  5.6  \\
3663.01  &  0.096  &  30.5  &  3.1  \\
3678.01  &  0.079  &  25.0  &  3.3  \\
3683.01  &  0.081  &  25.8  &  2.1  \\
3689.01  &  0.162  &  51.3  &  16.1  \\
6132.01  &  0.160  &  50.9  &  8.3  \\
\hline
\hline
\end{tabular}
\end{center}
\label{NoVarMinMass}
\end{table}%


\appendix

\section{Results from the spectroscopic observations}
\label{SpectroObs}

\subsection{General informations}

We present in this appendix the observation and their analysis we performed on each candidate. The radial velocities and their diagnoses are listed in Tables \ref{RVsingletable}, \ref{RVSB2table}, and \ref{RVSB3table}. For some candidates, we refer to the data validation (DV) summary produced by the \textit{Kepler} team. They are available at the NASA exoplanet archive\footnote{\url{http://exoplanetarchive.ipac.caltech.edu}}.

Some candidates turn out to be false positives and are actually member of the \textit{Kepler} EB catalog (Kirk et al, in prep.). This is the case for the following candidates: KOI-129.01, KOI-138.01, KOI-198.01, KOI-368.01, KOI-449.01, KOI-969.01, KOI-976.01, KOI-1020.01, KOI-1227.01, KOI-1232.01, KOI-1326.01, KOI-1452.01, KOI-1483.01, KOI-1645.01, KOI-1784.01, KOI-3411.01, KOI-3720.01, KOI-3721.01, KOI-3782.01, KOI-3783.01, KOI-3787.01, KOI-3811.01, KOI-5034.01, KOI-5086.01, KOI-5132.01, KOI-5436.01, KOI-5529.01, KOI-5708.01, KOI-5745.01, KOI-5976.01, KOI-6066.01, KOI-6175.01, KOI-6235.01, KOI-6460.01, KOI-6602.01, KOI-6800.01, KOI-6877.01, KOI-6933.01, KOI-7044.01, KOI-7054.01, KOI-7065.01, and KOI-7527.01.

Our observations also confirm the results of \citet{2015AJ....149...18K} for the candidates: KOI-969.01, KOI-1020.01, KOI-1137.01, KOI-1227.01, KOI-1326.01, KOI-1452.01, KOI-1645.01, KOI-1784.01, KOI-3721.01, and KOI-3782.01.



\subsection{KOI-129.01}
\label{129}

KOI-129.01 is a 24-day planet candidate reported for the first time in \citet{2014ApJS..210...19B} and flagged as a false positive. However, in the later candidate releases, this object was no longer flagged as a false positive. We obtained two SOPHIE HE RV which exhibit a variation at the level of about 10 \kms\, in phase with the \textit{Kepler} ephemeris. Assuming there is no significant drift in the data and a circular orbit at the transit ephemeris, we find that the RV semi-amplitude is K = 6.16 $\pm$ 0.04 \kms. Assuming a host mass of M$_{1}$ = 1.29$^{+0.28}_{-0.23}$ \Msun\, from \citet{2014ApJS..211....2H}, it gives a companion mass of M$_{2}$ = 0.11 $\pm$ 0.01 \Msun. The planet candidate KOI-129.01 is therefore a false positive and likely a very low mass star.

\subsection{KOI-138.01}
\label{139}

KOI-138.01 is a 49-day planet candidate reported for the first time in \citet{2011ApJ...736...19B}. We obtained five different epochs with SOPHIE HE. They present a large RV variation in \textit{anti-phase} with the \textit{Kepler} ephemeris. Assuming no significant drift in the data and fixing the orbital period to the one observed by \textit{Kepler}, we find that the orbit has a RV amplitude of K = 22.31 $\pm$ 0.08~\kms, an eccentricity of e = 0.33 $\pm$ 0.01, an argument of periastron of $\omega$ = 240.0 $\pm$ 0.4~\degr, and an epoch of periastron of T$_{p}$ = 2454922.61 $\pm$ 0.04. Assuming a host mass of M$_{1}$ = 1.42$^{_{+0.28}}_{^{-0.31}}$ \Msun\, from \citet{2014ApJS..211....2H}, the companion has a mass of M$_{2}$ = 0.63 $\pm$ 0.07 \Msun. The RV orbit also gives an epoch of secondary eclipse to be T$_{occ}$ = 2454973.542 $\pm$ 0.046, which coincides with the epoch of transit that \textit{Kepler} detected of T$_{0}$ = 2454973.766221 $\pm$ 3.24 10$^{-4}$. Therefore, the candidate KOI-138.01 is not a transiting planet but a secondary-only EB.

\subsection{KOI-198.01}
\label{198}

KOI-198.01 is a 87-day planet candidate reported for the first time in \citet{2014ApJS..210...19B}, and flagged as a false positive. It is no longer flagged as a false positive in the latter candidates releases. We obtained only two different epochs with SOPHIE HE. They show a RV span of about 10 \kms\, in \textit{anti-phase} with the \textit{Kepler} ephemeris. The large RV variation observed by SOPHIE is not compatible with a planet. We therefore conclude this candidate is a false positive. Given that the RV are in \textit{anti-phase} with the \textit{Kepler} ephemeris, we expected this candidate to be a secondary-only EB.

\subsection{KOI-201.01}
\label{201}

The giant-planet candidate KOI-201.01 was announced in \citet{2011ApJ...736...19B} with an orbital period of 4.2 days. The SOPHIE spectroscopic observations reported by \citet{2012AA...545A..76S} did not allow them to detected the Doppler signature of the planet. Recent observations with the HARPS-N spectrograph\footnote{OPTICON programme ID: OPT15A\_13 -- PI: H\'ebrard} allowed us to detect this signature and characterise the mass of the candidate, confirming also its planetary nature. The analysis of this system will be presented in a forthcoming paper (H\'ebrard et al., in prep.).

\subsection{KOI-221.01}
\label{221}

A giant-planet candidate was reported in \citet{2011ApJ...728..117B} with an orbital period of about 3 days. It was however found with a transit depth of about 0.37\%, which was below the minimum depth selected by \citet{2012AA...545A..76S}. With more \textit{Kepler} data, the transit depth was revised to be slightly above 0.4\% \citep{2015arXiv150202038M} and this candidate is thus within our selection criteria. With more data, it was also possible to find a Earth-size candidate at 6 days \citep{2014ApJS..210...19B}. We observed it twice with SOPHIE HE and find no significant RV, bisector nor FWHM variation. We fitted the RV with two circular orbits at the \textit{Kepler} ephemeris and find that K$_{221.01} <$ 99 \ms\, and K$_{221.02} <$ 156 \ms, at the 99\% level. Assuming a host mass of M$_{1}$ = 0.77$^{_{+0.11}}_{^{-0.07}}$ \Msun\, \citet{2014ApJS..211....2H}, we find that the candidates have masses of M$_{221.01} <$ 0.65 \Mjup\, and M$_{221.02} <$ 1.16 \Mjup, within a probability of 99\%. We can therefore exclude that these candidates are stars or brown dwarfs eclipsing the target star, but we can not rule out other false-positive scenarios.

\subsection{KOI-351.01 \& KOI-351.02}
\label{351}

The target star KOI-351 was found to host seven transiting planet candidates \citep{2014ApJ...781...18C} with orbital periods of 332d (KOI-351.01), 211d (KOI-351.02), 60d (KOI-351.03), 92d (KOI-351.04), 9d (KOI-351.05), 6d (KOI-351.06), and 125d (KOI-351.07). Only the two outermost planets are giant-planet candidates within our selection criteria. We obtained five RV with SOPHIE HE. They have a $rms$ of 16 \ms\, which is compatible with the uncertainties. The bisector and FWHM do not show significant variation with $rms$ of 27 \ms\, and 80 \ms\, (respectively). We modelled the RV with a 7-orbit model with fixed eccentricities to zero and ephemeris fixed at the \textit{Kepler} ones. We find upper-limits at the 99\% level for the amplitude of the planets of K$_{351.01} <$ 35.5 \ms, K$_{351.02} <$ 29.0 \ms, K$_{351.03} <$ 63.6 \ms, K$_{351.04} <$ 81.0 \ms, K$_{351.05} <$ 29.4 \ms, K$_{351.06} <$ 102.6 \ms, and K$_{351.07} <$ 29.1 \ms. 

Assuming a stellar host mass of M$_{1}$ = 0.99 $\pm$ 0.10 \Msun\, \citep{2014AJ....148...28S}, we derived upper-limits on the mass of these exoplanets, at the 99\% confidence interval, of M$_{351.01} <$ 1.16 \Mjup, M$_{351.02} <$ 0.82 \Mjup, M$_{351.03} <$ 1.17 \Mjup, M$_{351.04} <$ 1.76 \Mjup, M$_{351.05} <$ 0.29 \Mjup, M$_{351.06} <$ 1.04 \Mjup, and M$_{351.07} <$ 0.78 \Mjup. All these planets but the last one (KOI-351.07) were validated by the planet-likelihood multiplicity-boost \citep{2014ApJ...784...44L, 2014ApJ...784...45R}. Within the assumptions aforementioned, we can confirm that, if the objects are transiting the target star, they have a mass within the planetary range.

\subsection{KOI-368.01}
\label{368}

The target star KOI-368 was found to host a giant-planet candidate with a period of 110d \citep{2011ApJ...736...19B}. However, \citet{2013ApJ...776L..35Z} detected a clear secondary eclipse and conclude that the companion is a M dwarf. We observed this candidate host twice with SOPHIE HE and find a wide line profile (\vsini = 86.5 $\pm$ 0.6 \kms) which does not show significant RV variation. Because of its large rotation profile, we were not able to measure the bisector. We fitted these two measurements assuming a circular orbit and find that K $<$ 5.06 \kms, at the 99\% level. Assuming a host mass of M$_{1}$ = 2.3 $\pm$ 0.1 \Msun\, \citep{2013ApJ...776L..35Z}, we derived an upper-limit on the mass of the companion of M$_{2} <$ 0.21 \Msun, at the 99\% level and assuming a circular orbit. Our mass constraint is compatible with the late M-dwarf type claimed by \citet{2013ApJ...776L..35Z}. 

\subsection{KOI-372.01}
\label{372}

KOI-372.01 is a 125-day planet candidate reported for the first time in \citet{2011ApJ...736...19B}. The \textit{Kepler} light curve of this candidate reveals large photometric variability at the level of $\sim$ 1.5\% due to stellar activity. We observed it with SOPHIE HR and HE. We find a RV $rms$ of 24 \ms\, which is not compatible with the recent solution published by \citet{2015arXiv150404625M} of K = 132 $\pm$ 6 \ms. The RV and their in-depth analysis will be presented in Demangeon et al. (in prep.). We consider that the nature of this candidate is still unknown for the statistical analysis of this papier.

We derived the stellar atmospheric parameters that are reported in Table \ref{SpectroResults}. They correspond to a mass of M = 1.02 $\pm$ 0.04 \Msun, a radius of R = 1.13$^{_{+0.28}}_{^{-0.14}}$ \Rsun, and an age of 6.7$^{_{+1.8}}_{^{-3.2}}$ Gyr.

\subsection{KOI-449.01}
\label{449}

The planet candidate KOI-449.01 has an orbital period of $\sim$252 days and was announced in \citet{2014ApJS..210...19B} as a false positive. In later candidates lists, it was no longer flagged as a false positive. We observed this candidate three times with SOPHIE HE but detected no significant RV ($rms$ of 23 \ms), nor bisector or FWHM variation ($rms$ of 119 \ms\, and 52 \ms, respectively). By fitting a circular orbit at the \textit{Kepler} ephemeris, we find that K $<$ 1.1 \kms\, within a probability of 99\%. Assuming a host mass of M$_{1}$ = 1.02$^{_{+0.16}}_{^{-0.12}}$ \Msun\, \citep{2014ApJS..211....2H}, this corresponds to an upper-limit on the mass of this candidate of M$_{449.01} <$ 37.0 \Mjup. Therefore, within the circular approximation, we can exclude an EB as the source of the transit event. However, we can not exclude a transiting brown dwarf, nor a background source of false positive. 

This candidate is listed as an EB in \textit{Kepler} EB catalog (Kirk et al, in prep.) with a double period, showing some odd--even transits depth difference which suggest that this candidate is an EB in a circular orbit at twice the reported period. Since we detected no variation in the target star, we conclude this candidate is a CEB and not a transiting planet.

\subsection{KOI-464.01}
\label{464}

The candidate KOI-464.01 has an orbital period of $\sim$58 days, reported for the first time by \citet{2011ApJ...736...19B}. Another candidate transiting the same star at $\sim$5 days was also found with a transit depth of 800ppm. We observed it five times with SOPHIE HE and detected any clear variation neither in the RV ($rms$ of 27 \ms), nor in the bisector ($rms$ of 18 \ms), and FWHM ($rms$ of 47 \ms). We fitted these RV assuming two circular orbits at the \textit{Kepler} ephemeris. We find upper-limits on the amplitude of both candidates, at the 99\% confidence level of K$_{464.01} <$ 21.1 \ms\, and K$_{464.02} <$ 21.9 \ms. Assuming a host mass of M$_{1}$ = 0.97$^{_{+0.10}}_{^{-0.08}}$ \Msun\, \citep{2014ApJS..211....2H}, we find upper-limits on the mass of these candidate of M$_{464.01} <$ 0.68 \Mjup\, and M$_{464.02} <$ 0.29 \Mjup, at the 99\% level.

\subsection{KOI-531.01}
\label{531}

\begin{figure}[h]
\begin{center}
\includegraphics[width=\figw]{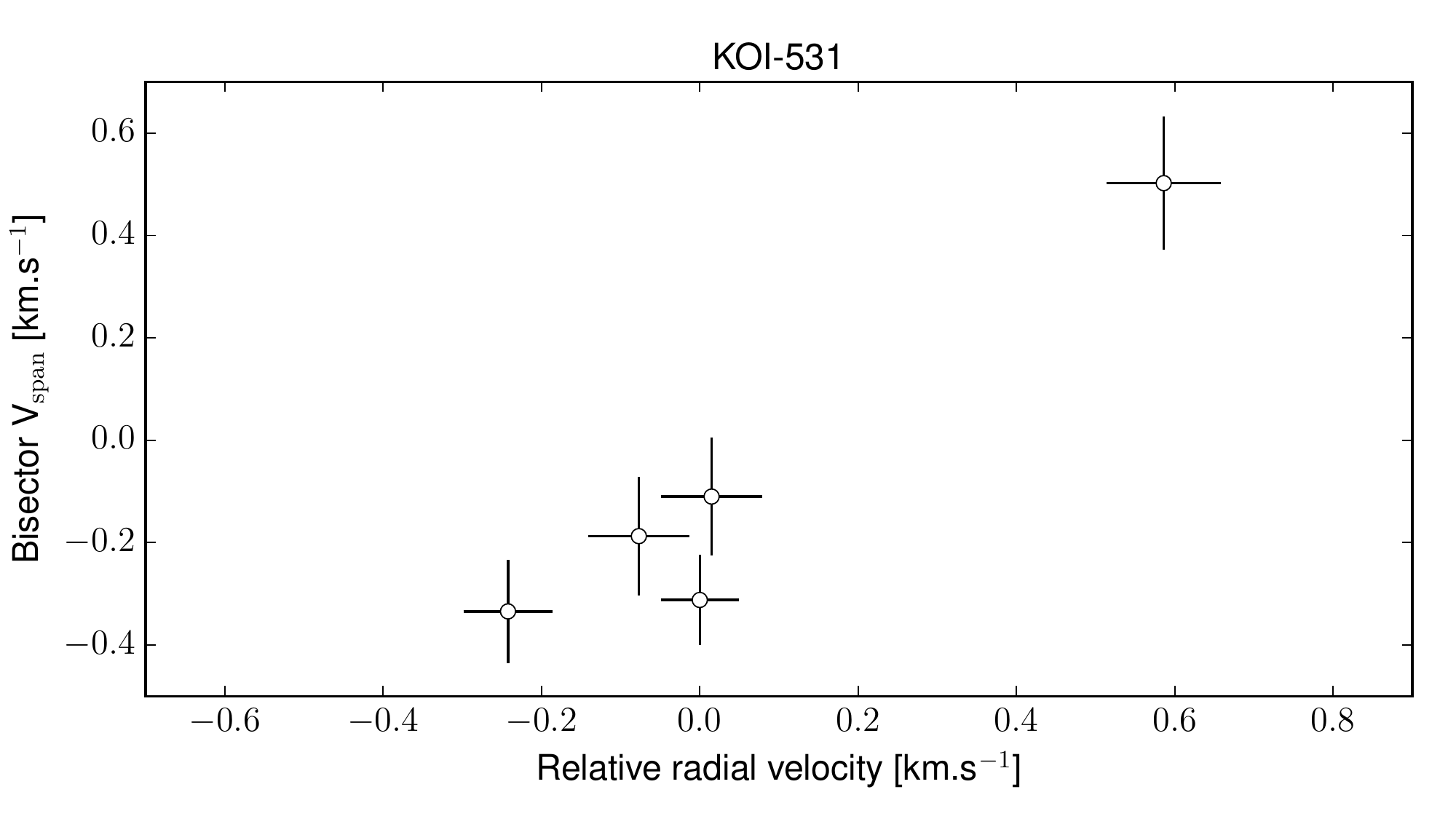}
\caption{Bisector V$_{\rm span}$ as a function of the RV for the target star KOI-531.}
\label{plot531}
\end{center}
\end{figure}

The transit candidate KOI-531.01 has an orbital period of $\sim$3.7 days. It was reported for the first time by \citet{2011ApJ...736...19B} with a wrong transit depth of 0.25\% due to a bad transit fit. Therefore, it was not included in the sample of \citet{2012AA...545A..76S}. The corrected transit depth being of 0.53\% \citep{2013ApJ...767...95D}, the candidate is now included in this sample. We secured five different epochs on this candidate with SOPHIE HE. The RV, bisector and FWHM present a large dispersion with a $rms$ of 280 \ms, 306 \ms, and 351 \ms\, for median uncertainties of 64 \ms, 115 \ms, and 160 \ms, respectively. The bisector is clearly correlated with the RV (see Fig. \ref{plot531}), with a Spearman correlation coefficient of $\rho =$ 0.70 $\pm$ 0.25. This candidate is clearly not a transiting planet, which would not have produced such large variation of the line-profile shape. Only a blended system, like a triple system, might explain the observed correlation between the RV and bisector \citep{2015MNRAS.451.2337S}. This actually confirms the recent multi-color observations with the GTC obtained by \citet{2015arXiv150607057C}.

\subsection{KOI-617.01}
\label{617}

The giant candidate KOI-617.01 was revealed by \citet{2011ApJ...728..117B}. It has an orbital period of $\sim$ 38 days. We secured four observations of this star with SOPHIE HE that revealed a clear SB2. By fitting the cross-correlation function with two Gaussians, we derived the RV of both components. We call the star A the component with the deepest line profile and star B the one with the faintest line profile in the CCF. We then fitted those RV with a combined Keplerian orbit, fixing only the orbital period to the transit one. We find an epoch of periastron of T$_{p} \approx$ 2456130.23, an eccentricity of $e \approx 0.23$, an argument of periastron of $\omega \approx$ 277 \degr, and RV amplitudes of K$_{A} \approx$ 39.56 \kms\, and K$_{B} \approx$ -42.71 \kms\, for stars A and B (respectively). This gives a mass ratio between the two stars of $q \approx 0.93$. Assuming a primary mass of M$_{1} \approx 1.056$ \Msun, \citep{2014ApJS..211....2H}, the secondary mass is of about M$_{2} \approx$ 0.98 \Msun. Note that the orbital ephemeris determined with the RV gives an epoch of primary and secondary eclipse of about 24550012.02 and 2455031.67, respectively. The transit epoch detected by \textit{Kepler} (2455031.60) is compatible with the secondary eclipse of this binary system. KOI-617 is thus a case of secondary-only EB, and not a transiting planet.

\subsection{KOI-620.01 \& KOI-620.02}
\label{620}

Two EGP candidates (KOI-620.01 and KOI-620.02) were detected on the target star KOI-620 by \citet{2011ApJ...728..117B} and \citet{2013ApJS..204...24B} with orbital periods of $\sim$45 and $\sim$130 days (respectively). Note that another set of transit was detected with a period of $\sim$85 days and a depth of 0.2\% \citep{2013ApJS..204...24B}, which is outside our selection criteria. Their transits exhibit transit timing variation which allowed \citet{2013MNRAS.428.1077S} to confirm their planetary nature. The same authors constrained the mass of the inner planet to be less than 3.23\Mjup. More recently, \citet{2014ApJ...783...53M} constrained the mass of the three planets in this system using their transit timing and found masses of a few Earth masses, leading to unexpected low density for these giant objects, with $\rho_{p} \leq 0.05$ g.cm$^{-3}$.

We obtained seven different epochs with SOPHIE HE of KOI-620. We detect no significant RV variation ($rms$ = 14 \ms). The line bisector does not show variations above the noise level ($rms$ = 24 \ms). The FWHM shows some variation at the level of $rms$ = 157 \ms, which is likely instrumental.

We fitted three circular orbits to these data fixing the ephemeris to the ones derived thanks to \textit{Kepler}. We find that K$_{620.01} <$ 45 \ms, K$_{620.02} <$ 91 \ms, and K$_{620.03} <$ 43 \ms\, at the 99\% level. Assuming a stellar mass of M$_{1}$ = 1.05$^{_{+0.17}}_{^{-0.14}}$ \Msun\, \citep{2014ApJS..211....2H}, those limits correspond to upper-limit on the planetary masses to M$_{620.01} <$ 0.85 \Mjup, M$_{620.02} <$ 2.43 \Mjup, and M$_{620.03} <$ 1.01 \Mjup\, with a 99\% confidence interval. Given the large uncertainties of our photon-noise limited spectroscopic observations, the derived upper-limit in mass are fully compatible with the mass constraints from \citet{2013MNRAS.428.1077S} and \citet{2014ApJ...783...53M}.

\subsection{KOI-969.01}
\label{969}

A giant-planet candidate with a period of 18 days was revealed in \citet{2014ApJS..210...19B} with a depth of 0.36\%. This candidate was not known at the time of the observation of \citet{2012AA...545A..76S}. Its transit depth was then revised to be of 0.45\% \citep{2015ApJS..217...16R}, which includes it in our sample. We observed it twice with SOPHIE HE that revealed a clear SB2 of two stars of similar flux. We fitted the cross-correlation function with a two-Gaussian profile to derive the RV of both stars. We call star A the one with the smallest RV variation and star B the one with the largest variation. The RV variation of both stars is anti-correlated, which gives a mass ratio of 99.6\%. The RV of the star A are in phase with the \textit{Kepler} ephemeris, indicating that the primary eclipse was detected in the light-curve. This candidate is not a transiting planet but an equal-mass EB. 

\subsection{KOI-976.01}
\label{976}

\begin{figure}[h]
\begin{center}
\includegraphics[width=\figw]{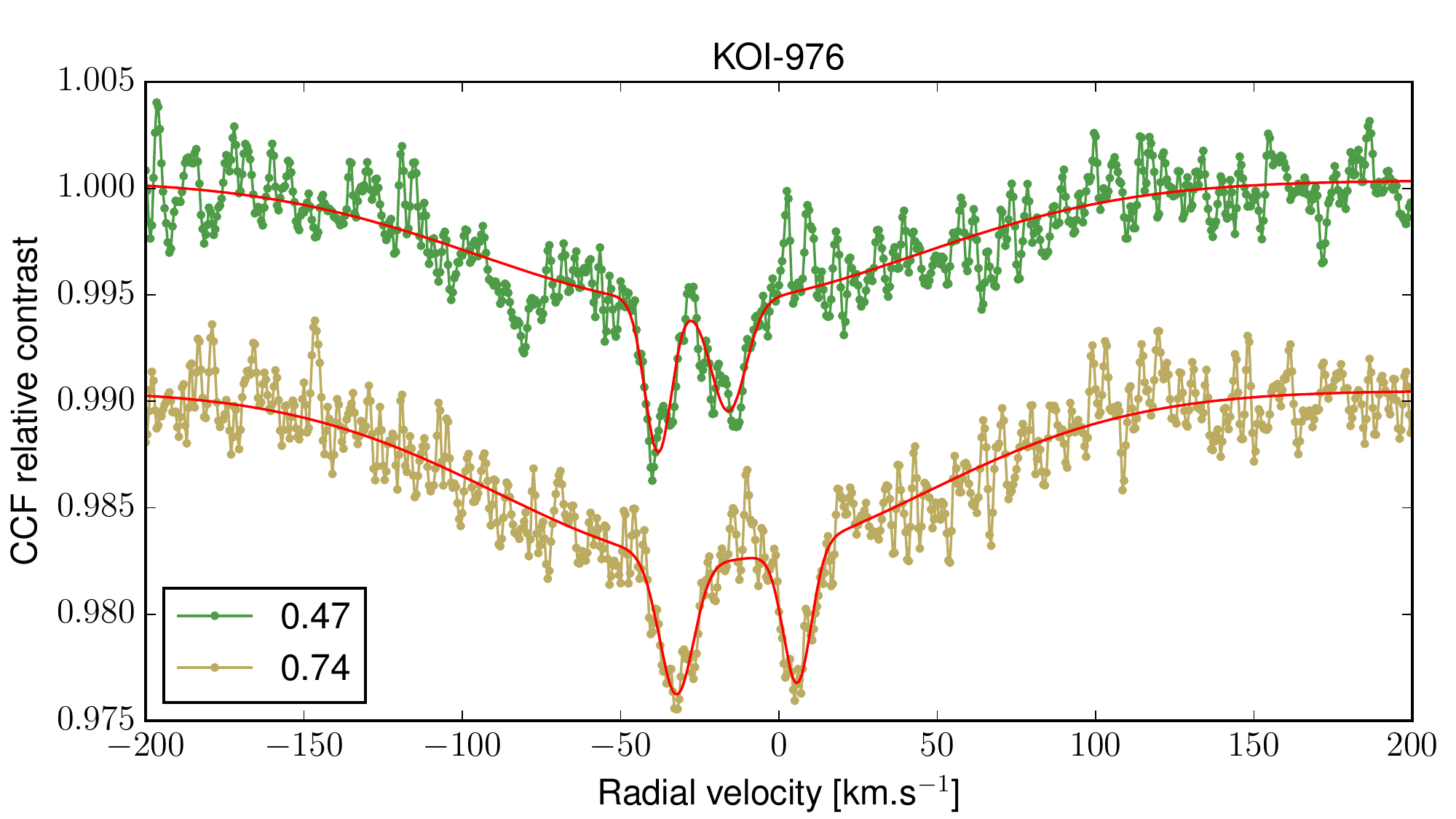}
\caption{Cross-correlated a function of the target star KOI-976 revealing at least three stellar components: a wide and shallow component and two narrower and deeper components. The legend indicates the orbital phase of the transiting candidate. An arbitrary offset in flux has been set between the two observations. The red lines are the three-Gaussian fits to the CCFs.}
\label{fig976}
\end{center}
\end{figure}

A transiting candidate has been reported on KOI-976 by \citet{2011ApJ...736...19B} with an orbital period of $\sim52$ days. We observed this star twice with SOPHIE HE. The cross-correlation function displayed in Fig. \ref{fig976} revealed a multiple stellar system, with at least three components. Given the large dilution produced by those stars, the true depth of this transiting candidate should be much larger than the observed one (2.67\%). Therefore, we conclude that this candidate is most likely a triple system or a background EB, and not a transiting planet. The derived RV of the three stellar components, by fitting three Gaussian profiles. 

\subsection{KOI-1020.01}
\label{1020}

The candidate KOI-1020.01 was revealed by \citet{2011ApJ...736...19B}. It has an orbital period of about 37 days. We observed this candidate six times with SOPHIE HE which revealed a clear SB2 with two lines of similar contrast and FWHM. We fitted the cross-correlation function with a two-Gaussian function and derived the RV of both stars. Since is it not possible to distinguish from the spectra which component is the brightest one, we arbitrarily called star A the most blue-shifted star, and star B the most red-shifted component at the first observation. We fitted the derived RV of both stars with a combined Keplerian orbit, fixing the period to the transit one. We find that the data are best fitted with an epoch of periastron of T$_{p}$ = 2456136.35, an eccentricity of $e=$ 0.43, an argument of periastron of $\omega=$ 51.2\degr, and amplitudes of K$_{A} = 45.9$\kms\, and K$_{B} = -46.1$\kms. Star A is therefore slightly more massive than star B, with a mass ratio of $ q \approx 99.5$\%. Those derived orbital parameters predict an epoch of primary eclipse of 2454996.9, which corresponds to the event detected by \textit{Kepler}, with an epoch of 2454997.1. The candidate KOI-1020 is thus a nearly equal-mass binary, and not a transiting planet. Given the large eccentricity of this system and its inclination, the secondary eclipse is not observable from the Earth.

\subsection{KOI-1089.01}
\label{1089}

The EGP candidate KOI-1089.01 has an orbital period of $\sim$ 87 days. It was revealed by \citet{2011ApJ...736...19B} together with a smaller planet candidate at 12.2 days. Using multi-color GTC observations, \citet{2014A&A...567A..14T} found that KOI-1089.01 is an EGP blended with another, unseen star. We secured eight epochs on this star with SOPHIE HE. We detect any significant RV variation in these data ($rms$ = 21 \ms). We find no significant variation in the bisector ($rms$ = 53 \ms) nor FWHM ($rms$ = 137 \ms). We analysed the SOPHIE RV together with the FIES data reported by \citet{2014A&A...567A..14T}. We modelled two Keplerian orbits fixing the ephemeris to the ones found by \textit{Kepler}. We find that, at the 99\% confidence interval, the amplitude of both transiting planet candidates is K$_{1089.01} <$ 67 \ms\, and K$_{1089.02} <$ 37 \ms. 

We derived the stellar atmospheric parameter that are reported in Table \ref{SpectroResults}.Those parameters give a stellar mass of M = 1.29$^{_{+0.21}}_{^{-0.10}}$ \Msun, a radius of R = 1.60$^{_{+0.68}}_{^{-0.36}}$ \Rsun, and an age of 3.1 $\pm$ 1.3 Gyr. Combining both results, the candidates have upper-mass limits of M$_{1089.01} <$ 1.12 \Mjup\, and M$_{1089.02} <$ 0.46 \Mjup, at the 99\% level. We can therefore exclude any massive object transiting KOI-1089 at $\sim$ 12 days, but we can not firmly rule out a background source of transit based on these data. We also improved the upper-limit on the mass of KOI-1089.01 reported by \citet{2014A&A...567A..14T}.

\subsection{KOI-1137.01}
\label{1137}

The candidate KOI-1137.01 was listed among the potential transiting planet in \citet{2014ApJS..210...19B}, with a transit depth of 1.5\%. In the latest candidate release \citep{2015arXiv150202038M}, its transit depth has been revised to 4.2\%. Since this is outside our selection criteria, we did not include it in the analysis. However, we observed it after the \citet{2014ApJS..210...19B} candidate release, and we report here its nature. We secured four observations of this candidates with SOPHIE HE. They exhibit a clear RV variation in phase with the \textit{Kepler} ephemeris ($rms$ = 270 \ms). They also show a clear variation of the line-profile bisector ($rms$ = 2.7 \kms) as well as in the FWHM ($rms$ = 640 \ms). Moreover, we clearly detect a second set of stellar lines in the cross-correlation function of two spectra. We conclude that this candidate is a triple system or a background EB, but not a transiting planet.

\subsection{KOI-1227.01}
\label{1227}

The candidate KOI-1227.01 was released by \citet{2011ApJ...736...19B}. It has an orbital period of 2.1 days, but was not included in the candidate sample from \citet{2012AA...545A..76S} due to its poor vetting statuts in \citet{2011ApJ...736...19B}. We observed it twice with SOPHIE HE which revealed a clear SB2 with both component of similar flux. We fitted the cross-correlation function with a two-Gaussian function. We call the star A (B) the one which show RV variation in phase (anti-phase, respectively) with the \textit{Kepler} ephemeris. Then, we fitted the RV of both stars with a combined circular orbit fixing the ephemeris to the transit ones. We find that the RV amplitude is K$_{A} \approx$ 72.67 \kms\, and K$_{B} \approx$ -71.99 \kms. 

It is quite surprising that K$_{A} >$ |K$_{B}$| (hence M$_{A} <$ M$_{B}$), since A varies in phase with the transit ephemeris and thus should be the most massive star in the system. The solution might be that this nearly equal-mass binary is orbiting at twice the period detected by \textit{Kepler}. In such situation, the primary and secondary eclipse would have the same depth and duration. So, assuming that the true period is 4.2 days (the epoch of primary eclipse is kept to T$_{0} \approx$ 2454966.576), the amplitudes are K$_{A} \approx$ 81.65 \kms\, and K$_{B} \approx$ -82.43 \kms. This gives a mass ratio of $q \approx$ 99.0\%, with M$_{A} >$ M$_{B}$. In any case, this candidate is not a transiting planet, but a nearly equal-mass binary.

\subsection{KOI-1230.01}
\label{1230}

KOI-1230.01 is a 166-day period candidate revealed in \citet{2011ApJ...736...19B}. The host was found to be a giant star, with a \logg\, of about 3 \cmss. We collected ten SOPHIE HE RVs. They present a large variation in phase with \textit{Kepler} ephemeris. Fitting these data with a Keplerian orbit with the ephemeris fixed at the transit ones, we find an amplitude K$_{1230.01} = 17.87 \pm 0.03$ \kms, an eccentricity $e = 0.6944 \pm 0.0009$, an argument of periastron of $\omega = 131.37 \pm 0.06$ \degr. Assuming a stellar host of M$_{1}$ = 1.78 $\pm$ 0.19 \Msun\, \citep{2014ApJS..211....2H}, the companion has a mass of M$_{2}$ = 0.59 $\pm$ 0.04 \Msun. This candidate is therefore an EB and not a transiting planet. This system will be further analysed in Bruno et al. (in prep.).

\subsection{KOI-1232.01}
\label{1232}

The giant-planet candidate KOI-1232.01 was revealed in \citet{2014ApJS..210...19B} and flagged as a false positive. It was not flagged as a false positive in the later candidate releases. It has an orbital period of 119.4 days. We secured three different epochs with SOPHIE HE. They revealed a large RV variation. Three measurements are not enough to fully constrain the orbit of such long period candidate. By assuming a circular orbit, no significant drift, and the transit ephemeris, we find that the RV amplitude is K = 13.74 $\pm$ 0.06 \kms. For this candidate, \citet{2014ApJS..211....2H} reported a host mass of M$_{1}$ = $0.60^{_{+0.07}}_{^{-0.03}}$ \Msun. This would give a companion mass of M$_{2}$ = 0.29 $\pm$ 0.02 \Msun. This candidate is therefore an EB and not a transiting planet.

\subsection{KOI-1271.01}
\label{1271}

A giant-planet candidate has been detected on the target star KOI-1271 by \citet{2013ApJS..204...24B} with a period of 162 days. It was found by \citet{2012ApJ...756..185F} to have large transit timing variation at the level of a few hours, but no other transiting candidate was found in the \textit{Kepler} light-curve. We secured 14 SOPHIE HE observations of this star. They have a RV $rms$ of 47 \ms, a bisector $rms$ of 79 \ms, and a FWHM $rms$ of 116 \ms. They are compatible with the typical uncertainty on this star.We analysed the RVs using one Keplerian orbit at the transit ephemeris. We find a hint of RV variation of K$_{1271.01}$ = 28 $\pm$ 17 \ms, (K$_{1271.01} <$ 77.5 \ms\, within the 99\% confidence interval) an eccentricity of $e = 0.17^{_{+0.22}}_{^{-0.13}}$, and an argument of periastron of $\omega = 197^{_{+74}}_{^{-120}}$ \degr.

We derived the stellar atmospheric parameters that are reported in Table \ref{SpectroResults}. We find a host mass of M = 1.33$^{_{+0.13}}_{^{-0.08}}$ \Msun, a radius of R = 1.57$^{_{+0.41}}_{^{-0.26}}$ \Rsun, and an age of 2.14$^{_{+0.55}}_{^{-0.84}}$ Gyr. Combining the results from the RV and stellar atmospheric results, we find that this candidate has a mass of M$_{1271.01}$ = 0.84 $\pm$ 0.49 \Mjup\, (i.e. an upper-limit at 99\% of 2.35 \Mjup). We can therefore rule out that this candidate is an eclipsing brown dwarf or binary. Given the transit timing variation and the hint of RV signal, we conclude that this candidate is likely a planet, without a firm establishment of its nature. Note that this candidate is in the \textit{Kepler} EB catalog (Kirk et al, in prep.) but we find no reason for that (the DV summary shows no odd -- even transits depth differences, nor significant centroids). Some confirmed exoplanets have already been misclassified as EB in this catalog \citep{2012AA...545A..76S}.

\subsection{KOI-1326.01}
\label{1326}

The giant-planet candidate was announced in \citet{2014ApJS..210...19B} with an orbital period of $\sim$ 53 days and a false positive flag. In later candidate releases, this candidate is no longer flagged as a false positive. We observed it twice with SOPHIE HE and find two line profiles in the cross-correlated function, revealing a clear SB2. We fitted a two-Gaussian function to the cross-correlation function. We call star A the one with the deepest line profile and star B, the one with the shallowest line profile. The two stars show anti-correlated RV variations with a slope of q = 85.8\%, corresponding to the mass ratio between the two stars. This confirms that star A is more massive than star B. However, the variation of star A is observed in \textit{anti-phase} with \textit{Kepler} ephemeris, revealing that the transit epoch match with the secondary eclipse of this binary system. This candidate is clearly not a transiting planet, but a nearly equal-mass secondary-only EB.

\subsection{KOI-1353.01}
\label{1353}

A giant-planet candidate has been announced by \citet{2011ApJ...736...19B} with a period of $\sim$125 days. Another transiting candidate was found by \citet{2013ApJS..204...24B} with a period of 34.5 days and a depth of about 400ppm. Finally, a third set of transit was discovered by the planet hunters community with a period of 66 days and reported in \citet{2014ApJ...795..167S}. These authors also performed a transit timing variation analysis of the candidates and found that the mass of the EGP was of M$_{1353.01}$ = 0.42 $\pm$ 0.05 \Mjup. Their derived masses for the inner and middle planets are 7.3 $\pm$ 6.8 \Mearth\, and 4.0 $\pm$ 0.9 \Mearth. They also reported that the orbits are nearly circular.

\begin{figure}[h]
\begin{center}
\includegraphics[width=\figw]{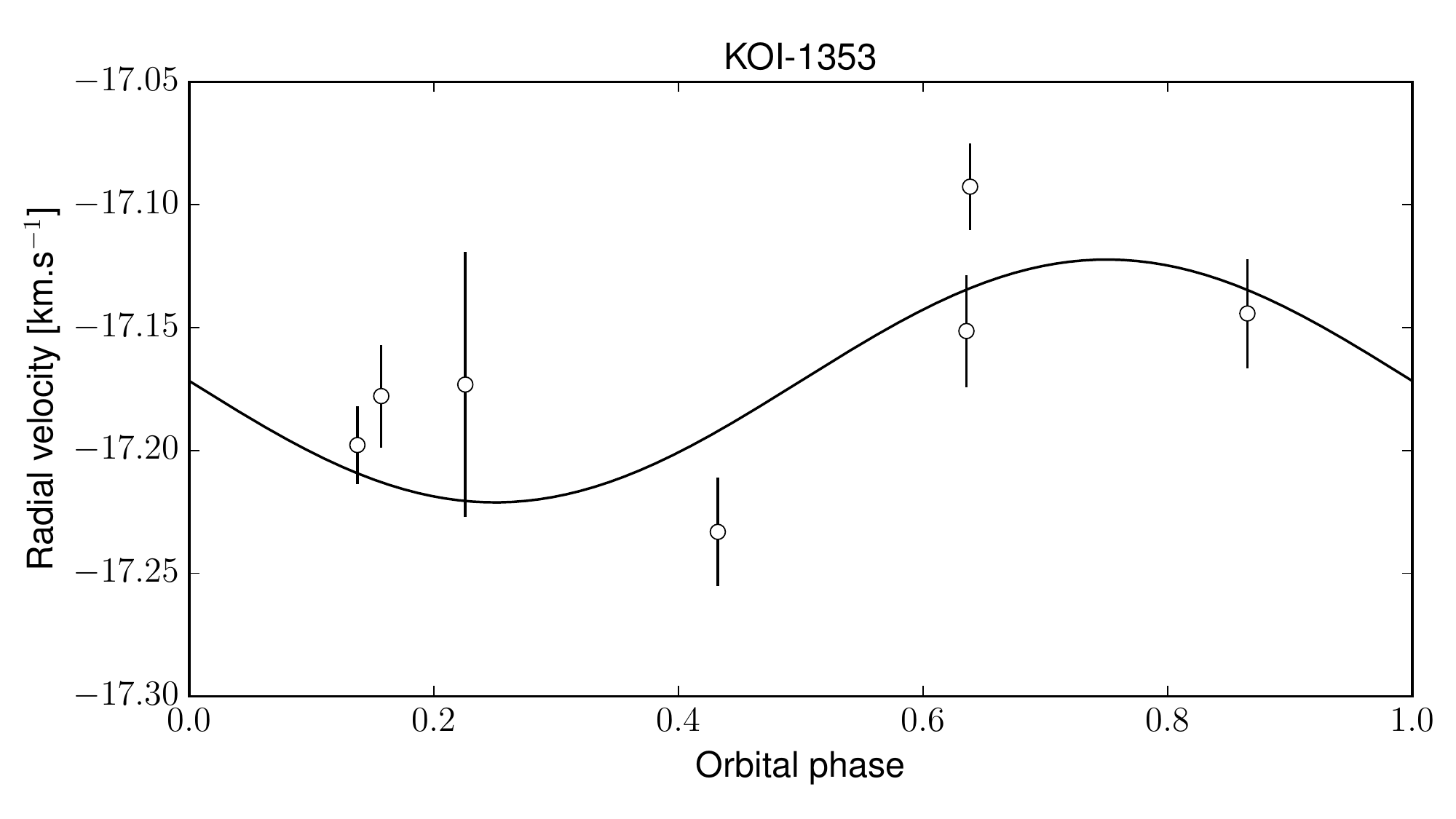}
\caption{Phase-folded RV curve of KOI-1353.03 together with the best circular model found.}
\label{fig1353}
\end{center}
\end{figure}

We observed this system with SOPHIE HE seven times. The RV, bisector and FWHM present $rms$ of 45 \ms, 42 \ms, and 117 \ms (respectively). We fitted these RV with a model of three circular orbits at the ephemeris provided by \textit{Kepler} (see Fig. \ref{fig1353}). We find upper-limit at the 99\% on the RV amplitudes of K$_{inner} <$ 25  \ms\, and K$_{middle} <$ 56 \ms\, for the inner (P=34d) and middle (P=66d) planets, respectively. For the outer, EGP, we detect an amplitude of K$_{outer} =$ 51 $\pm$ 11 \ms. At the 99\% level, this amplitude is K$_{outer} \in [23, 78]$ \ms.

We derived the stellar atmospheric parameters that are reported in Table \ref{SpectroResults}.  We find a host mass of M = 1.35$^{_{+0.11}}_{^{-0.07}}$ \Msun, a radius of R = 1.46$^{_{+0.27}}_{^{-0.15}}$ \Rsun, and an age of 1.71 $\pm$ 0.95 Gyr. Combining the RV results and our spectroscopic parameters, we find an upper-limit at the 99\% limit on the mass of inner and middle planets to be M$_{inner} <$ 0.52 \Mjup\, and M$_{middle} <$ 1.41 \Mjup, respectively. For the outer, EGP, we have a mass constraint of M$_{outer} =$ 1.55 $\pm$ 0.34 \Mjup\, (M$_{outer} \in [0.68, 2.43]$ \Mjup at the 99\% level). Thus, the EGP is found significantly (at 3.3-$\sigma$) more massive by RV than by transit timing variation, as already found by \citet{2014ApJ...783L...6W} for other planetary systems. Note that the star shows a photometric variability of more than 1.5\%, which could have impacted both our RV and the transit times measurements \citep{2013A&A...556A..19O, 2013MNRAS.430.3032B}. For the other planets, the masses derived by spectroscopy are fully compatible with the ones derived by \citet{2014ApJ...795..167S}.

\subsection{KOI-1391.01}
\label{1391}

A giant-planet candidate was found on the target star KOI-1391 by \citet{2011ApJ...736...19B} with an orbital period of almost 8 days. It was not observed by \citet{2012AA...545A..76S} because of its bad vetting status. We secured two observations with SOPHIE HE and find a large RV variation. Assuming a circular orbit, no significant drift and the \textit{Kepler} ephemeris, we find an amplitude of K = 32.93 $\pm$ 0.04 \kms. Assuming a stellar host of M$_{1}$ = 1.06$^{_{+0.19}}_{^{-0.13}}$ \Msun\, \citep{2014ApJS..211....2H}, the companion has a mass of M$_{2}$ = 0.39 $\pm$ 0.04 \Msun. This candidate is therefore an EB and not a transiting planet.

\subsection{KOI-1411.01}
\label{1411}

A giant-planet candidate with a period of $\sim$305 days was found by \citet{2014ApJS..210...19B} on the target star KOI-1411. We observed this star five times with SOPHIE HE. They present no significant RV variation in phase with the \textit{Kepler} ephemeris. The $rms$ are 15 \ms, 39 \ms, and 74 \ms\, for the RV, bisector and FWHM (respectively), which are compatible with the typical uncertainty on this star. We fitted the RV with a Keplerian orbit at the transit ephemeris. We find an upper-limit at 99\% on the amplitude K$_{1411.01}<$ 67 \ms. 

We derived the stellar atmospheric parameters that are reported in Table \ref{SpectroResults}. We find a host mass of M = 1.14$^{_{+0.15}}_{^{-0.10}}$ \Msun, a radius of R = 1.35$^{_{+0.58}}_{^{-0.27}}$ \Rsun, and an age of 5.0 $\pm$ 2.6 Gyr. Combining the RV results and the spectroscopic parameters, we have an upper-limit at the 99\% level on the mass of this candidate to be M$_{1411.01} <$ 2.13 \Mjup. We can therefore reject that this candidate is an eclipsing brown dwarf or binary but we can not firmly conclude on its nature.

\subsection{KOI-1426.02 \& KOI-1426.03}
\label{1426}

Three different sets of transits were detected in the \textit{Kepler} data of the target star KOI-1426 by \citet{2011ApJ...736...19B}. They have orbital periods of 39d (KOI-1426.01), 75d (KOI-1426.02), and 150d (KOI-1426.03). Only the two last ones are compatible with an EGP according to our selection criteria. Based on the planet-likelihood multiplicity boost described in \citet{2014ApJ...784...44L}, \citet{2014ApJ...784...45R} validated the planetary nature of the two innermost planets. However, the outer planet presenting a grazing transit (impact parameter of b $\approx$ 1.0), they did not validated it.

We observed this system six times with SOPHIE HE and find no significant variation. The $rms$ are 13 \ms, 33 \ms, and 100 \ms\, for the RV, bisector and FWHM (respectively), which is compatible with the typical photon noise for this target. We fitted the derived RV with a three-circular orbit model at the transit ephemeris. We don't have enough data to constrain the eccentricity of all the planets. We find upper-limits at the 99\% level on the amplitude of all the planets to K$_{inner} <$ 37 \ms, K$_{middle} <$ 21 \ms, and K$_{outer} <$ 39 \ms. Assuming a host mass of M$_{1}$ = 1.04$^{_{+0.20}}_{^{-0.12}}$ \Msun\, \citep{2014ApJS..211....2H}, we have an upper-limit on the mass of the candidates to M$_{inner} <$ 0.69 \Mjup, M$_{middle} <$ 0.45 \Mjup, and M$_{outer} <$ 1.03 \Mjup, with a confidence interval of 99\%. We can therefore rule out that any of these planet candidates is an EB nor transiting brown dwarf or even a massive EGP.

\subsection{KOI-1431.01}
\label{1431}

A giant-planet candidate was detected on the target star KOI-1431 by \citet{2013ApJS..204...24B}. It has an orbital period of about 345 days, which locates it in the habitable zone of its host star according to \citet{2013ApJ...770...90G}. We observed it seven times with SOPHIE HE. They do not show significant variation. The $rms$ are 13\ms, 18\ms, and 74\ms\, for the RV, bisector and FWHM (respectively), which are compatible with the uncertainties. We fitted these data using a Keplerian orbit at the transit ephemeris. We find an upper-limit at the 99\% limit on the amplitude of K$_{1431.01} <$ 24 \ms.

We derived the stellar atmospheric parameters that are reported in Table \ref{SpectroResults}. We find a host mass of M = 1.00$^{_{+0.07}}_{^{-0.04}}$ \Msun, a radius of R = 1.16$^{_{+0.38}}_{^{-0.18}}$ \Rsun, and an age of 8.9$^{_{+2.5}}_{^{-4.2}}$  Gyr. By combining both analyses, we find an upper-limit at the 99\% level on the mass of KOI-1431.01 of M$_{1431.01} <$ 0.73 \Mjup. We can therefore exclude scenarios of false positive invoking a brown dwarf or a star eclipsing KOI-1431. We can even rule out a massive EGP. However, we can not firmly establish its nature.

\subsection{KOI-1452.01}
\label{1452}

A transiting giant-planet candidate was found on the target star KOI-1452 with an orbital period of 1.15 days by \citet{2011ApJ...736...19B}. However, it was reported with a poor vetting flag and thus, not observed by \citet{2012AA...545A..76S}. We observed it twice with SOPHIE HE which revealed a SB2 with fast-rotating primary (\vsini\, = 36.2 $\pm$ 0.2 \kms) and a very faint secondary. We call star A the brightest component and star B the faintest one. We were able to derive the RV of both stars. We fitted the RV of both stars with circular orbits at the \textit{Kepler} ephemeris. We find that K$_{A} \approx$ 48.0 \kms\, and K$_{B} \approx$ -91.2 \kms, which gives a mass ratio of $q \approx$ 52.6\%. If the host has a mass of M$_{1} \sim$1.46 \Msun\, \citep{2014ApJS..211....2H}, thus the secondary has a mass of M$_{2} \sim$ 0.77 \Msun. Note that we detected a secondary eclipse in the \textit{Kepler} data with a depth of about 500ppm. Therefore, this candidate is not a transiting planet but an EB. 

\subsection{KOI-1465.01}
\label{1465}

\citet{2011ApJ...736...19B} reported a giant-planet candidate around the star KOI-1465 with an orbital period of almost 10 days. It was not observed by \citet{2012AA...545A..76S} because of its bad vetting status. We secured five observations with SOPHIE HE and find a large RV variation in phase with the \textit{Kepler} ephemeris. Assuming a circular orbit, no significant drift and the \textit{Kepler} ephemeris, we find an amplitude of K = 19.32 $\pm$ 0.03 \kms. Assuming a stellar host of M$_{1}$ = 0.94$^{_{+0.09}}_{^{-0.10}}$ \Msun\, \citep{2014ApJS..211....2H}, the companion has a mass of M$_{2}$ = 0.22 $\pm$ 0.02 \Msun. This candidate is therefore an EB and not a transiting planet.

\subsection{KOI-1483.01}
\label{1483}

We secured two SOPHIE HE observations on KOI-1483 which hosts a giant-planet candidate at $\sim$186 days \citep{2014ApJS..210...19B}. The RV exhibit a large variation in phase with the \textit{Kepler} ephemeris. Assuming a circular orbit, no significant drift and the \textit{Kepler} ephemeris, we find that the amplitude is K = 4.40 $\pm$ 0.02 \kms. Assuming a primary star mass of M$_{1}$ = 0.93 $\pm$ 0.10 \Msun\, \citep{2014ApJS..211....2H}, it gives a companion mass of M$_{2}$ = 0.12 $\pm$ 0.01 \Msun. This candidate is therefore an eclipsing low-mass star and not a transiting planet.
\subsection{KOI-1546.01}
\label{1546}

The target star KOI-1546 has been found to host a transiting giant-planet candidate with a period of 0.9d by \citet{2011ApJ...736...19B}. It was not observed by \citet{2012AA...545A..76S} because of its bad vetting status. We secured two observations with SOPHIE HE that revealed a hint of variation with K $\approx$ 77 \ms\, if the orbit is circular. However, the analysis of the line profile of the star also reveal a bisector variation ($\sim$ 250\ms), \textit{anti-correlated} with the RV as well as a large FWHM variation ($\sim$ 900 \ms) correlated with the RV. This is an evidence that the stellar host is blended with an unseen blended star which has a narrower line-profile width \citep{2015MNRAS.451.2337S}. We concluded that this system is either a triple system or a background EB, but not a transiting planet. Note that a stellar companion located at 0.6\arcsec\, and about 1 magnitude fainter in the $i$-band has been detected by \citet{2014A&A...566A.103L} in lucky imaging. If this companion is the host of the transit event, the companion might still be compatible with an inflated hot jupiter. Further observations are needed to confirm this.

\subsection{KOI-1574.01}
\label{1574}

Four different sets of transits were found in the \textit{Kepler} light curve of the target star KOI-1574: one giant-planet candidate at 114 days \citep[KOI-1574.01 ;]{2011ApJ...736...19B}, one Saturn-size candidate at 574 days \citep[KOI-1574.02 ;]{2014ApJS..210...19B}, and two Earth-size candidates at 5.8d and 9d \citep[KOI-1574.03 and KOI-1574.04, respectively ;][]{2015ApJS..217...16R}. Only the EGP at 114 days is within our selection criteria. The two large and long-orbital periods objects have already been confirmed thanks to the transiting timing variation analysis performed by \citet{2014A&A...561A.103O}. However, they reported a different period for the outermost planet of 192 d, i.e. one third of the period found by \citet{2014ApJS..210...19B}. They reported planetary masses for the two giant objects of M$_{1574.01} =$ 1.02 $\pm$ 0.03 \Mjup\, and M$_{1574.02} =$ 6.5 $\pm$ 0.8 \Mearth.

We secured five observations of KOI-1574 with SOPHIE HE. The data have a $rms$ of 12\ms\, which is compatible with the uncertainties. The bisector and FWHM have $rms$ of 44 \ms\, and 38 \ms, respectively. We fitted four circular orbits to the data fixing the orbital ephemeris to the transit ones. For KOI-1574.02, we choose the period of 572 days. We detect a hint of variation for the EGP at 114 days, with an amplitude of K$_{1574.01} =$ 41 $\pm$ 20 \ms, and an upper-limit of K$_{1574.01} <$ 90 \ms\, within the 99\% confidence interval. For the other planets, we find upper-limits at the 99\% level on their amplitudes of K$_{1574.02} <$ 39 \ms, K$_{1574.03} <$ 66 \ms, and K$_{1574.04} <$ 67 \ms.

Assuming a stellar host mass of M$_{1}$ = 1.08 $\pm$ 0.06 \Msun\, \citep{2014A&A...561A.103O}, it gives a mass constraint for the EGP in our sample of M$_{1574.01} =$ 1.05 $\pm$ 0.47 \Mjup\, (M$_{1574.01} <$ 2.25 \Mjup\, at the 99\% level). This value is fully compatible with the one derived by the transit timing analysis of \citet{2014A&A...561A.103O}. For the other planet candidates, we find upper-limits on the mass at the 99\% level of M$_{1574.02} <$ 1.66 \Mjup, M$_{1574.03} <$ 0.61 \Mjup, and M$_{1574.04} <$ 0.68 \Mjup. Assuming that KOI-1574.02 orbits with a period of 192d, as stated by \citet{2014A&A...561A.103O}, does not change significantly our constraints on the mass of KOI-1574.01. For the other planets, the upper-limits at 99\% change to M$_{1574.02} <$ 1.91 \Mjup, M$_{1574.03} <$ 0.58 \Mjup, and M$_{1574.04} <$ 1.10 \Mjup.

We conclude that none of these transiting candidates is an eclipsing brown dwarf or low-mass star eclipsing the target star. We confirm they have a mass within the planetary range.

\subsection{KOI-1645.01}
\label{1645}

The giant-planet candidate KOI-1645.01 was found by \citet{2014ApJS..210...19B} with an orbital period of 41 days. We observed it twice with SOPHIE HE and find a clear SB2. We fitted the cross-correlation function with a two-Gaussian profile to derive the RV of both stars. We call star A the one with the deepest line profile and star B the one with the shallowest line profile. The variation of both stars is anti-correlated, which gives a mass ratio of 87.6\%. The star A exhibits a RV variation in \textit{anti-phase} with the \textit{Kepler} ephemeris, revealing that the secondary eclipse was detected in the light-curve. This candidate is not a transiting planet but a secondary-only EB.

\subsection{KOI-1783.01}
\label{1783}

A giant-planet candidate was found to transit the host star KOI-1783 with a period of 134d by \citet{2013ApJS..204...24B}. We observed it twice with SOPHIE HE and find no significant RV variation. The bisector and FWHM variation are also compatible with their photon noise. By fitting a circular orbit with no drift at the transit ephemeris, we find an upper-limit at the 99\% level on the RV amplitude of K $<$ 81.3 \ms. 

We derived the stellar atmospheric parameters that are reported in Table \ref{SpectroResults}. We find a host mass of M = 1.57$^{_{+0.23}}_{^{-0.11}}$ \Msun, a radius of R = 1.84$^{_{+0.77}}_{^{-0.29}}$ \Rsun, and an age of 1.2 $\pm$ 0.5 Gyr. Combining the two analyses, we derive an upper-limit on the mass of this candidate of M$_{1783.01} <$ 2.83 \Mjup, within the 99\% confidence interval. We can therefore exclude a star or brown dwarf eclipsing the target star, but we can not firmly establish the planetary nature of this candidate.

\subsection{KOI-1784.01}
\label{1784}

A giant-planet candidate was revealed by \citet{2013ApJS..204...24B} with a period of 5 days and a depth of less than 0.4\%. It was not included in the sample of \citet{2012AA...545A..76S}. However, it was revised in \citet{2015ApJS..217...16R} with a depth of more than 0.4\%, which included it in our giant sample. We observed it twice with SOPHIE HE and find a clear SB2. We fitted the cross-correlation functions with a two-Gaussian function to derive the RV of both stars. We call star A the one with the deepest line profile and star B the one with the shallowest line profile. The RV of both stars are not correlated, which indicate that this system is not just an EB, but a more complex system, likely a triple. Assuming a circular orbit, the star B shows RV variation in phase with the \textit{Kepler} ephemeris with an amplitude of K$_{B} =$ 14.13 $\pm$ 0.46 \kms. Assuming a solar-mass for this star B, the companion would have a mass of about 0.12 \Msun. However, if the mass of star B is much lower, the companion could be in the brown dwarf regime. Therefore, this candidate is not a transiting planet but likely a triple system.

\subsection{KOI-1788.01}
\label{1788}

A giant-planet candidate transiting the target star KOI-1788 every 71 days (KOI-1788.01) has been reported by \citet{2013ApJS..204...24B}. The planet hunters community found another planet-candidate transiting this star, with an orbital period of nearly one year \citep[KOI-1788.02 ;][]{2013ApJ...776...10W}. We observed this target six times with SOPHIE HE. The derived RV have a $rms$ of 13 \ms. The bisector and FWHM present some variation, with $rms$ of 108 \ms\, and 76 \ms\, (respectively). These line-profile variation might be caused by the large activity of the star, which also imprint a photometric variability on the \textit{Kepler} light curve at the level of $\sim$ 4\%. We fitted these data with two circular orbits at the \textit{Kepler} ephemeris. We find upper-limits on the amplitude of both planet candidates, at the 99\% level, of K$_{1788.01} <$ 22.6 \ms\, and K$_{1788.02} <$ 77.2 \ms.

We derived the stellar atmospheric parameters that are reported in Table \ref{SpectroResults}. We find a host mass of M = 0.84$^{_{+0.32}}_{^{-0.09}}$ \Msun, a radius of R = 1.1$^{_{+1.6}}_{^{-0.3}}$ \Rsun, and an age of 12.5 $\pm$ 8.4 Gyr. Combining the two analyses, we find upper-limits on the mass of both candidates of M$_{1788.01} <$ 0.48 \Mjup\, and M$_{1788.02} <$ 3.0 \Mjup, within 99\% interval confidence. We can therefore exclude brown dwarfs or stars eclipsing the target star, but we can not firmly establish the nature of these candidates.

\subsection{KOI-2679.01}
\label{2679}

The giant-planet candidate KOI-2679.01 was revealed in \citet{2014ApJS..210...19B} with an orbital period of 111 days. We secured two observations with SOPHIE HE that revealed a unique and wide line profile, with \vsini\, = 29.8 $\pm$ 0.2 \kms. The RV do not show significant variation, nor the bisector and FWHM. We fitted a circular orbit at the \textit{Kepler} ephemeris and find that K$_{2679.01} <$ 1.41 \kms, at the 99\% level. Assuming a host mass of M$_{1}$ = 1.18$^{_{+0.27}}_{^{-0.17}}$ \Msun, the candidate mass has an upper-limit of M$_{2679.01} <$ 40.3 \Mjup, at the 99\% level. We can therefore exclude that KOI-2679.01 is an EB, but we can not firmly establish its nature.

\subsection{KOI-3411.01}
\label{3411}

A giant-planet candidate was found to transit the target star KOI-3411 with an orbital period of 27 days \citep{2015ApJS..217...16R}. We secured two observations with SOPHIE HE which revealed a clear SB2. We fitted the cross-correlation function with a two-Gaussian function. We call star A the line with the deepest profile and star B the one with the shallowest profile. The velocities of both stars are anti-correlated with a slope of 76.8\%, which corresponds to the mass ratio $q = M_{B} / M_{A}$. The variation of the star A, the most massive one in the system, is in phase with the \textit{Kepler} ephemeris. This means that the transit epoch observed by \textit{Kepler} corresponds to the primary eclipse of this binary. This candidate is not a transiting planet but a primary-only EB.

\subsection{KOI-3663.01}
\label{3663}

A giant-planet candidate was revealed by the planet hunters community in \citet{2013ApJ...776...10W} with an orbital period of 283 days. This candidate is located in the habitable zone. Based on statistical considerations, the same authors validated this candidate as a planet. Using Keck RV, they also excluded a 80 \Mjup companion at the 95.7\% probability. We secured four spectra of this host star with SOPHIE HE. The RV, bisector and FHWM have an $rms$ of 20 \ms, 37 \ms, and 141 \ms\, (respectively). We find a hint of correlation between the observed RV and the line-profile bisector. The Spearman correlation coefficient is 0.80 $\pm$ 0.28. More observations are needed to confirm this correlation and the planetary nature of this candidate.

\begin{figure}[h]
\begin{center}
\includegraphics[width=\figw]{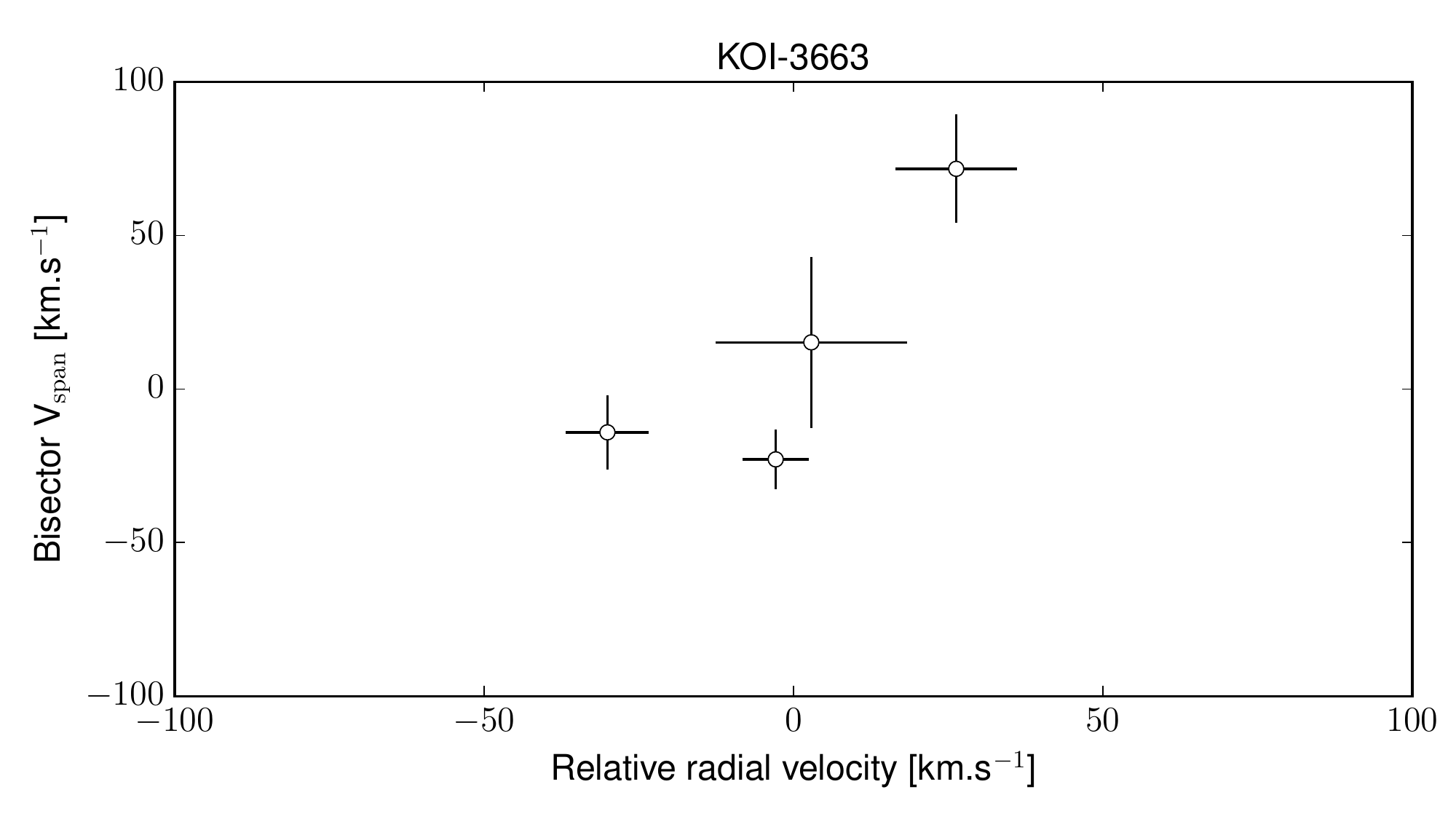}
\caption{Bisector V$_{\rm span}$ as a function of the RV for the target star KOI-3663.}
\label{plot3663}
\end{center}
\end{figure}

In their blend exclusion, \citet{2013ApJ...776...10W} ruled out the possibility that this candidate is a triple system based on the fact that a deep secondary eclipse would have been detected in the \textit{Kepler} light-curve. However, with an orbital period of 283 days, even a small eccentricity and a non-perfect orbital alignment with the line-of-sight might show either the primary or secondary eclipse of a binary. So, the absence of secondary eclipse does not firmly ruled out the scenario of a triple system. Our observed bisector correlation would indicate, if confirmed, that this candidate is blended with a binary star, thus likely in a triple system. Without the confirmation it is a triple system, we consider this candidate as a planet, as reported in \citet{2013ApJ...776...10W}.

We derived the stellar atmospheric parameters that are reported in Table \ref{SpectroResults}. We find a host mass of M = 1.05$^{_{+0.23}}_{^{-0.10}}$ \Msun, a radius of R = 1.37$^{_{+0.70}}_{^{-0.32}}$ \Rsun, and an age of 6.8 $\pm$ 4.5 Gyr.

\subsection{KOI-3678.01}
\label{3678}

The giant-planet candidate transiting the target star KOI-3678 with a period of 161 days was announced by \citet{2015ApJS..217...16R}. We observed it four times with SOPHIE HE. The observed $rms$ of the RV, bisector and FWHM are 23 \ms, 16 \ms, and 36 \ms, respectively. There is a hint of RV variation. Assuming a circular orbit, we find K = -31 $\pm$ 11 \ms, hence the RV are in \textit{anti-phase} with the \textit{Kepler} ephemeris. The target star exhibits a photometric variability at the level of about 1\% peak-to-valley. Considering a \vsini\, of 4.5 $\pm$ 1.2 \kms, we expect an activity-induced RV signal at the level of 45 \ms, which is compatible with our observed variation. No clear bisector nor FWHM correlation is found with the RV. For this reason, we did not attempt to model the RV. Assuming that any circular orbit with an amplitude of three times the $rms$ would have been clearly significantly detected, we can put some constraints to K$_{3678.01} <$ 70 \ms. Assuming a host mass of M$_{1}$ = 0.818$^{_{+0.15}}_{^{-0.06}}$ \Msun, we can constrain the mass of the transiting candidates to M$_{3678.01} <$ 1.43 \Mjup, within 99\% of probability. We can therefore exclude that a massive planet, brown dwarf or star is transiting / eclipsing the target star, but we can not firmly conclude on the nature of this candidate.

\subsection{KOI-3680.01}
\label{3680}

The giant-planet candidate KOI-3680.01 was revealed in \citet{2015ApJS..217...16R} with an orbital period of 141 days. We observed it with SOPHIE HE. The RV show a significant variation that correspond to a M $\sim$ 2.2 \Mjup\, planetary companion in an eccentric orbit. This planet will be further characterised in a forthcoming paper (H\'ebrard et al., in prep.).

\subsection{KOI-3681.01}
\label{3681}

The giant-planet candidate KOI-3681.01 was revealed in \citet{2015ApJS..217...16R} with an orbital period of 2018 days. We observed it with both SOPHIE HE and HR. The RV in both instrumental configuration show a significant variation that correspond to a M $\sim$ 4.4 \Mjup\, planetary companion in an eccentric orbit. This planet will be further characterised in a forthcoming paper.

\subsection{KOI-3683.01}
\label{3683}

A giant-planet candidate was found to transit the target star KOI-3683 with an orbital period of 214 days \citep{2015ApJS..217...16R}. We observed it twice with SOPHIE HE and find no significant variation at the level of the photon noise for the RV, bisector and FWHM. Assuming a circular orbit at the \textit{Kepler} ephemeris, we fitted these two RV and find that K$_{3683.01} <$ 53 \ms\, at the 99\% level. 

We derived the stellar atmospheric parameters that are reported in Table \ref{SpectroResults}. We derived a host mass of M = 1.49$^{_{+0.19}}_{^{-0.13}}$ \Msun, a radius of R = 1.70$^{_{+0.57}}_{^{-0.26}}$ \Rsun, and an age of 1.4 $\pm$ 0.7 Gyr. Combining the two analyses, we find that the candidate has a mass of M$_{3683.01} <$ 2.08 \Mjup, at the 99\% level. We can therefore exclude a star or brown dwarf eclipsing the target star, but we can not firmly establish its planetary nature based on these data.

\subsection{KOI-3685.01}
\label{3685}

\begin{figure}[h]
\begin{center}
\includegraphics[width=\figw]{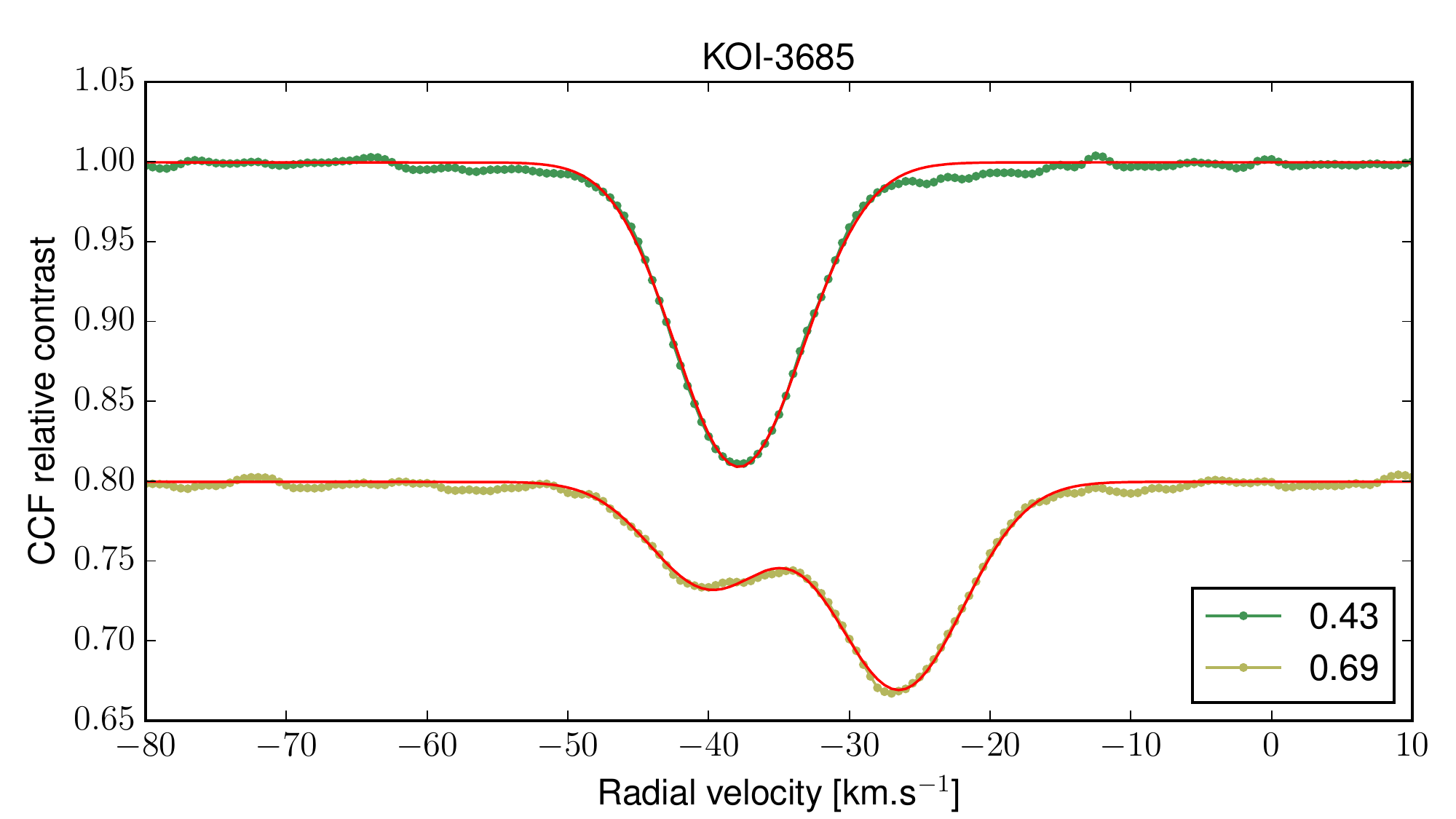}
\caption{Cross-correlated function of the target star KOI-3685 revealing two stellar components. The legend indicates the orbital phase of the transiting candidate. An arbitrary offset in flux has been set between the two observations. The red lines are the two-Gaussian fits to the CCFs.}
\label{fig3685}
\end{center}
\end{figure}

\citet{2015ApJS..217...16R} reported two transiting candidates around the target star KOI-3685: one giant-planet candidate with an orbital period of 209 days (KOI-3685.01) and a 1.6-\Rearth\, super-Earth at 7 days (KOI-3685.02). This system was not validated by \citet{2014ApJ...784...45R} because the giant-planet candidate has an impact parameter compatible with 1.0 with 1-$\sigma$ (i.e. a grazing transit). We secured two observations of this target star with SOPHIE HE. The cross-correlation functions reveal a clear SB2 that we fitted with a two-Gaussian function (see Fig. \ref{fig3685}). We call star A the star with the deepest line profile and star B the one with the shallowest line profile. Thus, star A is supposed to be the brightest component of the system. Note that for the first observation the two stars are blended. The RV uncertainties we estimated are probably under-estimated. 

If the RV variation of star B is real, both stars shows RV variation in \textit{anti-phase}. From the slope of the RV correlation between the two stars, it is possible to determine their mass ratio. So, if the observed variation are caused by the orbit of star B around star A, we find a mass ratio of $q \approx 13.7$. This means that star B would be about 14 times more massive than star A. Assuming a mass -- luminosity relation such as $\Delta L = q^{3.5}$, star B should be about 10$^{4}$ times more luminous than star A. This is hardly compatible with the fact that star A has a deeper line profile than star B. In such case, we would not be able to detect the spectral lines from star A. Thus, the RV variation of star A should be explained by another object in the system. Given their relatively close RV, it is likely that star A and B are bound, but are orbiting in a long period.

We fitted a circular orbit to the RV of star A at the ephemeris of KOI-3685.01 and KOI-3685.02. In both cases, the variation is seen in phase with the ephemeris, but the derived amplitudes are $K_{3685.01} \sim$ 7.7 \kms\, and $K_{3685.02} \sim$ 152 \kms\, for KOI-3685.01 and KOI-3685.02 (respectively). A RV of 152 \kms\, at 7 days would required a stellar host of more than 10 \Msun\, in order to keep the mass ratio $q \leq 1$. Thus, we concluded that the variation seen on star A is caused by the orbit of KOI-3685.01. Since the stars are blended, it is difficult to determine the mass of the host, KOI-3685 A. Assuming this star is a solar-like star, KOI-3685.01 would have a mass of about  M$_{3685.01}$ = 0.25 \Msun\, and is clearly in the stellar domain. Therefore, the system KOI-3685 is composed by at least three stars: A, B and .01 that we call now star C. In this system, it is likely that the star B orbits the binary AC. This system is already listed in the \textit{Kepler} EB catalog. 

In this complex system, the super-Earth candidate KOI-3685.02 could either transit the star A, B, or even C. In any case, its transit depth is severely diluted. The dilution is much greater if it transits star C. Using the asterodensity profiling \citep{2014MNRAS.440.2164K} and assuming that this candidate has a circular orbit, we find that the density of the transit host is $\rho =$ 0.18 $\pm$ 0.05 $\rho_{\odot}$. However, according to \citet{2014MNRAS.440.2164K}, the blend-effect which is expected to be strong in this case, makes that this value of $\rho$ is under-estimated. Therefore, using the asterodensity profiling of this candidate, we can only poorly constrain the density of the host to be $\rho >$ 0.18 $\rho_{\odot}$. This does not allow us to determine on which star this candidate transits. However, given the large dilution, this candidate is expected to be substantially larger than 1.6 \Rearth.

\subsection{KOI-3689.01}
\label{3689}

A giant-planet candidate was announced by \citet{2015ApJS..217...16R} to transit the target star KOI-3689 every 5.24 days. This candidate was not included in the sample of \citet{2012AA...545A..76S} because this target was not observed at that time. KOI-3689 was only observed by \textit{Kepler} during the quarter 10\footnote{The quarter 10 was observed between June and September 2011. The data were released in October 2012.}. We observed it twice with SOPHIE HE and find no variation in the RV, bisector and FWHM above the photon-noise floor. We fitted the RV with a circular orbit and find that K$_{3689.01} <$ 56 \ms\, within a probability of 99\%. 

We derived the stellar atmospheric parameters that are reported in Table \ref{SpectroResults}. We derived a host mass of M = 1.27 $\pm$ 0.20 \Msun, a radius of R = 1.41$^{_{+0.59}}_{^{-0.24}}$ \Rsun, and an age of 2.5$^{_{+2.0}}_{^{-1.1}}$ Gyr. Combining the two results, we find that the candidate has an upper-limit on its mass of M$_{3689.01} <$ 0.61 \Mjup, at the 99\% level. We can therefore exclude a massive planet, brown dwarf or a star transiting / eclipsing the target star. However, we can not rule a diluted transit on a background or companion star.

\subsection{KOI-3720.01}
\label{3720}

We observed twice with SOPHIE HE the target star KOI-3720. This star host a giant-planet candidate with an orbital period of 213 days \citep{2015ApJS..217...16R}. The cross-correlation function revealed a wide line profile with \vsini = 24.7 $\pm$ 0.1 \kms. The derived RV show a large variation that we fitted with a circular orbit. We find K$_{3720.01} =$ 4.80 $\pm$ 0.06 \kms. Assuming a host mass of  M$_{1}$ = 1.34$^{_{+0.29}}_{^{-0.25}}$ \Msun\, \citep{2014ApJS..211....2H}, we find that the companion has a mass of  M$_{2}$ = 0.18 $\pm$ 0.02 \Msun. This candidate is therefore not a transiting planet but clearly an EB. 

\subsection{KOI-3721.01}
\label{3721}

A giant-planet candidate was found to transit the target star KOI-3721 with an orbital period of 6.41 days \citep{2014ApJ...784...45R}. This candidate was not observed in the sample of \citet{2012AA...545A..76S} because it was only observed by \textit{Kepler} during the quarter 10, which was not available at that time. We observed it twice with SOPHIE HE. The observed cross-correlated functions revealed a clear SB2. We derived the RV of both stars by fitting a two-Gaussian function. We call star A the one with the deepest line profile and star B the one with the shallowest line profile. The star A do not show RV variation at the level of the photon noise. However, the star B shows a large RV variation. Fitting this variation with a circular orbit at the \textit{Kepler} ephemeris, we find that K$_{3721B} =$ -44.99 $\pm$ 0.87 \kms\, and a systemic RV of $\gamma_{3721B} =$ -0.22 $\pm$ 0.82 \kms. The RV amplitude being negative, the variation of the star B is in \textit{anti-phase} with the \textit{Kepler} ephemeris. Then, the RV of both stars are not correlated. So, if both the star A and B are bound, they are probably orbiting with a long orbital period and not at 6.41 days. 

Another object should thus be invoked to explained the observed transit and the RV of the star B. Two reasons might explain the variations in anti-phase of the star B: (1) the ephemeris are wrong and the true period is twice the observed one, (2) this system is a secondary-only EB. We tried fixing the orbital period at two times 6.41 days keeping the epoch of primary transit to the one reported in \citep{2014ApJ...784...45R} or shifted by half a double-period (hence, shifted by 6.41 days). In both cases, the RV observations are covering a very small fraction of the expected circular-orbit variation (orbital phases of 0.13 and 0.37 or phases of 0.63 and 0.87). So, fitting the RV of the star B at these new ephemeris gives huge values of K$_{3721B}$ to about $\pm$ 18000 \kms, which is unphysical for stars. We assumed that this scenario of a binary with twice the observed period is not reasonable. 

The only other scenario to explain the anti-correlated variation of the star B is to have a secondary-only EB, likely bound with another, brighter star. Even if having a secondary-only EB in a triple system is quite unlikely to occur, \citet{2013A&A...557A.139S} predicted that about ten \textit{Kepler} candidates could be of that type. It would required the binary KOI-3721~B to be eccentric with a period of 6.4 days, which is quite surprising with such orbital period. This is not impossible if the system is under a Lidov -- Kozai resonance \citep{1962P&SS....9..719L, 1962AJ.....67..591K}. Note that \citet{2008AJ....135..850D} already reported eccentric EBs among the transatlantic exoplanet survey data with orbital period as small as 2 days. 

It is clear that this candidate is not a transiting planet, but it is most likely a triple system. In this system, \citet{2015AJ....149...18K} detected three stellar components which confirm the triple system scenario.

\subsection{KOI-3780.01}
\label{3780}

The target star KOI-3780 was found to host a giant-planet candidate with a period of 28 days \citep{2015ApJS..217...16R}. We observed this target twice with SOPHIE HE and find a large RV variation, in phase with the \textit{Kepler} ephemeris. We fitted this variation with a circular orbit at the transit ephemeris and find an amplitude of K = 33.29 $\pm$ 0.23 \kms. Assuming a host mass of  M$_{1}$ = 1.20$^{_{+0.27}}_{^{-0.19}}$ \Msun, we find that the companion has a mass of  M$_{2}$ = 0.75 $\pm$ 0.08 \Msun. Therefore, this candidate is not a transiting planet but clearly an EB.

\subsection{KOI-3782.01}
\label{3782}

A giant-planet candidate with a period of 187 days was reported by \citet{2015ApJS..217...16R}. We observed it three times with SOPHIE HE which reveal a clear SB2. We fitted the cross-correlation functions with a two-Gaussian profile to derive the RV of both stars. We call star A the one with the deepest line profile and star B the one with the shallowest line profile. The RV of both stars are anti-correlated which gives a mass ratio of 81.4\%. The variation of star A is anti-correlated with the transit ephemeris, indicating that \textit{Kepler} detected the secondary eclipse of this binary system. This candidate is not a transiting planet but a secondary-only EB.

\subsection{KOI-3783.01}
\label{3783}

A giant-planet candidate was found to transit the target star KOI-3782 with an orbital period of 197 days \citep{2015ApJS..217...16R}. We observed it four times with SOPHIE HE. The cross-correlation functions revealed a wide line profile with a \vsini\, = 71.7 $\pm$ 0.1 \kms. We fitted them with a rotation profile as described in \citet{2012A&A...544L..12S}. For that fast-rotating star, we failed in fitting the V$_{\rm span}$ asymmetry diagnosis, because the profile of the star is clearly not Gaussian. We report instead the BIS diagnosis \citep[see][for a review on line-profile asymmetry diagnoses]{2015MNRAS.451.2337S}. Our RV have a $rms$ of 115 \ms\, with a typical photon noise of about 600 \ms. We fitted these RV with a Keplerian orbit at the transit ephemeris. We find that the RV amplitude is K$_{3783.01} <$ 4.49 \kms\, with a probability of 99\%. Assuming a host mass of  M$_{1}$ = 1.69$^{_{+0.35}}_{^{-0.22}}$ \Msun\, \citep{2014ApJS..211....2H}, we find a companion mass upper-limit of M$_{3783.01} <$ 0.13 \Msun\, at the 99\% level. The bisector does not show variation above the photon noise floor, relatively high for this target. Because of the fast rotation of the host star, our spectroscopic measurements are not able to rule out a low-mass star eclipsing the target star. We can not resolved the nature of this giant-transiting candidate. 

Recently, \citet{2015arXiv150601668B} characterised the host star KOI-3787 as a $\gamma$-Doradus or $\delta$-Scuti star. They also found that the transit event duration was too short for being host by this hot star and concluded it could be a false positive. According to their results, the host of the transit event is a star about eight times fainter than the target star. We don't have a S/N high enough to detect this contaminating star, which is most likely blended within the broad line profile of the target star. For these reasons, we consider this candidate as a chance-aligned EB and not as a transiting planet.

\subsection{KOI-3784.01}
\label{3784}

A giant-planet candidate was found to transit the target star KOI-3784 every 23.87 days \citep{2015ApJS..217...16R}. This candidate was not included in the sample of \citet{2012AA...545A..76S} because it was observed by \textit{Kepler} only during the quarter 10, which was not available at that time. We observed it twice with SOPHIE HE which revealed a clear SB2. We fitted the cross-correlated function with a two-Gaussian function and derived the RV of both stars. We call star A the one with the deepest line profile and star B the one with the shallowest line profile. The RV of both stars are anti-correlated. Using the slope of this anti-correlation, we measured a mass ratio of $q = M_{B} / M_{A} \approx$ 94.4\%. This confirms that the star A is the most massive component of this binary system. The RV variation of star A is in \textit{anti-phase} with the \textit{Kepler} ephemeris, indicating that the transit epoch corresponds to the secondary eclipse of this binary. This candidate is clearly not a transiting planet but a nearly equal-mass secondary-only EB.

\subsection{KOI-3787.01}
\label{3787}

We observed twice with SOPHIE HE the giant-planet candidate host KOI-3787, revealed by \citet{2015ApJS..217...16R} with an orbital period of 142 days. The derived RV show a large variation in phase with the \textit{Kepler} ephemeris. We fitted these data with a circular orbit at the transit period and epoch and find an amplitude of K = 5.00 $\pm$ 0.01 \kms. Assuming a host mass of  M$_{1}$ = 1.0$^{_{+0.18}}_{^{-0.11}}$ \Msun, we find that the companion has a mass of  M$_{2}$ = 0.14 $\pm$ 0.01 \Msun. Therefore, this candidate is not a transiting planet but a low-mass EB.

\subsection{KOI-3811.01}
\label{3811}

The target star KOI-3811 was found to host a giant-planet candidate that transits every 290 days \citep{2015ApJS..217...16R}. We observed it twice with SOPHIE HE. The cross-correlation function revealed a clear SB2 that we fitted with a two-Gaussian function. The star A (B) is the one with the deepest (shallowest) line profile. The two stars show anti-correlated RV variations, with a slope of q = 77.0\%. This corresponds to their mass ratio. The variation of the star A is in \textit{anti-phase} with \textit{Kepler} ephemeris, which means that the transit detected by \textit{Kepler} is the secondary eclipse of this eccentric system. We conclude that this candidate is not a transiting planet but a secondary-only EB.

\subsection{KOI-5034.01}
\label{5034}

A giant-planet candidate was found on the target star KOI-5034 with an orbital period of 283 days \citep{2015arXiv150202038M}. We observed it twice with SOPHIE HE. The cross-correlation function revealed a clear SB2 that we fitted with a two-Gaussian function. The RV of both stars are clearly anti-correlated with a slope close to 1.0, revealing a nearly-equal mass binary. However, given the different orbital phase observations (0.34 and 0.26), the two velocities show a relatively small variation for an equal-mass binary. This requires that the system is eccentric. Without more spectroscopic observations, we can not constrain the stellar masses and eccentricity of this system. This candidate is clearly not a transiting planet but an EB.

\subsection{KOI-5086.01}
\label{5086}

The candidate KOI-5086.01 was announced by \citet{2015arXiv150202038M} with an orbital period of 22 days. We secured three observations with SOPHIE HE that revealed a clear RV variation in \textit{anti-phase} with the \textit{Kepler} ephemeris. We concluded this system is a secondary-only EB. The light curve exhibits clear eclipse depth variation revealing that this EB is likely in a higher-order multiple stellar system. Since the RV variation is observed on the brightest star, we consider it as an EB.

\subsection{KOI-5132.01}
\label{5132}

A giant-planet candidate was announced by \citet{2015arXiv150202038M} on the target star KOI-5132 with an orbital period of 44 days. We observed it four times with SOPHIE HE. The RV clearly show a large variation in \textit{anti-phase} with the \textit{Kepler} ephemeris indicating a secondary-only EB. We fitted these data with a Keplerian orbit fixing the epoch of the secondary eclipse as the epoch of transit detected by \textit{Kepler} and fixing the orbital period to the transit one. We find a best-fit model with an eccentricity of $e =$ 0.18 and an amplitude of K = 25.12 \kms. Assuming a host mass of  M$_{1}$ = 1.98 \Msun\, \citep{2014ApJS..211....2H}, the companion has a mass of  M$_{2}$ = 0.81 \Msun, clearly in the stellar domain. Therefore, this candidate is not a transiting planet but a secondary-only EB.

\subsection{KOI-5384.01}
\label{5384}

The giant-planet candidate KOI-5384.01 was announced with an orbital period of almost 8 days \citep{2015arXiv150202038M}. We observed it twice with SOPHIE HE and find a wide line profile in the cross-correlation function. We fitted it with a rotation profile as described in \citet{2012A&A...544L..12S}. We found that \vsini\, = 43.6 $\pm$ 0.3 \kms. The BIS shows a significant variation of about 9 \kms. This is a hint for a blended stellar component, but we don't have enough data to confirm this. We fitted the derived RV with a circular orbit at the \textit{Kepler} ephemeris. We find an amplitude of K$_{5384.01} <$ 2.43 \kms\, within a probability of 99\%. Assuming a host mass of  M$_{1}$ = 1.05$^{_{+0.27}}_{^{-0.12}}$ \Msun\, \citep{2014ApJS..211....2H}, the candidate has a mass constraint of M$_{5384.01} <$ 27.0 \Mjup, at the 99\% level. We can therefore exclude a massive brown dwarf or a star eclipsing the target star, but we can not firmly establish the planetary nature of this candidate. Note that this candidate is listed in the \textit{Kepler} EB catalog (Kirk et al, in prep.) with a double period. It is also listed as a detached EB in \citet{2015AJ....149...68B}, but with the nominal period found by \citep{2015arXiv150202038M}. We conclude that this system is likely a CEB.

\subsection{KOI-5436.01}
\label{5436}

A giant-planet candidate with a period of 28 days was reported in \citet{2015arXiv150202038M}. We observed it twice with SOPHIE HE which reveal a clear SB2. We fitted the cross-correlation functions with a two-Gaussian profile to derive the RV of both stars. We call star A the one with the deepest line profile and star B the one with the shallowest line profile. The RV of both stars are anti-correlated, which indicates a mass ratio of 67.5\%. The variation of star A is in \textit{anti-phase} with the transit ephemeris, revealing that \textit{Kepler} observed the secondary eclipse of this binary system. This candidate is not a transiting planet but a secondary-only EB.

\subsection{KOI-5529.01}
\label{5529}

\begin{figure}[h]
\begin{center}
\includegraphics[width=\figw]{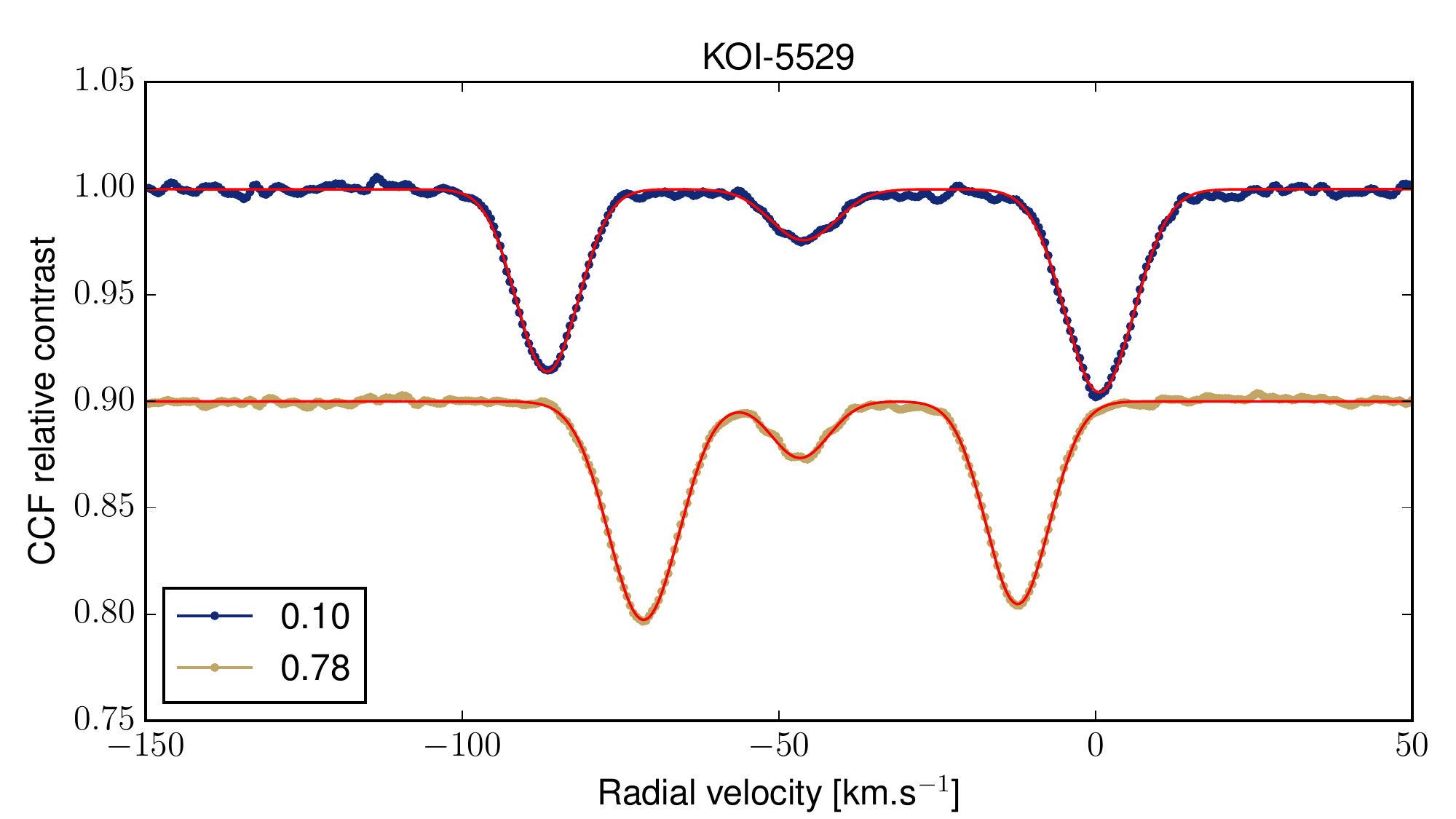}
\caption{Cross-correlated function of the target star KOI-5529 revealing three stellar components. The legend indicates the orbital phase of the transiting candidate. An arbitrary offset in flux has been set between the two observations. The red lines are the three-Gaussian fits to the CCFs.}
\label{fig5529}
\end{center}
\end{figure}

The \textit{Kepler} space telescope detected a giant-planet candidate transiting the star KOI-5529 every 70 days \citep{2015arXiv150202038M}. We observed it twice with SOPHIE HE. The cross-correlation function revealed three different sets of stellar lines that we fitted with a three-Gaussian profile (see Fig. \ref{fig5529}). We identify the stars A, B, and C as the deepest to the shallowest line profile. The stars A and B show large and anti-correlated RV variation, while star C shows a small and marginally significant variation. We interpreted these data as the signature of a triple system, with A and B orbiting with a period of 70 days, and C orbiting with a much longer orbital period. 

The slope of the correlation between the variations of A and B is of 96.9\%, with A the most massive star among the two. The variation of the star A is in \textit{anti-phase} with the \textit{Kepler} ephemeris, which reveals that \textit{Kepler} observed the secondary eclipse of this system. It is clear that this candidate is not a transiting planet but likely a triple system with a secondary-only EB.

\subsection{KOI-5708.01}
\label{5708}

A giant-planet candidate was found on the target star KOI-5708 by \citet{2015arXiv150202038M} with an orbital period of almost 8 days. This candidate was not known at the time of the observations of \citet{2012AA...545A..76S}, and thus, was not included in their sample. We observed it twice with SOPHIE HE. We find a large RV variation in phase with the \textit{Kepler} ephemeris. Assuming this variation is caused by the transiting object, we fitted a circular orbit and find an amplitude of K = 17.58 $\pm$ 0.02 \kms. Assuming a host mass of M$_{1}$ = 1.14$^{_{+0.26}}_{^{-0.22}}$ \Msun\, \citep{2014ApJS..211....2H}, it gives a companion mass of M$_{2}$ = 0.20 $\pm$ 0.03 \Msun. This candidate is clearly not a transiting planet but a EB.

\subsection{KOI-5745.01}
\label{5745}

The target star KOI-5745 hosts a giant-planet candidate with an orbital period of 11 days \citep{2015arXiv150202038M}. It was not included in the giant-candidate sample from \citet{2012AA...545A..76S} because it was discovered after their observations. We observed it twice with SOPHIE HE which revealed a clear SB2. We fitted the cross-correlation functions with a two-Gaussian profile. Note that the two line profiles are blended at the second epoch, so our photon noise uncertainties are likely underestimated. We call star A and B the brightest and faintest components of the system, identified based on their line-profile depth (the deepest is the brightest). The RV variations of both stars are anti-correlated, which allow us to measure their mass ratio $q = M_{B} / M_{A} = $ 91.1\%. The RV of the star A are in \textit{anti-phase} with the transit ephemeris which means that \textit{Kepler} detected the secondary eclipse of this system. It is not possible to fit this system using circular orbits, but we don't have enough point to measure the eccentricity and masses of the stars in this system. Therefore, this system is not a transiting planet but a secondary-only EB.

\subsection{KOI-5976.01}
\label{5976}

A giant-planet candidate was detected on the target star KOI-5976 with an orbital period of 2.7 days \citep{2015arXiv150202038M}. It was not included in the sample from \citet{2012AA...545A..76S} because it was not reported as a candidate in \citet{2013ApJS..204...24B}. We observed it eight times with SOPHIE HE. The RV, bisector and FHWM present $rms$ of 40 \ms, 48 \ms, and 99 \ms\, (respectively). We detected a hint of RV variation with K = 24$^{_{+23}}_{^{-15}}$ \ms\, assuming a circular orbit, with a large jitter of 41$^{_{+20}}_{^{-12}}$ \ms. We did not detect significant correlation between the RV and the bisector nor FWHM.

We derived the stellar atmospheric parameters that are reported in Table \ref{SpectroResults}. We find that these parameters correspond to a mass of M = 1.55$^{_{+0.83}}_{^{-0.55}}$ \Msun, a radius of R$_{1}$ = 7.8$^{_{+5.2}}_{^{-3.1}}$ \Rsun, and an age of 2.7$^{_{+9.1}}_{^{-1.9}}$ Gyr. The host is therefore clearly a giant star. Assuming that the transit occurs on the target star, with a depth of 1.3\%, the companion would have a radius of R$_{2}$ = 0.91$^{_{+0.61}}_{^{-0.36}}$ \Rsun. This is clearly not compatible with the expected radius of an EGP. Moreover, this large stellar radius corresponds to a circular orbit at 2.05$^{_{+1.6}}_{^{-0.94}}$ days. If the a companion is transiting this host star every 2.7 days, it is likely orbiting very close to the stellar surface, or even inside the star.

Since we did not detect a large RV variation on the target star, we conclude that this candidate is most likely a triple system with a giant primary star and not a transiting planet. 

\subsection{KOI-6066.01}
\label{6066}

A giant-planet candidate was announced by \citet{2015arXiv150202038M} to transit the target star KOI-6066 every 14 days. It was not reported by \citet{2013ApJS..204...24B} and thus not included in the sample of \citet{2012AA...545A..76S}. We observed it twice with SOPHIE HE and find a large RV variation in phase with the \textit{Kepler} ephemeris. We fitted them with a circular orbit at the transit ephemeris and find an amplitude of K = 23.00 $\pm$ 0.03 \kms. Assuming a host mass of M$_{1}$ = 1.23$^{_{+0.26}}_{^{+0.20}}$ \Msun\, \citep{2014ApJS..211....2H}, we find that the companion has a mass of M$_{2}$ = 0.56 $\pm$ 0.04 \Msun. Therefore, this candidate is not a transiting planet but an EB.

\subsection{KOI-6114.01}
\label{6114}

A giant-planet candidate transiting the target star KOI-6114 was first announced by \citet{2015arXiv150202038M} with an orbital period of 25 days. It was then flagged as a false-positive. Before this change of disposition in the archive, we secured two spectra with SOPHIE HE. The cross-correlation function revealed a shallow and narrow line with a large variation in phase with the \textit{Kepler} ephemeris and with K = 6.66 $\pm$ 0.14 \kms, assuming a circular orbit. We interpreted these data as a triple system with a very hot or fast rotating star which does not contribute to the cross-correlation function besides its continuum flux. This scenario is compatible with the effective temperature of the host found by \citet{2014ApJS..211....2H} of \teff\, = 9128$^{_{+273}}_{^{-402}}$ K. Therefore, we confirm this candidate is not a transiting planet but a triple system. Since it is now considered as a false-positive in the latest candidate release, we did not include it in our sample.

\subsection{KOI-6124.01}
\label{6124}

A giant-planet candidate was found on the target star KOI-6124 \citep{2015arXiv150202038M} with a period of 7 days. It was later on flagged as a false positive. In the mean time, we secured two spectra with SOPHIE HE that revealed a wide line profile with large RV variation. We estimated the \vsini\, of the host star to be \vsini\, = 51.3 $\pm$ 0.8 \kms. The RV show variation in \textit{anti-phase} with the transit ephemeris, indicating that \textit{Kepler} detected the secondary eclipse of this binary. We confirm this candidate is not a transiting planet but a secondary-only EB.

\subsection{KOI-6132.01}
\label{6132}

\citet{2015arXiv150202038M} reported two candidates transiting the target star KOI-6132: a giant-planet candidate at 33 days (KOI-6132.01) and a Neptune-size planet at 8 days (KOI-6132.02). Another set of transit was also detected with a period of 12 days (KOI-6132.03) but was rejected as false-positive by the same authors. We observed it six times with SOPHIE HE. We find no  significant RV variation. The $rms$ are 69 \ms, 64\ms, and 51\ms\, for the RV, bisector and FWHM. We fitted two circular orbits at the ephemeris of the two candidates and found that K$_{6132.01} < 109$ \ms\, and K$_{6132.02} < 120$ \ms\, at the 99\% level. Assuming a host mass of M$_{1}$ = 1.35$^{_{+0.25}}_{^{-0.27}}$ \Msun\, \citep{2014ApJS..211....2H}, it corresponds to companion mass of M$_{6132.01} < 2.25$ \Mjup\, and M$_{6132.02} < 1.50$ \Mjup, at the 99\% level. We can thus reject that these candidates are EB or brown dwarfs but we can not firmly establish their planetary natures.


\subsection{KOI-6175.01}
\label{6175}

A giant-planet candidate was found on the target star KOI-6175 with a period of 10 days \citep{2015arXiv150202038M}. We observed it twice with SOPHIE HE which revealed a large RV variation. By fitting this variation with a circular orbit at the \textit{Kepler} ephemeris, we find K = 20.13 $\pm$ 0.03 \kms. Assuming a host mass of M$_{1}$ = 1.38 $\pm$ 28 \Msun\, \citep{2014ApJS..211....2H}, the companion has a mass of M$_{2}$ = 0.29 $\pm$ 0.04 \Msun. Therefore, we conclude that this candidate is not a transiting planet but an EB.

\subsection{KOI-6235.01}
\label{6235}

A giant-planet candidate was found on the target star KOI-6235 with an orbital period of 2.05 days \citep{2015arXiv150202038M}. We observed it six times with SOPHIE HE. The $rms$ of the RV, bisector and FWHM are 67 \ms, 94 \ms, and 198 \ms\, (respectively). Those $rms$ are larger than the typical photon noise uncertainty on these measurements. We find a hint of RV variation with K = 89 $\pm$ 12 \ms\, assuming a circular orbit at the transit ephemeris. There is also a hint of correlation between the bisector and the RV. The FWHM also shows some variations. 

\begin{figure}[h]
\begin{center}
\includegraphics[width=\figw]{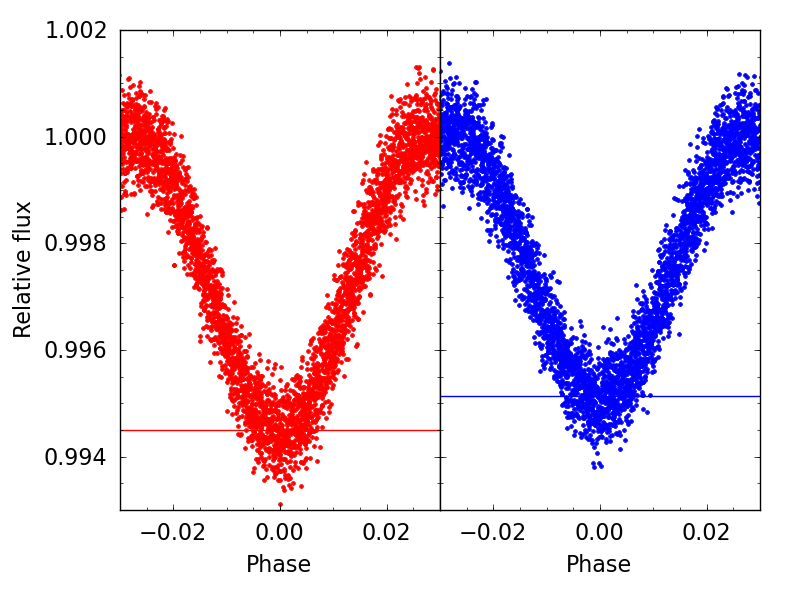}
\caption{Odd (left) and even (right) transits of the candidate KOI-6235.01.}
\label{fig6235}
\end{center}
\end{figure}

\citet{2014ApJS..211....2H} reported that the host is a giant star, with a radius of R$_{1}$ = 4.09$^{_{+2.76}}_{^{-1.12}}$ \Rsun. Therefore, with a transit depth of 0.5\%, the transiting companion would have a radius of R$_{2}$ = 0.38 $\pm$ 0.16 \Rsun. If the transit occurs on the target star, it would be too deep to be compatible with a planet. Moreover, by analysing the transits of this candidate, we found a significant odd -- even depth difference (see Fig. \ref{fig6235}), revealing that this candidate is clearly not a transiting planet but most likely a triple system with a giant primary and a nearly-equal mass EB with a period of 4.1 days. 

\subsection{KOI-6251.01}
\label{6251}

The giant-planet candidate KOI-6251.01 was found by \citet{2015arXiv150202038M} with an orbital period of 15 days. The first estimate of its transit depth was of 0.5\% \citep{2015arXiv150202038M}, but was later on revised to slightly above 3\%. This candidate was initially in our candidate list and observed twice with SOPHIE HE. Finally, since its transit depth is now outside our selection criteria, this candidate is no longer in our sample. We find a wide line profile in the cross-correlation function that we fitted with a rotation profile as in \citet{2012A&A...544L..12S} which gives a \vsini = 19.1 $\pm$ 0.3 \kms. The derived RV present a large variation in phase with the \textit{Kepler} ephemeris. By fitting a circular orbit at the transit ephemeris, we find an amplitude of K = 11.68 $\pm$ 0.70 \kms. Assuming a host mass of M$_{1}$ = 1.98$^{_{+0.17}}_{^{-0.47}}$ \Msun\, \citep{2014ApJS..211....2H}, the companion has a mass of M$_{2}$ = 0.21 $\pm$ 0.03 \Msun. Therefore, this candidate is not a transiting planet but an EB.

\subsection{KOI-6460.01}
\label{6460}

A giant-planet candidate was found by Coughlin et al. (in prep.) to transit the target star KOI-6460 every 1.2 days. This candidate was not reported in previous KOIs released and thus, was not included in the sample of \citet{2012AA...545A..76S}. The DV summary shows that this candidate is actually a slightly eccentric EB at twice the reported orbital period. The true primary eclipse depth is of about 20\%. We conclude this candidate is clearly not a transiting planet but an EB. For this reason, we did not observed it with SOPHIE.

\subsection{KOI-6602.01}
\label{6602}

A giant-planet candidate was found by Coughlin et al. (in prep.) to transit the target star KOI-6602 every 0.65 days. This candidate was not reported in previous KOIs released and thus, was not included in the sample of \citet{2012AA...545A..76S}. The DV summary shows that this candidate is actually an EB with a primary eclipse depth is of about 10\%. We conclude this candidate is clearly not a transiting planet but an EB. For this reason, we did not observed it with SOPHIE.

\subsection{KOI-6800.01}
\label{6800}

A giant-planet candidate was found by Coughlin et al. (in prep.) to transit the target star KOI-6800 every 2.54 days. This candidate was not reported in previous KOIs released and thus, was not included in the sample of \citet{2012AA...545A..76S}. The DV summary shows that this candidate is actually a slightly eccentric EB at twice the reported orbital period. We conclude this candidate is clearly not a transiting planet but an EB. For this reason, we did not observed it with SOPHIE.

\subsection{KOI-6877.01}
\label{6877}

A giant-planet candidate was found by Coughlin et al. (in prep.) to transit the target star KOI-6877 every 281 days. The host star is actually a red-giant star with a radius of almost R = 9.4 $\pm$ 0.1 \Rsun\, \citep{2013PhDT.......306B}. With a transit depth of about 0.8\%, the companion would have a radius compatible with a K dwarf and not with the one of an EGP. Moreover, the transit has a box shaped and could actually be the secondary eclipse of this EB. For this reason, we did not observed it with SOPHIE.

\subsection{KOI-6933.01}
\label{6933}

A giant-planet candidate was found by Coughlin et al. (in prep.) to transit the target star KOI-6933 every 7.2 days. A visual inspection of the DV summary reveal that this target is a clear EB with an eclipse depth of more than 40\%. We conclude this candidate is clearly not a transiting planet but an EB. For this reason, we did not observed it with SOPHIE.

\subsection{KOI-7044.01}
\label{7044}

A giant-planet candidate was found by Coughlin et al. (in prep.) to transit the target star KOI-7044 every 1.3 day. The host star is actually a red-giant star with a radius of about R = 13 \Rsun\, \citep{2014ApJS..211....2H}. With a transit depth of about 0.7\%, the companion would have a radius compatible with a G dwarf and not with the one of an EGP. Moreover, the DV summary shows a odds -- even transit depth difference indicating that this system is a CEB with twice the reported period. For this reason, we did not observed it with SOPHIE.

\subsection{KOI-7054.01}
\label{7054}

A giant-planet candidate was found by Coughlin et al. (in prep.) to transit the target star KOI-7054 every 0.53 day. This candidate was not reported in previous KOIs released and thus, was not included in the sample of \citet{2012AA...545A..76S}. A visual inspection of the light-curve showed that this candidate is actually an EB at twice the reported orbital period with a primary eclipse depth of about 50\%. We conclude this candidate is clearly not a transiting planet but an EB. For this reason, we did not observed it with SOPHIE.

\subsection{KOI-7065.01}
\label{7065}

A giant-planet candidate was found by Coughlin et al. (in prep.) to transit the target star KOI-7065 every 0.74 day. A visual inspection of the DV summary reveal that this target is a clear EB with a deep secondary eclipse. We conclude this candidate is clearly not a transiting planet but an EB. For this reason, we did not observed it with SOPHIE.





\subsection{KOI-7527.01}
\label{7527}

A giant-planet candidate was found by Coughlin et al. (in prep.) to transit the target star KOI-7527 every 1.3 days. This candidate was not reported in previous KOIs released and thus, was not included in the sample of \citet{2012AA...545A..76S}. A visual inspection of the light-curve revealed the presence of a clear secondary eclipse with a depth of about 0.8\%. We conclude this candidate is clearly not a transiting planet but an EB. For this reason, we did not observed it with SOPHIE.

\onecolumn

\section{SOPHIE data of the \textit{Kepler} targets}
\label{sophiedata}

\subsection{Single stars and SB1}

\begin{longtable}{cccccccccl}
\caption{SOPHIE RV of single stars}\\
\hline
\hline
BJD & RV & $\sigma_{\rm RV}$ & V$_{\rm span}$ & $\sigma_{\rm V_{\rm span}}$ & FWHM & $\sigma_{\rm FWHM}$ & Texp & S/N & Target\\
UTC & [\kms] & [\kms] & [\kms] & [\kms] & [\kms] & [\kms] & [s] & & \\
\hline
\endfirsthead
\caption{Continued.} \\
\hline
\hline
BJD & RV & $\sigma_{\rm RV}$ & V$_{\rm span}$ & $\sigma_{\rm V_{\rm span}}$ & FWHM & $\sigma_{\rm FWHM}$ & Texp & S/N  & Target \\
\hline
\endhead
\hline
\hline
\endfoot
\hline
\hline
\endlastfoot
2456869.41963  &  9.546  &  0.060  &  -0.454  &  0.108  &  27.180  &  0.150  &  2346  &  23.5 & KOI-129\\
2456933.33782  &  20.963  &  0.055  &  -0.421  &  0.098  &  27.042  &  0.137  &  2400  &  27.7 & KOI-129 \\
\hline
2456134.52471  &  -26.094  &  0.068  &  0.627  &  0.123  &  30.522  &  0.171  &  785  &  12.0 & KOI-138\\
2456157.52875  &  6.967  &  0.058  &  -0.239  &  0.104  &  29.826  &  0.144  &  3600  &  28.8 & KOI-138 \\
2456188.45296  &  -35.737  &  0.079  &  -0.054  &  0.143  &  31.192  &  0.198  &  900  &  11.5 & KOI-138\\
2456417.55490  &  -7.411  &  0.083  &  0.768  &  0.150  &  30.714  &  0.208  &  1800  &  7.4 & KOI-138\\
2456537.56957  &  -26.456  &  0.059  &  0.588  &  0.107  &  32.691  &  0.148  &  1802  &  8.4 & KOI-138\\
\hline
2456870.58675  &  -2.369  &  0.027  &  -0.017  &  0.049  &  9.954  &  0.068  &  1388  &  11.4  & KOI-198\\
\end{longtable}

\vspace{0.5cm}

\subsection{SB2}

\begin{longtable}{cccccccl}
\caption{SOPHIE RV of SB2 stars}\\
\hline
\hline
BJD & RV$_A$ & $\sigma_{\rm RV_A}$ & RV$_B$ & $\sigma_{\rm RV_B}$  &  Texp & S/N & Target\\
UTC & [\kms] & [\kms] & [\kms] & [\kms]  & [s] & & \\
\hline
\endfirsthead
\caption{Continued.} \\
\hline
\hline
BJD & RV$_A$ & $\sigma_{\rm RV_A}$ & RV$_B$ & $\sigma_{\rm RV_B}$  &  Texp & S/N & Target\\
\hline
\endhead
\hline
\hline
\endfoot
\hline
\hline
\endlastfoot
2456135.53044  &  -23.822  &  0.079  &  -107.521  &  0.137  &  900  &  8.2 & KOI-617\\
2456157.56114  &  -98.383  &  0.054  &  -27.082  &  0.087  &  1203  &  11.2  & KOI-617\\
2456417.53120  &  -81.286  &  0.111  &  -45.061  &  0.175  &  1800  &  7.5  & KOI-617\\
2456505.62902  &  -88.850  &  0.137  &  -37.426  &  0.216  &  892  &  7.2  & KOI-617\\
\hline
2457130.55767  &  -40.401  &  0.061  &  16.009  &  0.061  &  600  &  15.2 & KOI-969\\
2457136.58325  &  20.269  &  0.046  &  -44.876  &  0.046  &  600  &  20.0  & KOI-969\\
\hline
2456037.62360  &  -31.872  &  0.280  &  33.111  &  0.301  &  600  &  10.9 & KOI-1020\\
2456133.42558  &  58.992  &  0.178  &  -58.133  &  0.179  &  600  &  15.2  & KOI-1020\\
\end{longtable}%

\vspace{0.5cm}

\subsection{SB3}

\begin{longtable}{cccccccccl}
\caption{SOPHIE RV of SB3 stars}\\
\hline
\hline
BJD & RV$_{A}$ & $\sigma_{\rm RV_{A}}$ & RV$_{B}$ & $\sigma_{\rm RV_{B}}$ & RV$_{C}$ & $\sigma_{\rm RV_{C}}$ & Texp & S/N & Target\\
UTC & [\kms] & [\kms] & [\kms] & [\kms] & [\kms] & [\kms] & [s] & & \\
\hline
\endfirsthead
\caption{Continued.} \\
\hline
\hline
BJD & RV$_{A}$ & $\sigma_{\rm RV_{A}}$ & RV$_{B}$ & $\sigma_{\rm RV_{B}}$ & RV$_{C}$ & $\sigma_{\rm RV_{C}}$ & Texp & S/N & Target\\
\hline
\endhead
\hline
\hline
\endfoot
\hline
\hline
\endlastfoot
2455831.45089  &  -25.950  &  3.499  &  -38.169  &  0.704  &  -15.931  &  1.115  &  58  &  21.4 & KOI-976\\
2456476.48030  &  -21.515  &  2.643  &  -32.396  &  0.933  &  5.770  &  0.846  &  53  &  20.3  & KOI-976\\
\hline
2456867.61988  &  0.601  &  0.039  &  -86.584  &  0.040  &  -46.114  &  0.136  &  900  &  18.1 & KOI-5529\\
2457126.59385  &  -71.358  &  0.039  &  -12.303  &  0.040  &  -46.659  &  0.136  &  600  &  30.6 & KOI-5529
\label{RVSB3table}\\
\end{longtable}%

\vspace{0.5cm}

\section{SOPHIE data of the constant star HD185144}
\label{tableHD}

\begin{longtable}{ccc}
\caption{SOPHIE HE RV of the constant star HD185144 observed between 2011 and 2015.}\\
\hline
\hline
BJD & RV & $\sigma_{RV}$\\
UTC & [\kms] & [\kms]\\
\hline
\endfirsthead
\caption{Continued.} \\
\hline
BJD & RV & $\sigma_{RV}$\\
\hline
\endhead
\hline
\endfoot
\hline
\hline
\endlastfoot
2455751.426130  &  26.7605  &  0.0008 \\
2455751.427920  &  26.7614  &  0.0009 \\
2455751.429690  &  26.7594  &  0.0009 \\
$\vdots$ & $\vdots$ & $\vdots$\\
2457109.664570  &  26.7870  &  0.0013 \\
2457110.660530  &  26.7785  &  0.0012 \\
2457112.642380  &  26.7730  &  0.0008 
\label{table185144}
\end{longtable}%


\begin{thebibliography}{}
\bibitem[Adibekyan et al.(2012)]{2012A&A...547A..36A} Adibekyan, V.~Z., Delgado Mena, E., Sousa, S.~G., et al.\ 2012, \aap, 547, A36
\bibitem[Adibekyan et al.(2013)]{2013A&A...560A..51A} Adibekyan, V.~Z., Figueira, P., Santos, N.~C., et al.\ 2013, \aap, 560, A51
\bibitem[Almenara et al.(2015)]{2015A&A...575A..71A} Almenara, J.~M., Damiani, C., Bouchy, F., et al.\ 2015, \aap, 575, AA71
\bibitem[Andrews et al.(2013)]{2013ApJ...771..129A} Andrews, S.~M., Rosenfeld, K.~A., Kraus, A.~L., \& Wilner, D.~J.\ 2013, \apj, 771, 129 
\bibitem[Armitage \& Bonnell(2002)]{2002MNRAS.330L..11A} Armitage, P.~J., \& Bonnell, I.~A.\ 2002, \mnras, 330, L11
\bibitem[Asplund et al.(2009)]{2009ARA&A..47..481A} Asplund, M., Grevesse, N., Sauval, A.~J., \& Scott, P.\ 2009, \araa, 47, 481
\bibitem[Baglin et al.(2006)]{2006cosp...36.3749B} Baglin, A., Auvergne, M., Boisnard, L., et al.\ 2006, 36th COSPAR Scientific Assembly, 36, 3749
\bibitem[Bakos et al.(2004)]{2004PASP..116..266B} Bakos, G., Noyes, R.~W., Kov{\'a}cs, G., et al.\ 2004, \pasp, 116, 266
\bibitem[Bakos et al.(2010)]{2010ApJ...710.1724B} Bakos, G.~{\'A}., Torres, G., P{\'a}l, A., et al.\ 2010, \apj, 710, 1724
\bibitem[Baraffe et al.(2014)]{2014prpl.conf..763B} Baraffe, I., Chabrier, G., Fortney, J., \& Sotin, C.\ 2014, Protostars and Planets VI, 763
\bibitem[Baranne et al.(1996)]{1996A&AS..119..373B} Baranne, A., Queloz, D., Mayor, M., et al.\ 1996, \aaps, 119, 373
\bibitem[Barros et al.(2013)]{2013MNRAS.430.3032B} Barros, S.~C.~C., Bou{\'e}, G., Gibson, N.~P., et al.\ 2013, \mnras, 430, 3032
\bibitem[Barros et al.(2014)]{2014A&A...561L...1B} Barros, S.~C.~C., D{\'{\i}}az, R.~F., Santerne, A., et al.\ 2014, \aap, 561, L1
\bibitem[Batalha et al.(2010)]{2010ApJ...713L.103B} Batalha, N.~M., Rowe, J.~F., Gilliland, R.~L., et al.\ 2010, \apjl, 713, L103 
\bibitem[Batalha et al.(2013)]{2013ApJS..204...24B} Batalha, N.~M., Rowe, J.~F., Bryson, S.~T., et al.\ 2013, \apjs, 204, 24
\bibitem[Batalha(2014)]{2014PNAS..11112647B} Batalha, N.~M.\ 2014, Proceedings of the National Academy of Science, 111, 12647
\bibitem[Bayliss \& Sackett(2011)]{2011ApJ...743..103B} Bayliss, D.~D.~R., \& Sackett, P.~D.\ 2011, \apj, 743, 103
\bibitem[Beck(2013)]{2013PhDT.......306B} Beck, P.~G.\ 2013, Ph.D.~Thesis
\bibitem[Boisse et al.(2010)]{2010A&A...523A..88B} Boisse, I., Eggenberger, A., Santos, N.~C., et al.\ 2010, \aap, 523, A88
\bibitem[Bogn{\'a}r et al.(2015)]{2015arXiv150601668B} Bogn{\'a}r, Z., Lampens, P., Fr{\'e}mat, Y., et al.\ 2015, arXiv:1506.01668
\bibitem[Bonomo et al.(2010)]{2010A&A...520A..65B} Bonomo, A.~S., Santerne, A., Alonso, R., et al.\ 2010, \aap, 520, A65
\bibitem[Bonomo et al.(2012)]{2012A&A...538A..96B} Bonomo, A.~S., H{\'e}brard, G., Santerne, A., et al.\ 2012, \aap, 538, AA96
\bibitem[Bonomo et al.(2015)]{2015A&A...575A..85B} Bonomo, A.~S., Sozzetti, A., Santerne, A., et al.\ 2015, \aap, 575, A85
\bibitem[Borucki et al.(2009)]{2009Sci...325..709B} Borucki, W.~J., Koch, D., Jenkins, J., et al.\ 2009, Science, 325, 709
\bibitem[Borucki et al.(2011a)]{2011ApJ...728..117B} Borucki, W.~J., Koch, D.~G., Basri, G., et al.\ 2011a, \apj, 728, 117
\bibitem[Borucki et al.(2011b)]{2011ApJ...736...19B} Borucki, W.~J., Koch, D.~G., Basri, G., et al.\ 2011b, \apj, 736, 19
\bibitem[Bouchy et al.(2001)]{2001A&A...374..733B} Bouchy, F., Pepe, F., \& Queloz, D.\ 2001, \aap, 374, 733
\bibitem[Bouchy et al.(2005)]{2005A&A...431.1105B} Bouchy, F., Pont, F., Melo, C., et al.\ 2005, \aap, 431, 1105
\bibitem[Bouchy et al.(2009a)]{2009EAS....37..247B} Bouchy, F., Isambert, J., Lovis, C., Boisse, I., Figueira, P., H{\'e}brard, G., \& Pepe, F.\ 2009a, EAS Publications Series, 37, 247
\bibitem[Bouchy et al.(2009b)]{2009A&A...496..527B} Bouchy, F., Mayor, M., Lovis, C., et al.\ 2009b, \aap, 496, 527
\bibitem[Bouchy et al.(2009c)]{2009A&A...505..853B} Bouchy, F., H\'ebrard, G., Udry, S., et al.\ 2009c, \aap, 505, 853
\bibitem[Bouchy et al.(2011)]{2011A&A...533A..83B} Bouchy, F., Bonomo, A.~S., Santerne, A., et al.\ 2011, \aap, 533, AA83
\bibitem[Bouchy et al.(2013)]{2013A&A...549A..49B} Bouchy, F., D{\'{\i}}az, R.~F., H{\'e}brard, G., et al.\ 2013, \aap, 549, A49 
\bibitem[Bourrier et al.(2015)]{2015arXiv150404130B} Bourrier, V., Lecavelier des Etangs, A., H{\'e}brard, G., et al.\ 2015, \aap, 579, A55
\bibitem[Bradley et al.(2015)]{2015AJ....149...68B} Bradley, P.~A., Guzik, J.~A., Miles, L.~F., et al.\ 2015, \aj, 149, 68
\bibitem[Brown(2003)]{2003ApJ...593L.125B} Brown, T.~M.\ 2003, \apjl, 593, L125
\bibitem[Brown et al.(2011)]{2011AJ....142..112B} Brown, T.~M., Latham, D.~W., Everett, M.~E., \& Esquerdo, G.~A.\ 2011, \aj, 142, 112
\bibitem[Bruno et al.(2015)]{2015A&A...573A.124B} Bruno, G., Almenara, J.-M., Barros, S.~C.~C., et al.\ 2015, \aap, 573, AA124
\bibitem[Bryson et al.(2013)]{2013PASP..125..889B} Bryson, S.~T., Jenkins, J.~M., Gilliland, R.~L., et al.\ 2013, \pasp, 125, 889
\bibitem[Buchhave et al.(2014)]{2014Natur.509..593B} Buchhave, L.~A., Bizzarro, M., Latham, D.~W., et al.\ 2014, \nat, 509, 593
\bibitem[Burke et al.(2014)]{2014ApJS..210...19B} Burke, C.~J., Bryson, S.~T., Mullally, F., et al.\ 2014, \apjs, 210, 19
\bibitem[Burrows et al.(2007)]{2007ApJ...661..502B} Burrows, A., Hubeny, I., Budaj, J., \& Hubbard, W.~B.\ 2007, \apj, 661, 502
\bibitem[Cabrera et al.(2014)]{2014ApJ...781...18C} Cabrera, J., Csizmadia, S., Lehmann, H., et al.\ 2014, \apj, 781, 18
\bibitem[Cameron(2012)]{2012Natur.492...48C} Cameron, A.~C.\ 2012, \nat, 492, 48
\bibitem[Chabrier et al.(2014)]{2014prpl.conf..619C} Chabrier, G., Johansen, A., Janson, M., \& Rafikov, R.\ 2014, Protostars and Planets VI, 619
\bibitem[Charbonnel \& Palacios(2004)]{2004IAUS..215..440C} Charbonnel, C., \& Palacios, A.\ 2004, Stellar Rotation, 215, 440
\bibitem[Christiansen et al.(2013)]{2013ApJS..207...35C} Christiansen, J.~L., Clarke, B.~D., Burke, C.~J., et al.\ 2013, \apjs, 207, 35
\bibitem[Christiansen et al.(2015)]{2015arXiv150705097C} Christiansen, J.~L., Clarke, B.~D., Burke, C.~J., et al.\ 2015, \apj, 810, 95
\bibitem[Claret \& Bloemen(2011)]{2011A&A...529A..75C} Claret, A., \& Bloemen, S.\ 2011, \aap, 529, A75
\bibitem[Col{\'o}n et al.(2012)]{2012MNRAS.426..342C} Col{\'o}n, K.~D., Ford, E.~B., \& Morehead, R.~C.\ 2012, \mnras, 426, 342
\bibitem[Col{\'o}n et al.(2015)]{2015arXiv150607057C} Col{\'o}n, K.~D., Morehead, R.~C., \& Ford, E.~B.\ 2015, \mnras, 452, 3001
\bibitem[Cort{\'e}s et al.(2015)]{2015arXiv150602956C} Cort{\'e}s, C., Maciel, S.~C., Vieira, S., et al.\ 2015, arXiv:1506.02956
\bibitem[Courcol et al.(2015)]{2015arXiv150607144C} Courcol, B., Bouchy, F., Pepe, F., et al.\ 2015, \aap, 581, A38
\bibitem[Cox(2000)]{2000asqu.book.....C} Cox, A.~N.\ 2000, Allen's Astrophysical Quantities,
\bibitem[Csizmadia et al.(2015)]{2015arXiv150805763C} Csizmadia, S., Hatzes, A., Gandolfi, D., et al.\ 2015, arXiv:1508.05763
\bibitem[Dawson \& Murray-Clay(2013)]{2013ApJ...767L..24D} Dawson, R.~I., \& Murray-Clay, R.~A.\ 2013, \apjl, 767, L24
\bibitem[Dawson(2014)]{2014ApJ...790L..31D} Dawson, R.~I.\ 2014, \apjl, 790, L31
\bibitem[Dawson et al.(2014)]{2014ApJ...791...89D} Dawson, R.~I., Johnson, J.~A., Fabrycky, D.~C., et al.\ 2014, \apj, 791, 89
\bibitem[Deeg et al.(2010)]{2010Natur.464..384D} Deeg, H.~J., Moutou, C., Erikson, A., et al.\ 2010, \nat, 464, 384
\bibitem[Deleuil et al.(2014)]{2014A&A...564A..56D} Deleuil, M., Almenara, J.-M., Santerne, A., et al.\ 2014, \aap, 564, AA56
\bibitem[Delrez et al.(2014)]{2014A&A...563A.143D} Delrez, L., Van Grootel, V., Anderson, D.~R., et al.\ 2014, \aap, 563, A143
\bibitem[Demory \& Seager(2011)]{2011ApJS..197...12D} Demory, B.-O., \& Seager, S.\ 2011, \apjs, 197, 12
\bibitem[D{\'e}sert et al.(2011)]{2011ApJS..197...14D} D{\'e}sert, J.-M., Charbonneau, D., Demory, B.-O., et al.\ 2011, \apjs, 197, 14
\bibitem[D{\'e}sert et al.(2015)]{2015ApJ...804...59D} D{\'e}sert, J.-M., Charbonneau, D., Torres, G., et al.\ 2015, \apj, 804, 59
\bibitem[Devor et al.(2008)]{2008AJ....135..850D} Devor, J., Charbonneau, D., O'Donovan, F.~T., Mandushev, G., \& Torres, G.\ 2008, \aj, 135, 850
\bibitem[D{\'{\i}}az et al.(2012)]{2012A&A...538A.113D} D{\'{\i}}az, R.~F., Santerne, A., Sahlmann, J., et al.\ 2012, \aap, 538, A113
\bibitem[D{\'{\i}}az et al.(2013)]{2013A&A...551L...9D} D{\'{\i}}az, R.~F., Damiani, C., Deleuil, M., et al.\ 2013, \aap, 551, LL9
\bibitem[D{\'{\i}}az et al.(2014a)]{2014MNRAS.441..983D} D{\'{\i}}az, R.~F., Almenara, J.~M., Santerne, A., et al.\ 2014a, \mnras, 441, 983
\bibitem[D{\'{\i}}az et al.(2014b)]{2014A&A...572A.109D} D{\'{\i}}az, R.~F., Montagnier, G., Leconte, J., et al.\ 2014b, \aap, 572, AA109
\bibitem[Dong et al.(2014)]{2014ApJ...789L...3D} Dong, S., Zheng, Z., Zhu, Z., et al.\ 2014, \apjl, 789, L3
\bibitem[Dotter et al.(2008)]{2008ApJS..178...89D} Dotter, A., Chaboyer, B., Jevremovi{\'c}, D., et al.\ 2008, \apjs, 178, 89
\bibitem[Dressing \& Charbonneau(2013)]{2013ApJ...767...95D} Dressing, C.~D., \& Charbonneau, D.\ 2013, \apj, 767, 95
\bibitem[Dunham et al.(2010)]{2010ApJ...713L.136D} Dunham, E.~W., Borucki, W.~J., Koch, D.~G., et al.\ 2010, \apjl, 713, L136
\bibitem[Endl et al.(2011)]{2011ApJS..197...13E} Endl, M., MacQueen, P.~J., Cochran, W.~D., et al.\ 2011, \apjs, 197, 13
\bibitem[Endl et al.(2014)]{2014ApJ...795..151E} Endl, M., Caldwell, D.~A., Barclay, T., et al.\ 2014, \apj, 795, 151
\bibitem[Esteves et al.(2015)]{2015ApJ...804..150E} Esteves, L.~J., De Mooij, E.~J.~W., \& Jayawardhana, R.\ 2015, \apj, 804, 150
\bibitem[Faigler et al.(2013)]{2013ApJ...771...26F} Faigler, S., Tal-Or, L., Mazeh, T., Latham, D.~W., \& Buchhave, L.~A.\ 2013, \apj, 771, 26
\bibitem[Fischer \& Valenti(2005)]{2005ApJ...622.1102F} Fischer, D.~A., \& Valenti, J.\ 2005, \apj, 622, 1102
\bibitem[Fischer et al.(2012)]{2012MNRAS.419.2900F} Fischer, D.~A., Schwamb, M.~E., Schawinski, K., et al.\ 2012, \mnras, 419, 2900
\bibitem[Ford et al.(2012)]{2012ApJ...756..185F} Ford, E.~B., Ragozzine, D., Rowe, J.~F., et al.\ 2012, \apj, 756, 185
\bibitem[Fortney et al.(2011)]{2011ApJS..197....9F} Fortney, J.~J., Demory, B.-O., D{\'e}sert, J.-M., et al.\ 2011, \apjs, 197, 9
\bibitem[Fressin et al.(2013)]{2013ApJ...766...81F} Fressin, F., Torres, G., Charbonneau, D., et al.\ 2013, \apj, 766, 81
\bibitem[Gaidos(2013)]{2013ApJ...770...90G} Gaidos, E.\ 2013, \apj, 770, 90 
\bibitem[Gandolfi et al.(2013)]{2013A&A...557A..74G} Gandolfi, D., Parviainen, H., Fridlund, M., et al.\ 2013, \aap, 557, AA74
\bibitem[Gazzano et al.(2010)]{2010A&A...523A..91G} Gazzano, J.-C., de Laverny, P., Deleuil, M., et al.\ 2010, \aap, 523, A91
\bibitem[Gillon et al.(2013)]{2013A&A...552A..82G} Gillon, M., Anderson, D.~R., Collier-Cameron, A., et al.\ 2013, \aap, 552, A82
\bibitem[Gott et al.(2001)]{2001ApJ...549....1G} Gott, J.~R., III, Vogeley, M.~S., Podariu, S., \& Ratra, B.\ 2001, \apj, 549, 1
\bibitem[Guillot et al.(1996)]{1996ApJ...459L..35G} Guillot, T., Burrows, A., Hubbard, W.~B., Lunine, J.~I., \& Saumon, D.\ 1996, \apjl, 459, L35
\bibitem[Guillot(2005)]{2005AREPS..33..493G} Guillot, T.\ 2005, Annual Review of Earth and Planetary Sciences, 33, 493
\bibitem[Guillot et al.(2006)]{2006A&A...453L..21G} Guillot, T., Santos, N.~C., Pont, F., et al.\ 2006, \aap, 453, L21
\bibitem[Hadden \& Lithwick(2014)]{2014ApJ...787...80H} Hadden, S., \& Lithwick, Y.\ 2014, \apj, 787, 80
\bibitem[Halbwachs et al.(2003)]{2003A&A...397..159H} Halbwachs, J.~L., Mayor, M., Udry, S., \& Arenou, F.\ 2003, \aap, 397, 159
\bibitem[Hatzes \& Rauer(subm.)]{2015arXiv150605097H} Hatzes, A.~P., Rauer, H. \ 2015, arXiv:1506.05097
\bibitem[Havel et al.(2011)]{2011A&A...531A...3H} Havel, M., Guillot, T., Valencia, D., \& Crida, A.\ 2011, \aap, 531, A3
\bibitem[H{\'e}brard et al.(2010)]{2010A&A...516A..95H} H{\'e}brard, G., D{\'e}sert, J.-M., D{\'{\i}}az, R.~F., et al.\ 2010, \aap, 516, A95
\bibitem[H{\'e}brard et al.(2013)]{2013A&A...554A.114H} H{\'e}brard, G., Almenara, J.-M., Santerne, A., et al.\ 2013, \aap, 554, AA114 
\bibitem[H{\'e}brard et al.(2014)]{2014A&A...572A..93H} H{\'e}brard, G., Santerne, A., Montagnier, G., et al.\ 2014, \aap, 572, AA93 
\bibitem[Holman et al.(2010)]{2010Sci...330...51H} Holman, M.~J., Fabrycky, D.~C., Ragozzine, D., et al.\ 2010, Science, 330, 51 
\bibitem[Hou et al.(2009)]{2009ApJ...702.1199H} Hou, A., Parker, L.~C., Harris, W.~E., \& Wilman, D.~J.\ 2009, \apj, 702, 1199
\bibitem[Howard et al.(2010)]{2010Sci...330..653H} Howard, A.~W., Marcy, G.~W., Johnson, J.~A., et al.\ 2010, Science, 330, 653
\bibitem[Howard et al.(2012)]{2012ApJS..201...15H} Howard, A.~W., Marcy, G.~W., Bryson, S.~T., et al.\ 2012, \apjs, 201, 15
\bibitem[Howell et al.(2014)]{2014PASP..126..398H} Howell, S.~B., Sobeck, C., Haas, M., et al.\ 2014, \pasp, 126, 398
\bibitem[Huber et al.(2014)]{2014ApJS..211....2H} Huber, D., Silva Aguirre, V., Matthews, J.~M., et al.\ 2014, \apjs, 211, 2
\bibitem[Jenkins et al.(2010)]{2010ApJ...724.1108J} Jenkins, J.~M., Borucki, W.~J., Koch, D.~G., et al.\ 2010, \apj, 724, 1108
\bibitem[Jenkins et al.(2015)]{2015AJ....150...56J} Jenkins, J.~M., Twicken, J.~D., Batalha, N.~M., et al.\ 2015, \aj, 150, 56
\bibitem[Jontof-Hutter et al.(2014)]{2014ApJ...785...15J} Jontof-Hutter, D., Lissauer, J.~J., Rowe, J.~F., \& Fabrycky, D.~C.\ 2014, \apj, 785, 15
\bibitem[Johnson et al.(2010)]{2010PASP..122..905J} Johnson, J.~A., Aller, K.~M., Howard, A.~W., \& Crepp, J.~R.\ 2010, \pasp, 122, 905
\bibitem[Johnson et al.(2012)]{2012AJ....143..111J} Johnson, J.~A., Gazak, J.~Z., Apps, K., et al.\ 2012, \aj, 143, 111
\bibitem[Kipping(2013)]{2013MNRAS.434L..51K} Kipping, D.~M.\ 2013, \mnras, 434, L51
\bibitem[Kipping(2014)]{2014MNRAS.440.2164K} Kipping, D.~M.\ 2014, \mnras, 440, 2164
\bibitem[Koch et al.(2010)]{2010ApJ...713L.131K} Koch, D.~G., Borucki, W.~J., Rowe, J.~F., et al.\ 2010, \apjl, 713, L131
\bibitem[Kolbl et al.(2015)]{2015AJ....149...18K} Kolbl, R., Marcy, G.~W., Isaacson, H., \& Howard, A.~W.\ 2015, \aj, 149, 18
\bibitem[Kozai(1962)]{1962AJ.....67..591K} Kozai, Y.\ 1962, \aj, 67, 591 
\bibitem[Kurucz(1993)]{1993KurCD..13.....K} Kurucz, R.\ 1993, ATLAS9 Stellar Atmosphere Programs and 2 km/s grid.~Kurucz CD-ROM No.~13.~ Cambridge, Mass.: Smithsonian Astrophysical Observatory, 1993., 13,
\bibitem[Lagarde et al.(2012)]{2012A&A...543A.108L} Lagarde, N., Decressin, T., Charbonnel, C., et al.\ 2012, \aap, 543, A108
\bibitem[Latham et al.(2010)]{2010ApJ...713L.140L} Latham, D.~W., Borucki, W.~J., Koch, D.~G., et al.\ 2010, \apjl, 713, L140
\bibitem[Latham et al.(2011)]{2011ApJ...732L..24L} Latham, D.~W., Rowe, J.~F., Quinn, S.~N., et al.\ 2011, \apjl, 732, L24
\bibitem[Lopez et al.(2012)]{2012ApJ...761...59L} Lopez, E.~D., Fortney, J.~J., \& Miller, N.\ 2012, \apj, 761, 59
\bibitem[Lidov(1962)]{1962P&SS....9..719L} Lidov, M.~L.\ 1962, \planss, 9, 719 
\bibitem[Lillo-Box et al.(2014)]{2014A&A...566A.103L} Lillo-Box, J., Barrado, D., \& Bouy, H.\ 2014, \aap, 566, A103
\bibitem[Lissauer et al.(2011)]{2011ApJS..197....8L} Lissauer, J.~J., Ragozzine, D., Fabrycky, D.~C., et al.\ 2011, \apjs, 197, 8
\bibitem[Lissauer et al.(2012)]{2012ApJ...750..112L} Lissauer, J.~J., Marcy, G.~W., Rowe, J.~F., et al.\ 2012, \apj, 750, 112
\bibitem[Lissauer et al.(2014)]{2014ApJ...784...44L} Lissauer, J.~J., Marcy, G.~W., Bryson, S.~T., et al.\ 2014, \apj, 784, 44
\bibitem[Lund et al.(2014)]{2014A&A...570A..54L} Lund, M.~N., Lundkvist, M., Silva Aguirre, V., et al.\ 2014, \aap, 570, A54
\bibitem[Ma \& Ge(2014)]{2014MNRAS.439.2781M} Ma, B., \& Ge, J.\ 2014, \mnras, 439, 2781
\bibitem[Mancini et al.(2015)]{2015arXiv150404625M} Mancini, L., Lillo-Box, J., Southworth, J., et al.\ 2015, arXiv:1504.04625
\bibitem[Masuda(2014)]{2014ApJ...783...53M} Masuda, K.\ 2014, \apj, 783, 53 
\bibitem[Mayor \& Queloz(1995)]{1995Natur.378..355M} Mayor, M., \& Queloz, D.\ 1995, \nat, 378, 355
\bibitem[Mayor et al.(subm.)]{2011arXiv1109.2497M} Mayor, M., Marmier, M., Lovis, C., et al.\ 2011, submitted to A\&A, arXiv:1109.2497
\bibitem[Miller \& Fortney(2011)]{2011ApJ...736L..29M} Miller, N., \& Fortney, J.~J.\ 2011, \apjl, 736, L29
\bibitem[Mordasini et al.(2009a)]{2009A&A...501.1139M} Mordasini, C., Alibert, Y., \& Benz, W.\ 2009a, \aap, 501, 1139
\bibitem[Mordasini et al.(2009b)]{2009A&A...501.1161M} Mordasini, C., Alibert, Y., Benz, W., \& Naef, D.\ 2009b, \aap, 501, 1161
\bibitem[Mordasini et al.(2012a)]{2012A&A...541A..97M} Mordasini, C., Alibert, Y., Benz, W., Klahr, H., \& Henning, T.\ 2012a, \aap, 541, A97
\bibitem[Mordasini et al.(2012b)]{2012A&A...547A.112M} Mordasini, C., Alibert, Y., Georgy, C., et al.\ 2012b, \aap, 547, A112
\bibitem[Morgan \& Keenan(1973)]{1973ARA&A..11...29M} Morgan, W.~W., \& Keenan, P.~C.\ 1973, \araa, 11, 29
\bibitem[Mortier et al.(2013)]{2013A&A...551A.112M} Mortier, A., Santos, N.~C., Sousa, S., et al.\ 2013, \aap, 551, A112
\bibitem[Mortier et al.(2014)]{2014A&A...572A..95M} Mortier, A., Sousa, S.~G., Adibekyan, V.~Z., Brand{\~a}o, I.~M., \& Santos, N.~C.\ 2014, \aap, 572, AA95
\bibitem[Morton \& Johnson(2011)]{2011ApJ...738..170M} Morton, T.~D., \& Johnson, J.~A.\ 2011, \apj, 738, 170
\bibitem[Morton(2012)]{2012ApJ...761....6M} Morton, T.~D.\ 2012, \apj, 761, 6
\bibitem[Moutou et al.(2009)]{2009A&A...498L...5M} Moutou, C., H{\'e}brard, G., Bouchy, F., et al.\ 2009, \aap, 498, L5
\bibitem[Moutou et al.(2013)]{2013A&A...558L...6M} Moutou, C., Bonomo, A.~S., Bruno, G., et al.\ 2013, \aap, 558, LL6
\bibitem[Moutou et al.(2013)]{2013Icar..226.1625M} Moutou, C., Deleuil, M., Guillot, T., et al.\ 2013, \icarus, 226, 1625
\bibitem[Mullally et al.(2015)]{2015arXiv150202038M} Mullally, F., Coughlin, J.~L., Thompson, S.~E., et al.\ 2015, arXiv:1502.02038
\bibitem[M{\"u}ller et al.(2013)]{2013A&A...560A.112M} M{\"u}ller, H.~M., Huber, K.~F., Czesla, S., Wolter, U., \& Schmitt, J.~H.~M.~M.\ 2013, \aap, 560, A112
\bibitem[Nayakshin(2014)]{2014arXiv1411.5264N} Nayakshin, S.\ 2014, arXiv:1411.5264 
\bibitem[Nayakshin(2015)]{2015arXiv150207585N} Nayakshin, S.\ 2015, arXiv:1502.07585
\bibitem[O'Donovan et al.(2006)]{2006ApJ...651L..61O} O'Donovan, F.~T., Charbonneau, D., Mandushev, G., et al.\ 2006, \apjl, 651, L61 
\bibitem[Ofir et al.(2014)]{2014A&A...561A.103O} Ofir, A., Dreizler, S., Zechmeister, M., \& Husser, T.-O.\ 2014, \aap, 561, AA103
\bibitem[Oshagh et al.(2013)]{2013A&A...556A..19O} Oshagh, M., Santos, N.~C., Boisse, I., et al.\ 2013, \aap, 556, AA19
\bibitem[Owen \& Wu(2013)]{2013ApJ...775..105O} Owen, J.~E., \& Wu, Y.\ 2013, \apj, 775, 105
\bibitem[P{\'a}l et al.(2008)]{2008ApJ...680.1450P} P{\'a}l, A., Bakos, G.~\'A., Torres, G., et al.\ 2008, \apj, 680, 1450
\bibitem[Pecaut \& Mamajek(2013)]{2013ApJS..208....9P} Pecaut, M.~J., \& Mamajek, E.~E.\ 2013, \apjs, 208, 9
\bibitem[Pepe et al.(2002)]{2002A&A...388..632P} Pepe, F., Mayor, M., Galland, F., et al. \ 2002, \aap, 388, 632
\bibitem[Perruchot et al.(2008)]{2008SPIE.7014E..17P} Perruchot, S., Kohler, D., Bouchy, F., et al.\ 2008, \procspie, 7014
\bibitem[Perruchot et al.(2011)]{2011SPIE.8151E..37P} Perruchot, S., Bouchy, F., Chazelas, B., et al.\ 2011, \procspie, 8151, 
\bibitem[Petigura et al.(2013)]{2013PNAS..11019273P} Petigura, E.~A., Howard, A.~W., \& Marcy, G.~W.\ 2013, Proceedings of the National Academy of Science, 110, 19273
\bibitem[Pollacco et al.(2006)]{2006PASP..118.1407P} Pollacco, D.~L., Skillen, I., Collier Cameron, A., et al.\ 2006, \pasp, 118, 1407
\bibitem[Raghavan et al.(2010)]{2010ApJS..190....1R} Raghavan, D., McAlister, H.~A., Henry, T.~J., et al.\ 2010, \apjs, 190, 1
\bibitem[Ranc et al.(2015)]{2015A&A...580A.125R} Ranc, C., Cassan, A., Albrow, M.~D., et al.\ 2015, \aap, 580, A125
\bibitem[Rauer et al.(2014)]{2014ExA....38..249R} Rauer, H., Catala, C., Aerts, C., et al.\ 2014, Experimental Astronomy, 38, 249
\bibitem[Ricker et al.(2015)]{2015JATIS...1a4003R} Ricker, G.~R., Winn, J.~N., Vanderspek, R., et al.\ 2015, Journal of Astronomical Telescopes, Instruments, and Systems, 1, 014003
\bibitem[Rowe et al.(2014)]{2014ApJ...784...45R} Rowe, J.~F., Bryson, S.~T., Marcy, G.~W., et al.\ 2014, \apj, 784, 45
\bibitem[Rowe et al.(2015)]{2015ApJS..217...16R} Rowe, J.~F., Coughlin, J.~L., Antoci, V., et al.\ 2015, \apjs, 217, 16
\bibitem[Rowe \& Thompson(2015)]{2015arXiv150400707R} Rowe, J.~F., \& Thompson, S.~E.\ 2015, arXiv:1504.00707
\bibitem[Sanchis-Ojeda et al.(2013)]{2013ApJ...775...54S} Sanchis-Ojeda, R., Winn, J.~N., Marcy, G.~W., et al.\ 2013, \apj, 775, 54
\bibitem[Santerne et al.(2011a)]{2011A&A...528A..63S} Santerne, A., D{\'{\i}}az, R.~F., Bouchy, F., et al.\ 2011a, \aap, 528, A63
\bibitem[Santerne et al.(2011b)]{2011A&A...536A..70S} Santerne, A., Bonomo, A.~S., H{\'e}brard, G., et al.\ 2011b, \aap, 536, AA70
\bibitem[Santerne et al.(2012a)]{2012A&A...544L..12S} Santerne, A., Moutou, C., Barros, S.~C.~C., et al.\ 2012a, \aap, 544, L12
\bibitem[Santerne et al.(2012b)]{2012AA...545A..76S} Santerne, A., D{\'{\i}}az, R.~F., Moutou, C., et al.\ 2012b, \aap, 545, AA76
\bibitem[Santerne(2012)]{SanternePhDThesis} Santerne, A. \ 2012, Ph.D.~Thesis, Aix-Marseille University
\bibitem[Santerne et al.(2013a)]{2013A&A...557A.139S} Santerne, A., Fressin, F., D{\'{\i}}az, R.~F., et al.\ 2013a, \aap, 557, AA139
\bibitem[Santerne et al.(2013b)]{2013sf2a.conf..555S} Santerne, A., D{\'{\i}}az, R.~F., Almenara, J.-M., et al.\ 2013b, SF2A-2013: Proceedings of the Annual meeting of the French Society of Astronomy and Astrophysics, 555
\bibitem[Santerne et al.(2014)]{2014A&A...571A..37S} Santerne, A., H{\'e}brard, G., Deleuil, M., et al.\ 2014, \aap, 571, AA37
\bibitem[Santerne et al.(2015)]{2015MNRAS.451.2337S} Santerne, A., D{\'{\i}}az, R.~F., Almenara, J.-M., et al.\ 2015, \mnras, 451, 2337 
\bibitem[Santos et al.(2001)]{2001A&A...373.1019S} Santos, N.~C., Israelian, G., \& Mayor, M.\ 2001, \aap, 373, 1019
\bibitem[Santos et al.(2002)]{2002A&A...392..215S} Santos, N.~C., Mayor, M., Naef, D., et al.\ 2002, \aap, 392, 215
\bibitem[Santos et al.(2013)]{2013A&A...556A.150S} Santos, N.~C., Sousa, S.~G., Mortier, A., et al.\ 2013, \aap, 556, AA150
\bibitem[Schmitt et al.(2014a)]{2014AJ....148...28S} Schmitt, J.~R., Wang, J., Fischer, D.~A., et al.\ 2014a, \aj, 148, 28
\bibitem[Schmitt et al.(2014b)]{2014ApJ...795..167S} Schmitt, J.~R., Agol, E., Deck, K.~M., et al.\ 2014b, \apj, 795, 167
\bibitem[Schneider et al.(2011)]{2011A&A...532A..79S} Schneider, J., Dedieu, C., Le Sidaner, P., Savalle, R., \& Zolotukhin, I.\ 2011, \aap, 532, A79
\bibitem[Slawson et al.(2011)]{2011AJ....142..160S} Slawson, R.~W., Pr{\v s}a, A., Welsh, W.~F., et al.\ 2011, \aj, 142, 160
\bibitem[Sneden(1973)]{1973PhDT.......180S} Sneden, C.~A.\ 1973, Ph.D.~Thesis,
\bibitem[Sousa et al.(2008)]{2008A&A...487..373S} Sousa, S.~G., Santos, N.~C., Mayor, M., et al.\ 2008, \aap, 487, 373
\bibitem[Sousa et al.(2011)]{2011A&A...533A.141S} Sousa, S.~G., Santos, N.~C., Israelian, G., Mayor, M., \& Udry, S.\ 2011, \aap, 533, A141
\bibitem[Sousa et al.(2015)]{2015A&A...577A..67S} Sousa, S.~G., Santos, N.~C., Adibekyan, V., Delgado-Mena, E., \& Israelian, G.\ 2015, \aap, 577, A67
\bibitem[Southworth(2008)]{2008MNRAS.386.1644S} Southworth, J.\ 2008, \mnras, 386, 1644
\bibitem[Southworth(2011)]{2011MNRAS.417.2166S} Southworth, J.\ 2011, \mnras, 417, 2166
\bibitem[Steffen et al.(2013)]{2013MNRAS.428.1077S} Steffen, J.~H., Fabrycky, D.~C., Agol, E., et al.\ 2013, \mnras, 428, 1077
\bibitem[Shporer et al.(2011)]{2011AJ....142..195S} Shporer, A., Jenkins, J.~M., Rowe, J.~F., et al.\ 2011, \aj, 142, 195
\bibitem[Shporer et al.(2014)]{2014ApJ...788...92S} Shporer, A., O'Rourke, J.~G., Knutson, H.~A., et al.\ 2014, \apj, 788, 92
\bibitem[Torres et al.(2005)]{2005ApJ...619..558T} Torres, G., Konacki, M., Sasselov, D.~D., \& Jha, S.\ 2005, \apj, 619, 558
\bibitem[Torres et al.(2012)]{2012ApJ...757..161T} Torres, G., Fischer, D.~A., Sozzetti, A., et al.\ 2012, \apj, 757, 161
\bibitem[Triaud(2011)]{2011A&A...534L...6T} Triaud, A.~H.~M.~J.\ 2011, \aap, 534, L6 
\bibitem[Tsantaki et al.(2013)]{2013A&A...555A.150T} Tsantaki, M., Sousa, S.~G., Adibekyan, V.~Z., et al.\ 2013, \aap, 555, A150
\bibitem[Tingley et al.(2014)]{2014A&A...567A..14T} Tingley, B., Parviainen, H., Gandolfi, D., et al.\ 2014, \aap, 567, AA14
\bibitem[Udry et al.(2003)]{2003A&A...407..369U} Udry, S., Mayor, M., \& Santos, N.~C.\ 2003, \aap, 407, 369
\bibitem[Wang et al.(2013)]{2013ApJ...776...10W} Wang, J., Fischer, D.~A., Barclay, T., et al.\ 2013, \apj, 776, 10
\bibitem[Wang et al.(2015)]{2015ApJ...799..229W} Wang, J., Fischer, D.~A., Horch, E.~P., \& Huang, X.\ 2015, \apj, 799, 229
\bibitem[Wang \& Fischer(2015)]{2015AJ....149...14W} Wang, J., \& Fischer, D.~A.\ 2015, \aj, 149, 14
\bibitem[Weiss et al.(2013)]{2013ApJ...768...14W} Weiss, L.~M., Marcy, G.~W., Rowe, J.~F., et al.\ 2013, \apj, 768, 14
\bibitem[Weiss \& Marcy(2014)]{2014ApJ...783L...6W} Weiss, L.~M., \& Marcy, G.~W.\ 2014, \apjl, 783, LL6
\bibitem[Wilson(1941)]{1941ApJ....93...29W} Wilson, O.~C.\ 1941, \apj, 93, 29
\bibitem[Winn et al.(2010)]{2010ApJ...718L.145W} Winn, J.~N., Fabrycky, D., Albrecht, S., \& Johnson, J.~A.\ 2010, \apjl, 718, L145
\bibitem[Wright et al.(2012)]{2012ApJ...753..160W} Wright, J.~T., Marcy, G.~W., Howard, A.~W., et al.\ 2012, \apj, 753, 160
\bibitem[Wright et al.(2013)]{2013ApJ...770..119W} Wright, J.~T., Roy, A., Mahadevan, S., et al.\ 2013, \apj, 770, 119
\bibitem[Zhou \& Huang(2013)]{2013ApJ...776L..35Z} Zhou, G., \& Huang, C.~X.\ 2013, \apjl, 776, LL35
\end{thebibliography}
\end{document}